\documentclass[twocolumn]{aastex63}

\newcommand\rcr{$R_{\rm CR}$}
\newcommand\cotwo{\mbox{CO($J=2{-}1$)}}

\usepackage{amsmath, amstext}
\usepackage[figure,figure*]{hypcap}
\usepackage{times}
\usepackage[figuresright]{rotating}
\usepackage{afterpage}

\received{\today}
\revised{\today}
\accepted{\today}

\submitjournal{AJ}

\shorttitle{PHANGS Pattern Speeds}
\shortauthors{Williams et al.}

\graphicspath{{./}{figures/}}

\begin{document}


\title{Applying the Tremaine-Weinberg Method to Nearby Galaxies:\\Stellar Mass-Based Pattern Speeds, and Comparisons with ISM Kinematics}





\correspondingauthor{Thomas G. Williams}
\email{williams@mpia.de}

\author[0000-0002-0012-2142]{Thomas G. Williams}
\affiliation{Max Planck Institut f{\"u}r Astronomie, K{\"o}nigstuhl 17, 69117 Heidelberg, Germany}

\author[0000-0002-3933-7677]{Eva Schinnerer}
\affiliation{Max Planck Institut f{\"u}r Astronomie, K{\"o}nigstuhl 17, 69117 Heidelberg, Germany}

\author{Eric Emsellem}
\affiliation{European Southern Observatory, Karl-Schwarzschild-Str. 2, 85748 Garching, Germany}  
\affiliation{Universit{\'e} Lyon 1, ENS de Lyon, CNRS, Centre de Recherche Astrophysique de Lyon UMR5574, 69230 Saint-Genis-Laval, France}

\author[0000-0002-6118-4048]{Sharon Meidt}
\affiliation{Sterrenkundig Observatorium, Universiteit Gent, Krijgslaan 281 S9, B-9000 Gent, Belgium}

\author[0000-0002-0472-1011]{Miguel Querejeta}
\affiliation{Observatorio Astron{\'o}mico Nacional (IGN), C/Alfonso XII 3, Madrid E-28014, Spain}

\author[0000-0002-2545-5752]{Francesco Belfiore}
\affiliation{INAF -- Osservatorio Astrofisico di Arcetri, Largo E. Fermi 5, I-50157, Firenze, Italy}

\author{Ivana~Be\v{s}li{\'c}}
\affiliation{Argelander-Institut f\"{u}r Astronomie, Universit\"{a}t Bonn, Auf dem H\"{u}gel 71, 53121 Bonn, Germany}

\author{Frank Bigiel}
\affiliation{Argelander-Institut f\"{u}r Astronomie, Universit\"{a}t Bonn, Auf dem H\"{u}gel 71, 53121 Bonn, Germany}

\author[0000-0002-5635-5180]{M{\'e}lanie Chevance}
\affiliation{Astronomisches Rechen-Institut, Zentrum f\"{u}r Astronomie der Universit\"{a}t Heidelberg, M\"{o}nchhofstra\ss e 12-14, 69120 Heidelberg, Germany}

\author[0000-0002-5782-9093]{Daniel A. Dale} 
\affiliation{Department of Physics \& Astronomy, University of Wyoming, 1000 East University Avenue, Laramie, WY 82070, USA}

\author[0000-0001-6708-1317]{Simon C. O. Glover}
\affiliation{Universit\"{a}t Heidelberg, Zentrum f\"{u}r Astronomie, Institut f\"{u}r Theoretische Astrophysik,  Albert-Ueberle-Str. 2, 69120 Heidelberg, Germany}

\author[0000-0002-3247-5321]{Kathryn Grasha}
\affiliation{Research School of Astronomy and Astrophysics, Australian National University, Canberra, ACT 2611, Australia}

\author[0000-0002-0560-3172]{Ralf S.~Klessen}
\affiliation{Universit\"{a}t Heidelberg, Zentrum f\"{u}r Astronomie, Institut f\"{u}r Theoretische Astrophysik,  Albert-Ueberle-Str. 2, 69120 Heidelberg, Germany}
\affiliation{Universit\"{a}t Heidelberg, Interdisziplin\"{a}res Zentrum f\"{u}r Wissenschaftliches Rechnen, Im Neuenheimer Feld 205, 69120 Heidelberg, Germany}

\author[0000-0002-8804-0212]{J.~M.~Diederik Kruijssen}
\affiliation{Astronomisches Rechen-Institut, Zentrum f\"{u}r Astronomie der Universit\"{a}t Heidelberg, M\"{o}nchhofstra\ss e 12-14, 69120 Heidelberg, Germany}

\author[0000-0002-2545-1700]{Adam K. Leroy}
\affiliation{Department of Astronomy, The Ohio State University, 140 West 18$^{\rm th}$ Avenue, Columbus, OH 43210, USA}

\author{Hsi-An Pan}
\affiliation{Max Planck Institut f{\"u}r Astronomie, K{\"o}nigstuhl 17, 69117 Heidelberg, Germany}

\author[0000-0003-3061-6546]{J{\'e}r{\^o}me Pety}
\affiliation{IRAM, 300 rue de la Piscine, F-38406 Saint Martin d'H{\'e}res, France} 

\affiliation{15 LERMA, Observatoire de Paris, PSL Research University, CNRS, Sorbonne Universit{\'e}s, 75014 Paris, France}

\author{Ismael Pessa}
\affiliation{Max Planck Institut f{\"u}r Astronomie, K{\"o}nigstuhl 17, 69117 Heidelberg, Germany}

\author[0000-0002-5204-2259]{Erik Rosolowsky}
\affiliation{4-183 CCIS, University of Alberta, Edmonton, Alberta, Canada}

\author[0000-0002-2501-9328]{Toshiki Saito}
\affiliation{Max Planck Institut f{\"u}r Astronomie, K{\"o}nigstuhl 17, 69117 Heidelberg, Germany}

\author[0000-0002-6363-9851]{Francesco Santoro}
\affiliation{Max Planck Institut f{\"u}r Astronomie, K{\"o}nigstuhl 17, 69117 Heidelberg, Germany}

\author{Andreas Schruba}
\affiliation{Max-Planck-Institut f\"ur Extraterrestrische Physik, Giessenbachstra{\ss}e~1, D-85748 Garching bei M\"unchen, Germany}

\author[0000-0001-6113-6241]{Mattia C. Sormani}
\affiliation{Universit\"{a}t Heidelberg, Zentrum f\"{u}r Astronomie, Institut f\"{u}r Theoretische Astrophysik,  Albert-Ueberle-Str. 2, 69120 Heidelberg, Germany}

\author[0000-0003-0378-4667]{Jiayi Sun} 
\affiliation{Department of Astronomy, The Ohio State University, 140 West 18$^{\rm th}$ Avenue, Columbus, OH 43210, USA}

\author[0000-0002-7365-5791]{Elizabeth J. Watkins}
\affiliation{Astronomisches Rechen-Institut, Zentrum f\"{u}r Astronomie der Universit\"{a}t Heidelberg, M\"{o}nchhofstra\ss e 12-14, 69120 Heidelberg, Germany}

\begin{abstract}
We apply the Tremaine-Weinberg method to 19 nearby galaxies using stellar mass surface densities and velocities derived from the PHANGS-MUSE survey, to calculate (primarily bar) pattern speeds ($\Omega_{\rm P}$). After quality checks, we find that around half (10) of these stellar mass-based measurements are reliable. For those galaxies, we find good agreement between our results and previously published pattern speeds, and use rotation curves to calculate major resonance locations (co-rotation radii and Lindblad resonances). We also compare these stellar-mass derived pattern speeds with H$\alpha$ (from MUSE) and \cotwo\ emission from the PHANGS-ALMA survey. We find that in the case of these clumpy ISM tracers, this method erroneously gives a signal that is simply the angular frequency at a representative radius set by the distribution of these clumps ($\Omega_{\rm clump}$), and that this $\Omega_{\rm clump}$ is significantly different to $\Omega_{\rm P}$ ($\sim$20\% in the case of H$\alpha$, and $\sim$50\% in the case of CO). Thus, we conclude that it is inadvisable to use ``pattern speeds'' derived from ISM kinematics. Finally, we compare our derived pattern speeds and co-rotation radii, along with bar properties, to the global parameters of these galaxies. Consistent with previous studies, we find that galaxies with a later Hubble type have a larger ratio of co-rotation radius to bar length, more molecular-gas rich galaxies have higher $\Omega_{\rm P}$, and more bulge-dominated galaxies have lower $\Omega_{\rm P}$. Unlike earlier works, however, there are no clear trends between the bar strength and $\Omega_{\rm P}$, nor between the total stellar mass surface density and the pattern speed.
\end{abstract}


\section{Introduction}\label{sec:intro}

A fundamental, and currently open question in galaxy formation and evolution is how spiral arms and bars are formed and how they evolve. A natural question to ask is how long-lived these structures are, but the answer to this question remains elusive; even ``transient'' structures in terms of galaxies may persist for millions of years \citep[e.g.][]{2005Bournaud,2012Grand}, and so are far beyond the timescales that we are able to observe. The most common theory for the formation of these structures \citep[][although see \citealt{2013Sellwood} and sect.~6.4.2 of \citealt{2008BinneyTremaine} for alternative theories]{1963Lindblad,1966LinShu} is that density waves propagating through galaxies act on the gas, forming stars either along spiral shock lines (forming grand-design spiral arms). Bars can form from disk instabilities, even in the absence of gas \citep[see sect.~6.3 of][] {2008BinneyTremaine}. A key prediction of this density wave theory is that the waves propagating through these morphological features will have a roughly invariant angular velocity across a large range of galactic radii. These angular velocities are referred to as pattern speeds, $\Omega_{\rm P}$ \citep[whether this is true in the case of spiral arms is disputed, see review by][]{2014Dobbs}. 

The pattern speed of a spiral arm or a bar is a key parameter of the structure, and is associated with the evolution of the galaxy it is present within. For instance, a bar can only grow self-consistently if it lies within the co-rotation radius \citep[i.e. where the stars move at the same speed as the density wave;][]{1980Contopoulos}. Density waves driving spiral arms have been shown to trigger star formation \citep[e.g.][]{1993Rand, 1996Knapen}, and the interface of a bar and a spiral arm can also trigger massive starburst events \citep[e.g.][]{2017Beuther}, so galaxy evolution determines $\Omega_{\rm P}$. Furthermore, when combined with the rotation curve of a galaxy, these speeds will set the location of resonances within a galaxy (the co-rotation radii, and Lindblad resonances), which in turn maintain and regulate the density wave \citep{1970Lin}, and can have significant effects on the distributions of stars \citep[e.g.][]{2019Fragkoudi}.

However, the density wave speed is not directly observable, so we must turn to indirect methods to infer this parameter. There are a number of ways to do this. For example, by identifying resonance locations one can predict the radius of co-rotation, and thus the pattern speed \citep[e.g.][]{1989Elmegreen,1996Elmegreen}. This method is limited by an uncertain conversion from resonance radius to co-rotation radius \cite[e.g.][]{2003Kranz}, and by the fact that given the particular pattern speed and galaxy rotation curve, certain resonances may not exist. Another option is to match simulations (where the pattern speed is directly known) to galaxies \cite[e.g.][]{1993GarciaBurillo,1999RautiainenSalo, 2001aWeiner, 2001bWeiner,2015Sormani}. The selection of the ``best'' model here is somewhat qualitative, as direct comparisons between observations and simulations are difficult. This method has the benefit of being more direct, but requires a suite of tailored simulations for each individual object (selecting a value of the stellar mass-to-light ratio, the shape of the dark matter halo and its mass, etc.), and so is only feasible for small samples of galaxies.

Due to these difficulties, studies are typically limited to a single galaxy (or a very small number of them). This means that the literature contains heterogeneous measures of pattern speeds using different methods. Given that these different methods have different systematics (or, perhaps are more sensitive to different pattern speeds within the galaxy), making direct comparisons between these various literature values is difficult. Furthermore, applying the same method to different kinematic tracers may yield differing results \citep[e.g. stars and H{\sc i}, see][]{1998Westpfahl}. To draw statistically robust conclusions about the distributions of pattern speeds in the local galaxy population, it is important to have homogeneous measures of these quantities -- not only in technique, but in tracer, too.

Ideally, to extend studies of pattern speeds (and resonance locations) to large samples of galaxies, we desire a method of pattern speed determination that is widely applicable, which is data (rather than model or simulation) driven, and as quantitative as possible (i.e.\ with minimal reliance on by-eye feature classification). For this, the Tremaine-Weinberg \citep{1984TremaineWeinberg} method stands out as one of the most popular approaches, due to its minimal model assumptions, and the fact that it is purely based on the observed kinematics. We will describe this method in more detail in Sect.~\ref{sec:tw_method}. As a brief historical introduction, this method was originally designed for long-slit spectroscopy and applied to measure bar pattern speeds of dust-free, early-type galaxies. It is sensitive to non-axisymmetry along a number of (ideally infinitely long) slits positioned along the major axis of the galaxy. By taking the slope of the intensity-weighted velocity with respect to the intensity-weighted position for a number of slits, the velocity component mis-alignment with the line of nodes (i.e. the major axis of the galaxy) can be quantified. The commonly used form of the method was first formulated by \cite{1995MerrifieldKuijken}, and has since been applied to a number of other galaxies, including late-type galaxies \citep[e.g.][amongst others]{1999Gerssen,2002Debattista,2003Aguerri,2003Corsini,2007Corsini,2019Guo}. With the advent of integral field unit (IFU) spectroscopy, it is now possible to apply the Tremaine-Weinberg method to derived maps of the stellar mass and stellar velocities \citep[see, e.g.][]{2004DebattistaWilliams,2019Guo}, and with the Gaia satellite it has been applied to stars in the Milky Way \citep{2019Sanders}.

The method has also been used in conjunction with ionized or neutral gas tracers, with the goal of probing non-axisymmetric structure further out in the disk, either with H$\alpha$ \citep[e.g.][]{2006Emsellem,2009Fathi} or occasionally CO \citep[e.g.][]{2004RandWallin,2004Zimmer}. The underlying assumption, in such cases, is that this gas roughly obeys continuity when there is little chemical transformation between the gas and other phases of the ISM (i.e. molecular to atomic hydrogen, or dust), and that the chemical abundance of the tracers remains constant.  In the case of CO, \cite{2004RandWallin} argued that it should remain a valid tracer when the ISM is molecule-dominated (so there is little conversion from molecular gas to and from atomic gas) and when the star formation rate is low to moderate \citep[so that there is little molecular gas lost to star formation, or expelled by feedback;][]{2020Chevance}. However, as discussed in \cite{2004RandWallin}, the clumpiness of ISM tracers like CO and H$\alpha$ can pose a further issue: clumpy, highly asymmetric disks introduce a fake signal in Tremaine-Weinberg integrals.  This effect becomes especially pronounced at high resolution, when CO and H$\alpha$ morphology becomes characterised by clumpiness organised around, for example, H{\sc ii} regions, GMCs, sharp spiral arms (e.g. \citealt{2018Kreckel,2019Schinnerer}; S.~Meidt et al.\ in prep.) We revisit this in more detail in the context of our sample in Sect.~\ref{sec:clumpy_ism}. Because of these potential shortcomings, fully sampled stellar kinematics are preferable for applying the Tremaine-Weinberg method.

Recent works have combined a number of literature values of pattern speeds to attempt to perform statistical analyses \citep{2020Cuomo}. Our study complements and builds upon this earlier work by measuring pattern speeds homogeneously for a number of galaxies. In this work, we apply the Tremaine-Weinberg method to two of the Physics at High Angular resolution in Nearby GalaxieS (PHANGS\footnote{\url{phangs.org}}) surveys, namely to observations taken using the Multi Unit Spectroscopic Explorer (MUSE) instrument on the VLT (referred to as PHANGS-MUSE; P.I.\ E.~Schinnerer; E.~Emsellem et al.\ in prep.), and the Atacama Large Millimeter/Submillimeter Array (ALMA) instrument \citep[referred to as PHANGS-ALMA; P.I.\ E.~Schinnerer, and from pilot proposals with P.I.s G.~Blanc and A.~K.\ Leroy;][]{PHANGSSurvey}. Our work focuses on pattern speeds from the stellar mass surface density ($\Sigma_{\ast}$) from MUSE, but we also study the application of this method to ISM tracers -- ionised gas (H$\alpha$) from MUSE, and cold, molecular gas [\cotwo] emission from ALMA. With 19 galaxies mapped as part of PHANGS-MUSE, and 84 with PHANGS-ALMA, this gives us an unprecedented opportunity not only to measure pattern speeds homogeneously for a large sample of galaxies at high ($\lesssim100$\,pc) resolution, but also to perform vital cross-checks between different kinematic tracers for many galaxies, which are currently poorly explored in the literature. This work derives pattern speeds, along with resonance locations for these galaxies, which are tabulated in Table~\ref{table:pattern_speeds}, and are also made available online in a machine-readable format\footnote{The code used in this work, as well as a .fits table, are available online at \url{https://github.com/thomaswilliamsastro/phangs_pattern_speeds}}. For future works that may have improved distance or orientation measurements, this table also includes these parameters as used in our work, to allow for simple rescaling in subsequent works. 

The structure of this paper is as follows: we provide an overview of the PHANGS programmes and data products (Sect.~\ref{sec:data}), before summarising the Tremaine-Weinberg method, our tests of its efficacy on the data, and our application to the entire PHANGS-MUSE and PHANGS-ALMA data sets (Sect.~\ref{sec:calculating_pattern_speeds}). In Sect.~\ref{sec:clumpy_ism}, we present our tests showing that applying the Tremaine-Weinberg method to ISM tracers can yield erroneous signals, and showing that this is generally an issue in our sample. We present an overview of our derived pattern speeds in the context of previous work, and make comparison between our pattern speeds and previously derived pattern speeds for the same galaxies in the literature (Sect.~\ref{sec:derived_pattern_speeds}). We then calculate the radii of major resonances for the entire PHANGS sample (Sect.~\ref{sec:resonance_radii}). We study the potential correlations between these derived dynamical properties of the galaxy and some of its global parameters (such as Hubble type, molecular gas fraction; Sect.~\ref{sec:correlations}). We discuss the implications of these results (Sect.~\ref{sec:discussion}), before presenting a summary of our work, along with future prospects for large studies of pattern speeds in the future (Sect.~\ref{sec:conclusions}).

\section{Data}\label{sec:data}

This work uses data from two of the PHANGS large programmes: PHANGS-MUSE and PHANGS-ALMA. For details on the data reduction and product creation, we refer the readers to E.~Emsellem et al.\ (in prep.) for PHANGS-MUSE, and \cite{PHANGSPipeline} and \cite{PHANGSSurvey} for the PHANGS-ALMA data processing pipeline and survey description, respectively. We provide here only a brief description of the data products we use in the proceeding sections, which are the surface brightness and velocity maps for each tracer (MUSE stellar mass and velocity, MUSE H$\alpha$, and ALMA CO, collectively referred to as ``kinematic maps''), along with associated error maps.

Along with these kinematic maps, we use orientation parameters (the galaxy position angle, inclination, and systemic velocities) from \cite{2020Lang}. We also make use of distances from \citet[][and references therein]{2020Anand} to calculate physical pattern speeds in ${\rm km\,s^{-1}\,kpc^{-1}}$, rather than in ${\rm km\,s^{-1}\,arcsec^{-1}}$. Finally, we make use of catalogues from the Spitzer Survey of Stellar Structure in Galaxies \citep[S$^4$G;][]{2010Sheth} -- in particular, bar strengths \citep{2016DiazGarcia}, bar orientations \citep{2015HerreraEndoqui}, and bulge to total flux ratios and disk scale lengths \citep{2015Salo}.

\subsection{PHANGS-MUSE}\label{sec:phangs_muse}

Our primary sample in this study consists of MUSE optical spectroscopy for 19 galaxies as part of the PHANGS-MUSE data release (DR)~2.0 (E.~Emsellem et al., in prep.). The reduction is performed using standard MUSE recipes (e.g.\ wavelength and flux calibration, cosmic ray rejection, mosaicking), run through {\sc pymusepipe}\footnote{\url{https://github.com/emsellem/pymusepipe}}. These reduced products are then run through the MUSE data analysis pipeline (DAP), which is run in three stages: firstly, stellar kinematics are measured (the stellar velocity and higher order moments); next, the properties of stellar populations are estimated (e.g.\ age, stellar metallicity, stellar mass). Both of these stages are performed on Voronoi binned data to a stellar continuum signal-to-noise ratio (S/N) of 35, to maximise reliability. The fit is performed via pPXF \citep{2004CappellariEmsellem,2017Cappellari}, and makes use of E-MILES \citep{2016Vazdekis} simple stellar population models of eight ages (0.15 - 14 Gyr) and four metallicities ([Z/H] = [-1.5,-0.35,0.06,0.4]). Only the wavelength range 4850-7000\AA\, are used in the fit, in order to avoid strong sky residuals in the reddest part of the MUSE spectral range. Finally, for individual spaxels the properties of emission lines are measured (fluxes and kinematics), via a simultaneous fit of continuum and emission lines also performed via pPXF. The DAP fits only a single stellar population, so the stellar masses and velocities are an average of young and older stars. We primarily want to apply the method to the old stellar population, as we do not expect young stars to obey continuity. As the stellar mass is dominated by old stars (typically, the mass-weighted age per spaxel is of the order of Gyr), and the velocities of young and old stars are typically similar \citep{2020RosadoBelza, 2020Shetty}, therefore, averaging these values over stellar populations will not bias our results. Finally, for individual spaxels, the properties of emission lines are measured (fluxes and kinematics). We make use of the stellar mass surface density and velocity maps (from the first two processing stages), and the H$\alpha$ flux and velocity maps (from the final processing stage).

\subsection{PHANGS-ALMA}\label{sec:phangs_alma}

We make use of $^{12}$\cotwo\ maps (hereafter ``CO'' for brevity) from the PHANGS-ALMA survey. Whilst the observations and data reduction are detailed separately \citep{PHANGSSurvey,PHANGSPipeline}, we provide a brief summary of the products used in this study.

These CO maps are provided as part of the internal PHANGS-ALMA DR v3.4. This includes a total of 84 galaxies (of which two are omitted due to no detected emission in the cubes). We use cubes combined with all available data from the 12m and 7m arrays, and total power (TP) observations. This means that we maximise the resolution of the observations, whilst still being sensitive to extended emission on all scales. In total, 69 of our galaxies have 12m+7m+TP data, 7~have 12m+7m, 7~have 7m+TP, and 1~has 7m only. Typically, the data without 12m antenna configurations are for the nearest galaxies, so the spatial resolution remains reasonably consistent. As shown in the right-hand panel of Fig. \ref{fig:omega_clump}, we do not expect the particular antenna configuration to have a strong impact on our results.

The 12m+7m data are imaged simultaneously, using a multi-scale clean followed by a single-scale clean. This combined cube is then corrected for primary beam pickup, and convolved to have a round synthesised beam. The TP data are imaged separately using the method described in \cite{2020Herrera}, and these two cubes are then combined in Fourier space (``feathered''). The cubes are then collapsed into standard moment~0 (integrated intensity) and moment~1 (intensity-weighted velocity), along with error maps. There are two types of masks used to create these maps: firstly, strict masks that grow out any voxels with $\mathrm{S/N} > 4$ in 2 or more successive channels down to voxels with $\mathrm{S/N} > 2$ in 2 or more successive channels. Secondly, there are broad masks that take the union of the highest resolution strict mask from a strict mask generated at a spatial resolution of 500pc, which captures lower level, extended emission. For more details of this masking, see \citet{PHANGSPipeline}. We opt to use broad, rather than strict maps, to maximise our completeness (at the cost of increased noise). We find a negligible impact on our results using these different masking schemes in most cases. This choice only matters for the low surface brightness galaxies in the sample, for which we typically find a poor fit in any case. 

We use maps at their highest resolution ($1.3^{+0.4}_{-0.2}$\,arcsec, corresponding to $100^{+31}_{-35}$\,pc, whereby the native resolution maps are smoothed to a common circular beam for each galaxy. Combined with the MUSE stellar and H$\alpha$ maps, we have three different kinematic tracers, which affords us useful cross-checks between derived pattern speeds for the sample of 19 galaxies where the observations overlap.

\section{Calculating Pattern Speeds}\label{sec:calculating_pattern_speeds}

\begin{figure*}[t]
\plotone{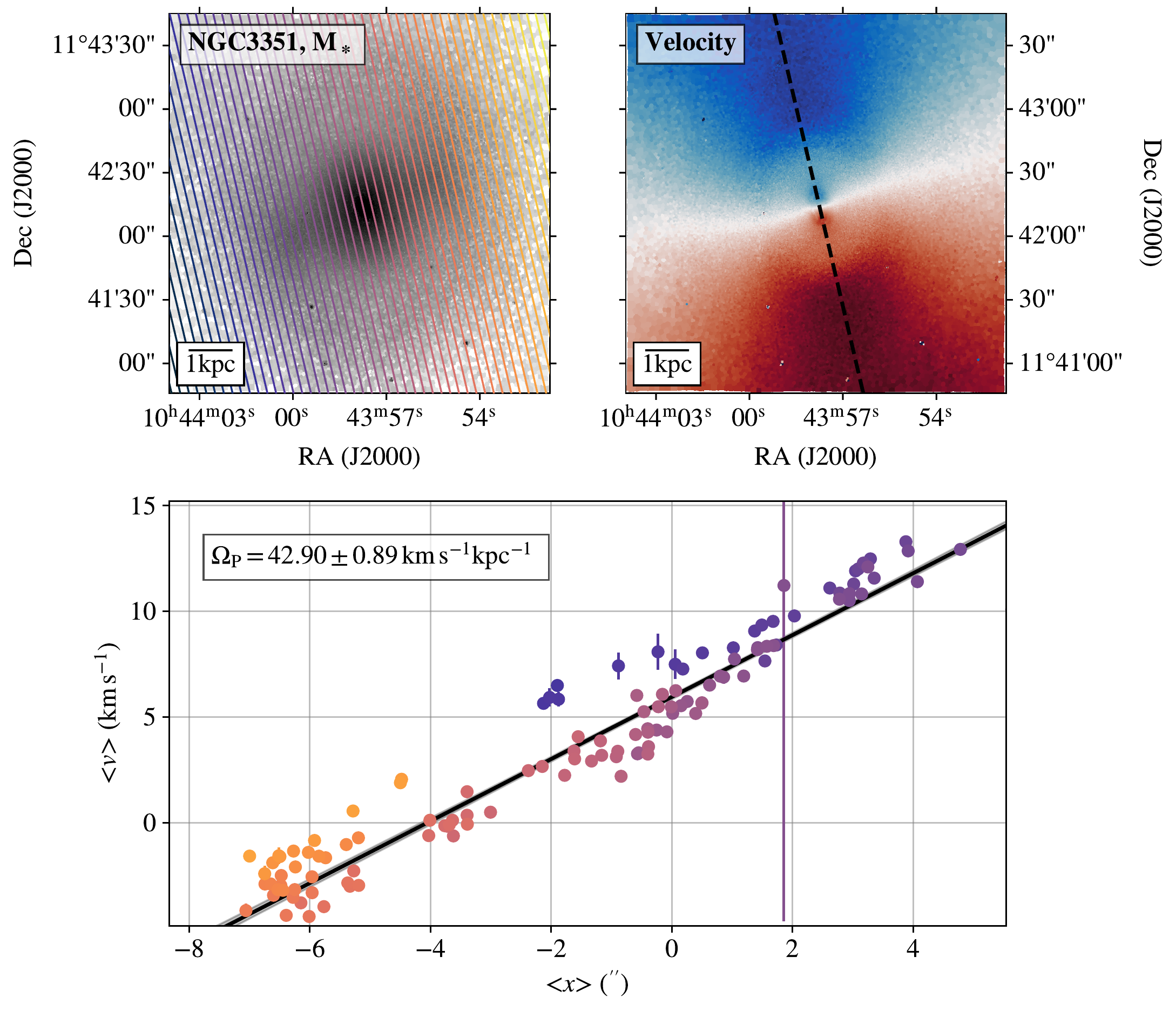}
\caption{{\it Top left}: Stellar mass surface density map of NGC~3351 shown in greyscale, with Tremaine-Weinberg integral slits of 1\arcsec~width, oriented parallel to the major axis overlaid. Only one in every four slits is shown, due to the slit density, and are coloured according to their position along the kinematic minor axis.For this galaxy, the quality flag, $Q = 1$ (see Sect. \ref{sec:quality_flagging}). {\it Top right}: Stellar velocity map for the same galaxy, with a dashed black line showing the kinematic major axis, passing through the galaxy centre. {\it Bottom}: intensity-weighted velocity ($\langle v \rangle$) versus intensity-weighted position ($\langle x \rangle$) for each of the slits (the colour corresponds to the slit colour in the above top-left panel). The black line shows the best fit, and the grey shaded region the errors on the fit (in this case, this region is extremely small). One point has an extremely large uncertainty in this panel, and the error bar extends across the entire range of $\langle v \rangle$ shown. \label{fig:ngc3351_tw_integral}}
\end{figure*}

\begin{figure*}[t]
\plottwo{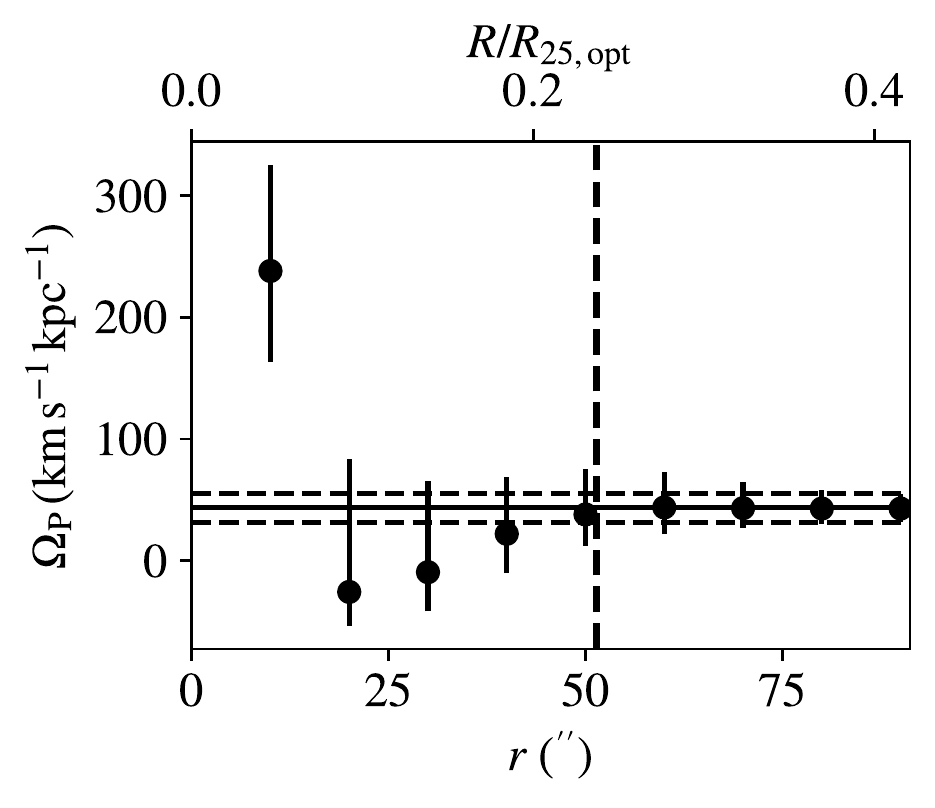}{{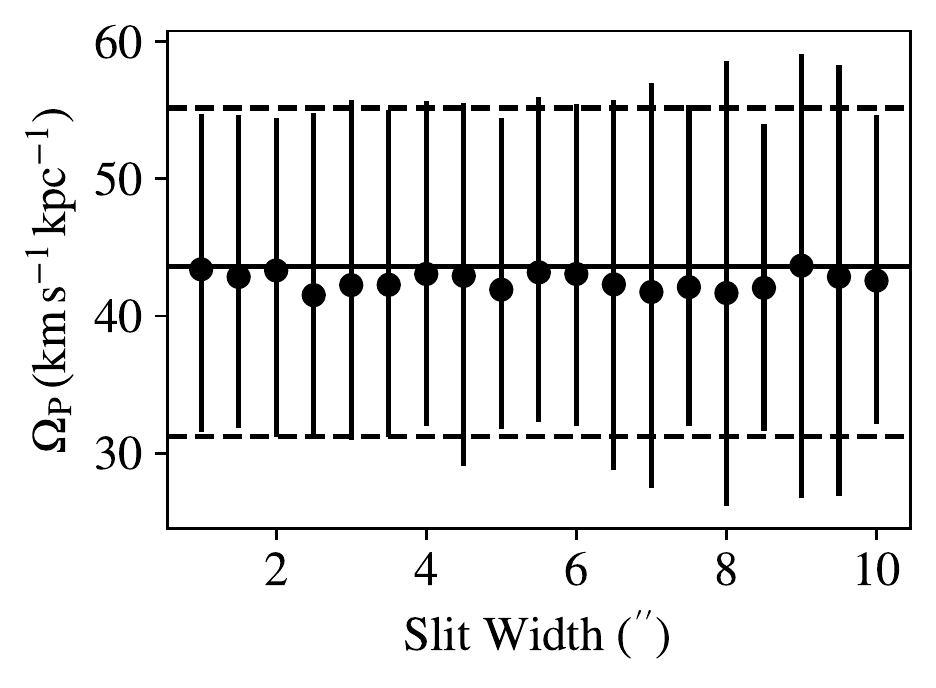}}
\caption{{\it Left}: Recovered pattern speed, $\Omega_{\rm P}$, versus Tremaine-Weinberg integral slit length, $r$, for MUSE stellar mass surface density observations of NGC~3351 (at this distance, $1\arcsec$ corresponds to ${\sim}50$\,pc). The vertical dashed black line indicates the bar extent in this galaxy. {\it Right}: Recovered pattern speed versus Tremaine-Weinberg integral slit width for the same galaxy (the resolution of this map is $\sim$1\arcsec). As the pattern speed measured for each slit width is not independent from other slit widths (they use the same data, and larger slit widths combine information from narrower slits), the scatter in each measurement is much smaller than the typical uncertainty. In both cases, the horizontal black line indicates our fiducial pattern speed, and the dashed black lines either side of it the uncertainty on this measurement. \label{fig:ngc3351_slit_length_width}}
\end{figure*}

In this section, we present a brief overview of the Tremaine-Weinberg method (Sect.~\ref{sec:tw_method}). We detail our treatments of the various uncertainties associated with the method (Sect.~\ref{sec:tw_errors}), and then investigate the effect of slit lengths and widths on the recovered pattern speeds (Sect.~\ref{sec:slit_length_width}). We then apply the method to our data set (Sect.~\ref{sec:data_prep}), and perform a~posteriori checks on the pattern speeds (Sect.~\ref{sec:quality_flagging}). We stress that we only provide one pattern speed; for barred galaxies, we expect this to be the pattern speed of the bar, and for non-barred galaxies this will be a pattern speed for the spiral arms. As an umbrella term, we will refer to these measured pattern speeds as ``primary'' pattern speeds throughout this work, as these structures will dominate the non-axisymmetry the Tremaine-Weinberg method is sensitive to when we are in a regime with multiple pattern speeds present (we discuss this further in Sect.~\ref{sec:tw_method}).

\subsection{The Tremaine-Weinberg Method}\label{sec:tw_method}

The Tremaine-Weinberg method is a model independent method to calculate a pattern speed within a galaxy. It has three, minimal assumptions:
\begin{enumerate}
    \item The galaxy disk is flat;
    \item The disk contains a single, well-defined pattern speed;
    \item The tracer obeys the continuity equation.
\end{enumerate}
Assumption (1) is justified as our observations cover mainly the inner disks of galaxies \citep[typically around 1 $R_{25}$;][]{2020Lang}, and the detected emission is less radially extended than, e.g. H{\sc i}. Thus, we do not expected these observations will be sensitive to disk warping. We therefore assume that this assumption holds true across all of our sample.

On point (2), prior studies have has shown that many galaxies host multiple pattern speeds, corresponding to different morphological features \citep[e.g. bars and spiral arms within a galaxy;][]{2008bMeidt, 2018Beckman}. In this case, assumption (2) may be invalid. We would expect two different (but linked) phenomena here. For slits that contain no information from the bar, we would expect to measure a different pattern speed than for slits with bar information, if the pattern speeds are different between spiral arms and bars. This is reflected in our quality flagging  (Sect.~\ref{sec:quality_flagging}), and we believe this to only  affect a small subset of our sample. 

Furthermore, as we make our slits as long as possible, some slits may pass through both bars and spiral arms. If these structures have different pattern speeds, then any pattern speed we measure will be some weighted average of these two distinct values. We expect this measurement to strongly deviate from the bar pattern speed only when the spatial and kinematic non-axisymmetry of the secondary pattern (i.e.\ spiral arms) is strong relative to the primary pattern speed, and also when this secondary pattern extends over a considerable portion of the slits \citep[for examples of this, see scenarios considered by][]{2008aMeidt}. For most of the galaxies in our sample, this does not appear to be the case; outer spirals tend to be both substantially dimmer and exhibit weaker streaming motions compared to the bars in our sample: whereas residual velocities from the fitted rotation curves by \citet{2020Lang} are typically 20\,km\,s$^{-1}$ or greater in our sample for bars, they reach a maximum of only ${\sim}10$\,km\,s$^{-1}$ in the spiral arms. Attempting to quantify the bias this may present would require detailed simulations of galaxies where we know a ground truth for the pattern speed. This is beyond the scope of this work, but will be revisited in later studies. For the present work, we assume that the secondary pattern is weak compared to the bar pattern, and so the bias is small -- much like in earlier works applying the Tremaine-Weinberg method to stellar kinematics \citep[e.g.][]{2015Aguerri, 2019Guo, 2020GarmaOehmichen}.

Whilst assumption~(3) (the tracer obeys the continuity equation) is approximately valid for the old stellar mass surface density distribution that we are primarily sensitive to with our MUSE observations, it may not be formally valid given the clumpy nature of CO and H$\alpha$ emission, the phase transitions from H{\i} to H$_2$ to H{\sc ii}, and the way these phases participate in the star formation process. We investigate this further in Sect.~\ref{sec:clumpy_ism}, and find that this is generally an issue within our sample. For this reason, the bulk of our analysis is performed on pattern speeds measured from stellar kinematics.

If the continuity equation is valid, then
\begin{equation}\label{eq:continuity_eq}
    \frac{\partial \Sigma}{\partial t} + \frac{\partial}{\partial x}(\Sigma v_x) + \frac{\partial}{\partial y}(\Sigma v_y) = 0 , 
\end{equation}
where $(v_x, v_y)$ is the mean velocity of the tracer at $(x, y)$, and $\Sigma$ the surface brightness at $(x, y)$. In polar coordinates $(r, \phi)$, we assume that the surface brightness ($\Tilde{\Sigma}$) in a frame rotating with an angular pattern speed, $\Omega_{\rm P}$:
\begin{equation}
    \Sigma = \Tilde{\Sigma}(r, \phi - \Omega_{\rm P} t),
\end{equation}
and from this we can write
\begin{equation}
    \frac{\partial \Sigma}{\partial t} = -\Omega_{\rm P} \frac{\partial \Tilde{\Sigma}}{\partial \phi} = \Omega_{\rm P} \left(y \frac{\partial \Sigma}{\partial x} - x \frac{\partial \Sigma}{\partial y} \right).
\end{equation}
Substituting this into Eq.~\ref{eq:continuity_eq} and integrating over both $x$ and $y$ \citep[see][]{1984TremaineWeinberg}, we can obtain
\begin{equation}
    \Omega_{\rm P} \sin(i) = 
    \frac{
    \int_{-\infty}^\infty h(y) 
    \int_{-\infty}^\infty \Sigma v_{\rm LOS}(x,y)\, {\rm d}x {\rm d}y
    }
    {\int_{-\infty}^\infty h(y)
    \int_{-\infty}^\infty \Sigma x\, {\rm d}x {\rm d}y
    },
\end{equation}
where $i$ is the inclination of the galaxy, $v_{\rm LOS}$ the line-of-sight velocity, and $h(y)$ a weight function. In this work, we take $h(y)$ to be a boxcar function from $y-{\rm d}y$ to $y+{\rm d}y$, to represent a pseudo-slit parallel to the line of nodes. Throughout this work, we will refer to ``slits'' and ``integrals'' interchangeably in this context. This equation can be simplified to
\begin{equation}
    \Omega_{\rm P} \sin(i) = \frac{\langle v \rangle}{\langle x \rangle}\,,
\end{equation}
by recognising that the integrals in the numerator and denominator are simply the intensity-weighted velocity and position along a slit, respectively. In this formalism, non-zero values of $\langle v \rangle/\langle x \rangle$ are caused by non-axisymmetric structure within the slit. Thus, taking a number of slits and plotting $\langle v \rangle$ versus $\langle x \rangle$ yields a straight line with a slope equivalent to $\Omega_{\rm P} \sin(i)$.

There are some limitations to this method given the data. Firstly, the integrals formally should extend from $-\infty$ to $\infty$, whereas in reality this is not the case due to the limited field of view of the observations. However, assuming the disk is axisymmetric at large $x$, we can instead integrate from $-x_0$ to $x_0$ (where this is set by the extent of the field of view of the observations). Secondly, this method is less effective for galaxies that are face-on (due to loss of kinematic information), or edge-on (due to loss of photometric information). Finally, for galaxies with bars we expect our primary pattern speed to be the bar pattern speed. If the bar is oriented along the galaxy minor axis, the integrals will tend to cancel out and no pattern speed will be measured.

\subsection{Uncertainties on the Pattern Speeds}\label{sec:tw_errors}

In order to provide robust estimates on the pattern speed uncertainties, we take multiple sources of error into account when calculating them. Firstly, we account for errors in both $\langle x \rangle$ and $\langle v \rangle$ separately, based on the error in each pixel along the slit, and summed in quadrature. These are then propagated into the fitting routine using Orthogonal Distance Regression (ODR), which allows us to effectively account for errors in both $\langle x \rangle$ and $\langle v \rangle$. We use the {\tt scipy} implementation of ODR ({\sc scipy.odr}). We find that, given the high S/N of our observations and the large numbers of pixels along each slit, the formal errors on these ODR fits are very small. As an example, one such fit is shown in Fig.~\ref{fig:ngc3351_tw_integral}, for NGC~3351. In Appendix \ref{app:all_tw_plots}, we show this visualisation for all MUSE galaxies. Even our highest quality fits have some points that lie off the fitted line. These can be caused by multitude of reasons, including unmasked foreground stars, small variations in the galaxy position angle with galactocentric radius, or (particularly at the edges of the observations) insufficiently long slits. However, particularly for the fits with a quality flag $Q=1$ (see Sect. \ref{sec:quality_flagging}), the fits tend to look excellent.

We also account for errors in the position angle of the galaxy, and the galaxy centre. We do this via a Monte-Carlo method, perturbing the line of nodes and the galaxy centre by the measured errors \citep[recorded in][]{2020Lang}. We use 1000 bootstraps, measure the pattern speed for each of these iterations, and quote the pattern speed as the median of this distribution, with the associated errors as the 16$^{\rm th}$ and 84$^{\rm th}$ percentiles, as we find these errors tend to be asymmetric. These are listed in Table \ref{table:pattern_speeds}. We find the errors on our pattern speeds to be $11^{+15}_{-3}\%$ (the median percentage error and 16$^{\rm th}$ and 84$^{\rm th}$ percentiles), and that it is the error in the position angle that dominates, as has been shown in previous work \citep{2003Debattista,2020GarmaOehmichen}. We do not include uncertainties from the inclination or distance in our uncertainty for the pattern speeds. This is because many of our comparisons are to other quantities that are inclination- and distance-dependent (see Sect. \ref{sec:correlations}). 

\subsection{Effects of slit length and width on recovered pattern speeds}\label{sec:slit_length_width}

Finally, we investigate the effects of both the slit length and slit width on recovered pattern speeds. These can affect the pattern speeds in two ways: firstly, the slits must be sufficiently long as to reach a sufficient radius at which the disk is roughly axisymmetric (and the effect from morphological features no longer prominent). If this is not achieved, the measurement will be biased. Secondly, if the slits are too wide, we may have insufficient slits covering the bar to retrieve a reliable pattern speed measurement. We use a number of slit lengths, from 10\arcsec\ to 150\arcsec\ ($\sim$1 to 15\,kpc, dependent on the distance and inclination of the galaxy), and slit widths from 0.5\arcsec\ to 10\arcsec\ (in general, the resolution of our data is $\sim$1\arcsec, but the pixels oversample the beam; this corresponds to $\sim$50\,pc to 1\,kpc). 1\arcsec\ corresponds to between around 25\,pc for the nearer targets in the sample, and 120\,pc for the farthest. The results of this experiment for NGC~3351 are shown in Fig.~\ref{fig:ngc3351_slit_length_width}. We find that the measured pattern speeds tend to converge as the slits become longer (typically, slightly longer than the bar). We also find that the slit width has a minimal impact on the recovered pattern speed (right panel of Fig.~\ref{fig:ngc3351_slit_length_width}). With larger slit widths, we have fewer points to fit, and typically the uncertainty in the pattern speed becomes slightly larger. Motivated by these results, we opt to make each individual slit as long as possible, and 1\arcsec~wide to approximately match the resolution of the data. For surveys with larger number of galaxies, but with physical resolutions of $\sim$kpc scales, rather than ${\sim}100$\,pc scales \citep[e.g. MaNGA;][]{2015Bundy}, efforts to measure pattern speeds in these galaxies should still produce reliable results \citep[right panel of Fig.~\ref{fig:ngc3351_slit_length_width}; see also work by][]{2019Guo}. We perform these slit length and slit width diagnostics for all galaxies, and these form a critical component of our diagnostic assessments (Sect.~\ref{sec:quality_flagging}).

\subsection{Data Preparation}\label{sec:data_prep}

We perform a number of pre-processing steps to the data before calculating the Tremaine-Weinberg integrals. These are:
\begin{enumerate}
    \item {\bf Mask foreground stars.} For the MUSE data only, foreground stars can be a contaminant. These are clearly recognisable in the data with extreme (positive or negative) velocities (with respect to the systemic velocity of the galaxy). We therefore remove any pixels with $|v| > 300\,{\rm km\,s}^{-1}$. This cut is arbitrary, and may need to be tailored for other data sets. However, we find that the recovered pattern speeds with masked or non-masked stars are not significantly different;
    \item {\bf Subtract systemic velocity.} For the ALMA data only, the systemic velocity is not subtracted in the moment~1 maps. We estimate the systemic velocity using {\sc PAFit} \citep{2006Krajnovic}, and subtract it from the ALMA velocity field (moment~1) map. This has the effect of centring the $\langle v \rangle$ integrals. As we are primarily interested in the slope of $\langle v \rangle$ versus $\langle x \rangle$, this value is not particularly important, but is included so that the line approximately passes through the origin;
    \item {\bf Remove integrals that do not cross the bar.} For galaxies with bars, we attempt to isolate the bar pattern speed from any other pattern speeds present in the galaxy (i.e. to minimise the effects of potentially different pattern speeds in the spiral arms). This is done by removing any slits that do not at least partially cover the bar. We note that these slits may cross multiple features with multiple patterns, but as the slit lengths need to be suitably long to reach the axisymmetric disk, this is unavoidable;
    \item {\bf Symmetrise the integral.} Finally, after the other pre-processing steps, we make sure each integral goes from $-x_0$ to $x_0$. To do this, we find the maximum distance from the minor axis where both sides of each slit still contain data. Given the field of view of the observations, and previous bar masking, this may lead to slits that are different lengths from each other. It is critically important that each slit is symmetric, as if they are not, this asymmetry can induce a false signal in the integrals.
\end{enumerate}
Having performed this pre-processing, we place a number of slits across the surface brightness and velocity maps, and calculate the intensity-weighted position and velocity of all the pixels within each slit. We do this for all 19 galaxies covered by MUSE, and all 84 of our galaxies covered by ALMA. We find that all of the MUSE galaxies can be fitted, but due to no detected CO emission in the cube, the ALMA data for IC~5332 and NGC~3239 cannot be fitted. This leaves us with a total of 82 ALMA galaxies, and 18 of these overlap with MUSE observations. IC~5332 is present in both samples, but as no emission is detected in the ALMA cube, we only present measurements from the MUSE observations.

\subsection{Quality Flagging}\label{sec:quality_flagging}

\begin{figure}[t]
\plotone{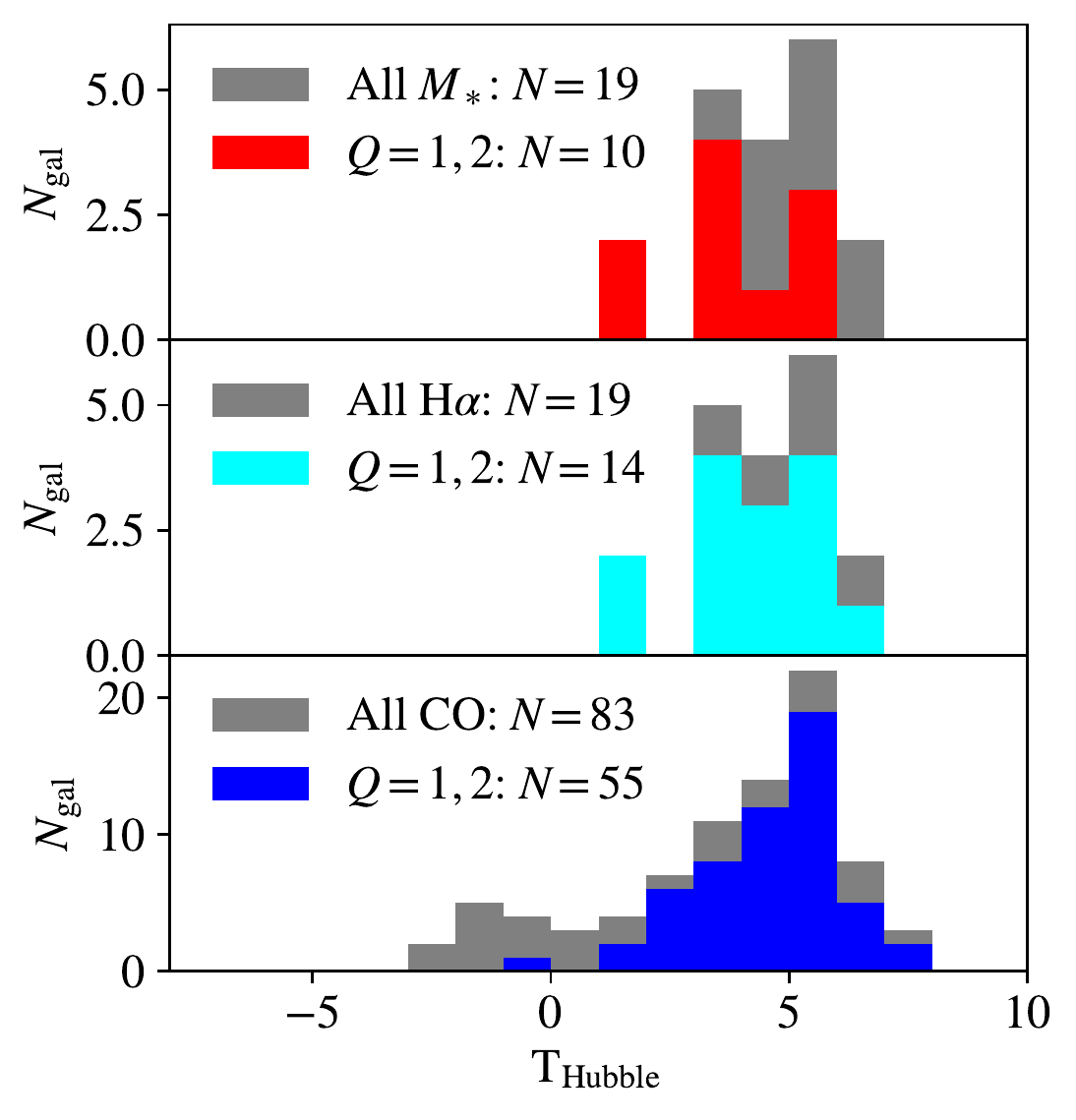}
\caption{Distribution of well-constrained Tremaine-Weinberg values as a function of Hubble morphological types for {\it top}: MUSE-$M_\ast$ measured values; {\it middle}: MUSE-H$\alpha$ measured values; and {\it bottom}: ALMA-CO measured values. In each case, grey indicates all the values for each tracer (i.e. the underlying population), and the coloured bars the well-constrained sample: red for MUSE-$M_\ast$, cyan for MUSE-H$\alpha$, and blue for ALMA-CO. \label{fig:quality_flag_morph_type}}
\end{figure}

Having applied the Tremaine-Weinberg method for all the galaxies (with all available tracers) in our sample, we perform a~posteriori quality flagging on these values. We include all measured values in Table~\ref{table:pattern_speeds} along with these quality flags~($Q$). We take a minimalist approach to the quality flagging, to provide clear delineation between the flags, which are defined as follows:
\begin{enumerate}
    \item Single, well defined slope: integrals have converged, pattern speed is stable with decreasing slit width;
    \item Clear multiple slopes visible in the $\langle v \rangle/\langle x \rangle$ plot, but otherwise the fit would be a quality flag~(1);
    \item Poor fit: integral has not converged, points do not form a clear, well-defined slope;
    \item Data of insufficient quality to calculate a reliable slope.
\end{enumerate}
Our flags span this entire range, from 1 to 4 (see Table \ref{table:pattern_speeds}). We note that these quality flags simply represent the quality of the fits, and will not take into account the issues present in the ISM tracers discussed in Sect.~\ref{sec:clumpy_ism}. We do include these in the table with the caveats mentioned in this later section, both because we will make comparison between the tracers in Sect.~\ref{sec:derived_pattern_speeds}, and because future work will investigate these values in the context of other methods.

This flagging is performed independently on each tracer by three of the authors, and our final flag is the mode of these three flags. In the case of disagreement between all three authors, we instead take the highest flag value of any author, to be as conservative as possible. We opt for the mode to ensure there are no fractional flags which may cause confusion as to their definition. In this work going forwards, we will define ``well-constrained'' Tremaine-Weinberg values as having $Q=1$ or~2, and only use these in our analyses. The only galaxy we use with a quality flag of 2 is NGC~3627, for which points towards the south-west appear to have a different slope (see Fig. \ref{fig:app_ngc3627}).

The distribution of well-constrained values is shown as a function of Hubble morphological type \citep[from the HyperLeda database;][]{2014Makarov} in Fig.~\ref{fig:quality_flag_morph_type}. Many of the galaxies flagged as having poorly measured values are from earlier Hubble types. This is likely due to their lack of prominent morphological features, and so no strong signal in the Tremaine-Weinberg plots. For the MUSE stellar mass values, we find that the fit is well-constrained for 10 of the 19~galaxies. Of these, 9~have bars present, and 1 (NGC~0628) does not. The PHANGS-MUSE sample contains 15~barred galaxies and 4~non-barred, so we preferentially measure bar pattern speeds. For some with a quality flag of 3 or~4, we detect no evidence for a pattern (IC~5332 and NGC~5068), but in these cases the velocity field is very irregular (and so the position angles change with radius).

\section{Application to ISM Tracers}\label{sec:clumpy_ism}

\begin{figure*}[ht]
\plottwo{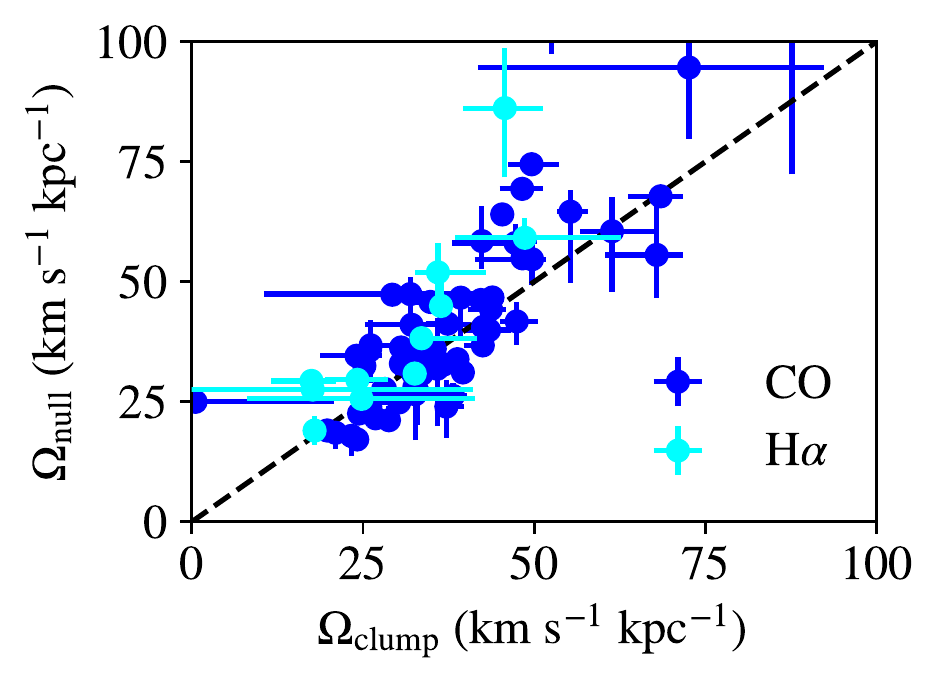}{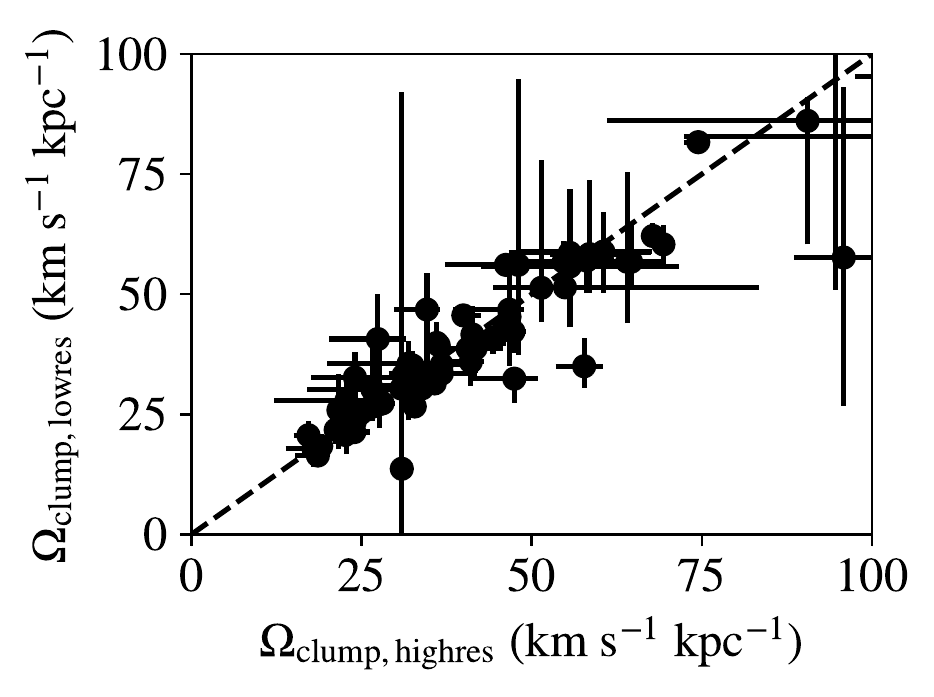}
\caption{{\it Left: }Comparison between the Tremaine-Weinberg signal generated an axisymmetric surface brightness and velocity distribution with incomplete map coverage ($\Omega_{\rm null}$), and from the actual ALMA-CO (blue) and MUSE-H$\alpha$ (cyan) data ($\Omega_{\rm clump}$). This test has been carried out for all ALMA galaxies where a rotation curve has been fitted \cite[][65 galaxies]{2020Lang}, and to the 13 galaxies in the MUSE DR1.0. The ${1\!:\!1}$ relation is shown as a dashed black line. {\it Right:} Comparison between $\Omega_{\rm clump}$ measured for high-resolution data (generally 12m+7m+TP) and lower resolution data (7m+TP). \label{fig:omega_clump}}
\end{figure*}

Whilst we base the majority of this study on pattern speeds derived from stellar kinematics, the PHANGS-ALMA data set in particular offers a much larger sample size. The pattern speed for the Milky Way bar derived from gas dynamics \citep{2015Sormani} has been found to match well with that derived from stellar dynamics \citep{2019Sanders}, and so using the ISM may provide useful, independent pattern speed measurements. We therefore turn to the question of whether ISM tracers are intrinsically compromised due to their incomplete coverage, and clumpiness. If each slit goes through a single clump, even if there is no pattern present, applying the Tremaine-Weinberg method will give non-zero integrals, and this can lead us to measure a false pattern speed. In reality, in these data, each slit will pass through many of these clumps, which are not positioned randomly (i.e. they cluster on spiral arms, or bars). We thus critically examine whether we can disentangle the true signal from the effect of incomplete coverage.

To test this, for each galaxy we take an axisymmetrised velocity and surface brightness profile, based on the CO maps. For the velocity field, we project the fits from \cite{2020Lang} into the frame of the galaxy, and for the CO data we take the average surface brightness within a number of annuli. We then mimic the incomplete coverage of these maps by blanking pixels in these axisymmetrised profiles where emission is not detected, and apply the method described in Sect.~\ref{sec:tw_method} to these. In this case, there is no pattern speed present in these maps, and any deviation is simply due to the incomplete CO coverage and the non-axisymmetry of the CO morphology. We refer to this as $\Omega_{\rm null}$, and compare this to the ``pattern speed'' measured from the actual CO data, as shown in the left panel of Fig.~\ref{fig:omega_clump}. We refer to this as $\Omega_{\rm clump}$. As we can see, the points tend to lie close to the ${1\!:\!1}$ relation, indicating that much of the signal we measure using the Tremaine-Weinberg method may simply be due to the clumpy nature of the tracer itself. We also find a similar trend repeating this test for H$\alpha$. The angular velocity that we measure with this experiment is some function of the underlying rotation curve of the galaxy and the sampling of this rotation curve by the tracer's distribution of ``clumps''. In the right panel of Fig.~\ref{fig:omega_clump}, we compare $\Omega_{\rm clump}$ from the higher resolution ($\sim$1\arcsec) and lower resolution ($\sim$5\arcsec) ALMA data. The values calculated are very similar, and we find that repeating the $\Omega_{\rm null}$ test with these values calculated from lower resolution data yields much the same results as the higher resolution data. This indicates that this effect is endemic to the use of CO, and not simply due to the high resolution of our data.

As an initial exploration into this, we take some simple density-wave spiral models following \cite{2008BinneyTremaine}, with a known pattern speed, and blank regions to mimic the clumpiness of the employed tracer. We find that more spatially extended distributions of clumps (probing further out in galactocentric radius), produce lower $\Omega_{\rm clump}$ than more centrally concentrated arrangements. We believe that this leads to the results in Sect.~\ref{sec:tracer_comparison_om_p}, where $\Omega_{\rm clump}$ and $\Omega_{\rm P}$ appear systematically offset. As CO extends further out in galaxies hosting larger bars (the outermost CO-emitting radius tabulated by \citealt{2020Lang} increases roughly with increasing $R_{\rm bar}$). Secondly, longer bars tend to have lower pattern speeds than shorter bars in similar-sized galaxies (i.e. $\mathcal{R}\sim1.2$). As applied to PHANGS-ALMA, we will thus find lower $\Omega_{\rm clump}$ in galaxies with lower bar pattern speeds. Note that this may not always hold for other tracers, e.g. for much more extended H{\sc i} disks, when the extent of the tracer is not tied to the morphology of the dynamical feature in question. However, it does appear that this is the case for both our ALMA-CO and MUSE-H$\alpha$ data. A~full exploration of the links between $\Omega_{\rm clump}$ and $\Omega_{\rm P}$ is beyond the scope of this work, and will require detailed simulations of galaxies.

\section{Derived Pattern Speeds}\label{sec:derived_pattern_speeds}

\begin{deluxetable*}{ccccccccc}
\tablecaption{Pattern speeds and co-rotation radii for the ten well-constrained stellar mass pattern speeds.}
\label{table:well_constrained_pattern_speeds}
\tablehead{\colhead{Galaxy} & \colhead{PGC} & \colhead{D} & \colhead{$i$} & \colhead{PA} & \colhead{Bar?} & \colhead{$\Omega_{\rm P}$} & \colhead{Q} & \colhead{$R_{\rm CR}$}\\ \colhead{ } & \colhead{ } & \colhead{$\mathrm{Mpc}$} & \colhead{$\mathrm{{}^{\circ}}$} & \colhead{$\mathrm{{}^{\circ}}$} & \colhead{ } & \colhead{$\mathrm{\frac{km}{kpc\,s}}$} & \colhead{ } & \colhead{$\mathrm{kpc}$}}
\startdata
NGC0628 & 5974 & 9.84 & 8.9 & 20.7 & 0 & $31.1^{+4.0}_{-2.9}$ & 1 & $4.5\pm2.0$ \\
NGC1087 & 10496 & 15.85 & 42.9 & 359.1 & 1 & $31.9^{+3.2}_{-1.6}$ & 1 & $4.3\pm1.1$ \\
NGC1433 & 13586 & 12.11 & 28.6 & 199.7 & 1 & $20.0^{+2.4}_{-1.9}$ & 1 & $6.3\pm0.4$ \\
NGC1512 & 14391 & 17.13 & 42.5 & 261.9 & 1 & $22.3^{+5.1}_{-5.5}$ & 1 & $5.4\pm2.2$ \\
NGC1672 & 15941 & 19.4 & 42.6 & 134.3 & 1 & $22.7^{+0.6}_{-0.6}$ & 1 & $6.1\pm1.6$ \\
NGC2835 & 26259 & 12.38 & 41.3 & 1.0 & 1 & $34.6^{+4.3}_{-3.1}$ & 1 & $2.8\pm0.6$ \\
NGC3351 & 32007 & 9.96 & 45.1 & 193.2 & 1 & $43.6^{+11.6}_{-12.4}$ & 1 & $3.4\pm0.8$ \\
NGC3627 & 34695 & 11.32 & 57.3 & 173.1 & 1 & $29.1^{+20.6}_{-8.1}$ & 2 & $1.9\pm0.1$ \\
NGC4303 & 40001 & 16.99 & 23.5 & 312.4 & 1 & $43.5^{+5.3}_{-10.0}$ & 1 & $3.8\pm2.5$ \\
NGC7496 & 70588 & 18.72 & 35.9 & 193.7 & 1 & $16.8^{+6.5}_{-12.1}$ & 1 & $4.3\pm2.0$
\enddata
\end{deluxetable*}

\begin{figure}[t]
\plotone{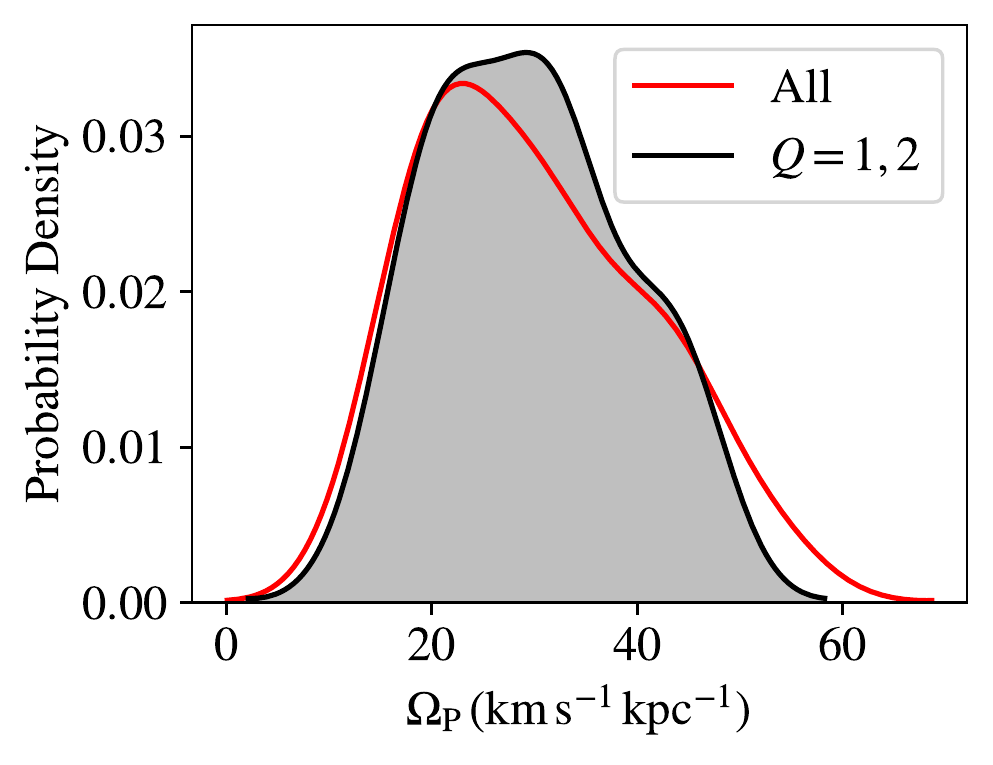}
\caption{The distribution of pattern speeds derived from MUSE stellar mass maps for 19~galaxies. The red line shows the Kernel Density Estimate (KDE) distribution for all calculated pattern speeds \citep[using the optimal bandwidth formula from][]{Silverman86}, the shaded black line the distribution of well-constrained pattern speeds. The KDE is normalised such that the integral of the distribution is equal to one, so the shape of each distribution, rather than the absolute scaling, is what should be considered here. \label{fig:pattern_speed_distribution}}
\end{figure}

\begin{figure}[t]
\plotone{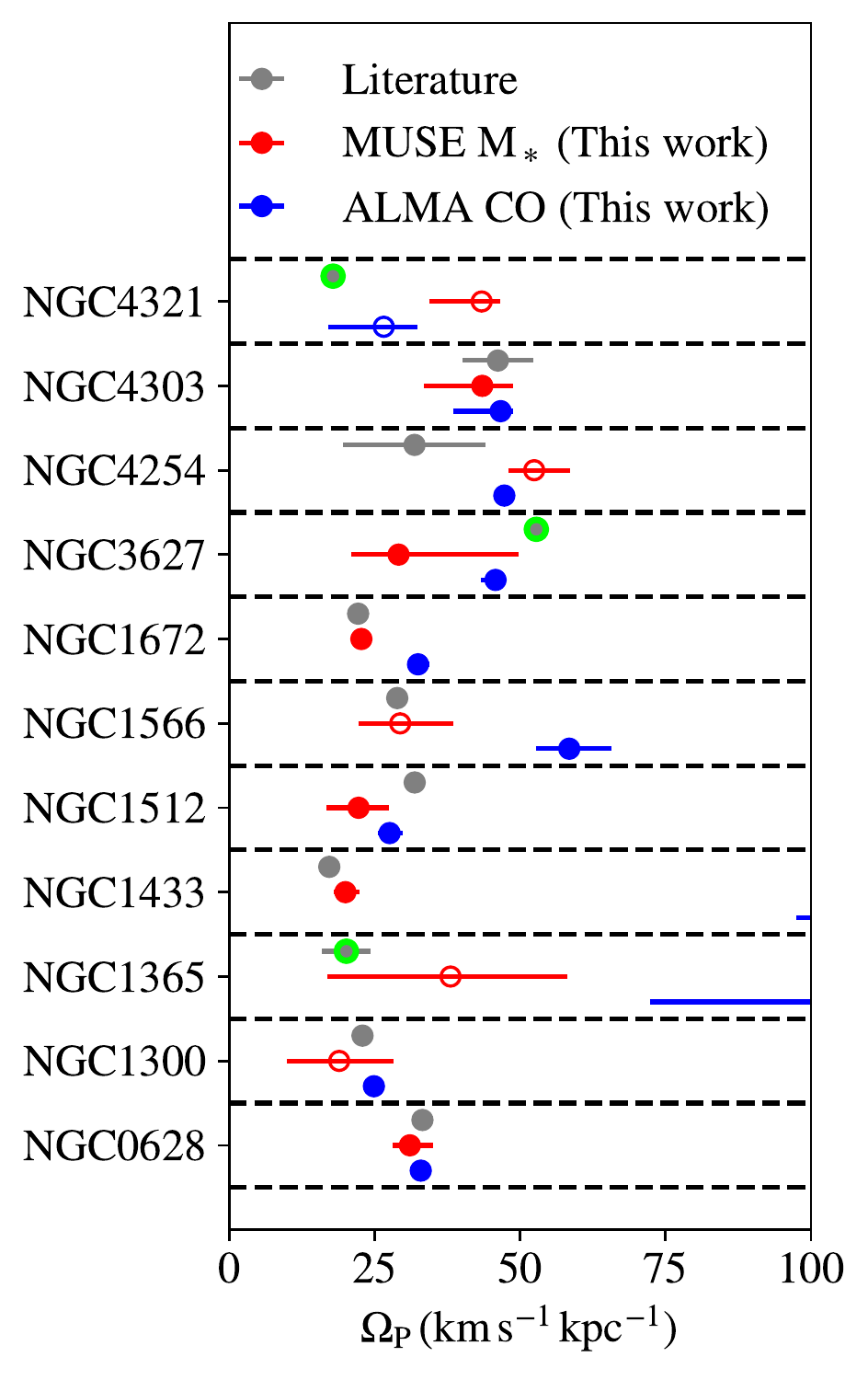}
\caption{Comparison between derived pattern speeds for MUSE stellar mass maps (red symbols), ALMA CO-based $\Omega_{\rm clump}$ (blue symbols) and previously published literature values (grey symbols). If the pattern speed is not well measured ($Q=3$ or~4), the dot is unfilled. For literature pattern speeds from applying Tremaine-Weinberg to ISM tracers, we outline these points in light green. \label{fig:pattern_speed_comparison_literature}}
\end{figure}

\begin{figure}[t]
\plotone{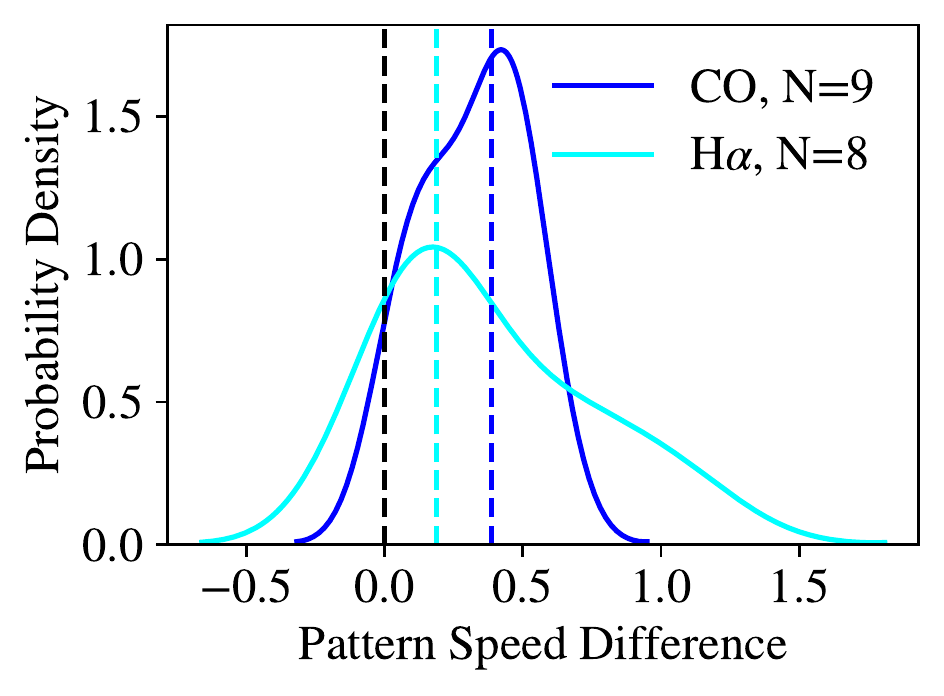}
\caption{KDE plot showing the relative pattern speed difference (Eq.~\ref{eq:pattern_speed_difference}) for both CO (blue line) and H$\alpha$ (cyan line) with respect to the stellar mass pattern speed. The dashed black line indicates where $\Omega_{\rm P} = \Omega_{\rm clump}$. The vertical dashed blue and cyan lines show the median of the distributions for CO and H$\alpha$, respectively. \label{fig:ism_mstar_comparison}}
\end{figure}

With our final selection of well-constrained Tremaine-Weinberg values in hand (Table~\ref{table:well_constrained_pattern_speeds}), we show the distribution of the 19 MUSE stellar mass pattern speeds in Fig.~\ref{fig:pattern_speed_distribution}. Note that we use KDE plots throughout this work to improve readability, by avoiding the `steps' we would see in histograms for these small number of data points. However, any statistics are calculated from the data themselves, rather than the distribution of the KDE. For the values we consider well-constrained, the average speed is $30^{+10}_{-9}\,{\rm km\,s^{-1}\,kpc^{-1}}$ (this is the median of the sample, along with the 16$^{\rm th}$ and 84$^{\rm th}$ percentiles of the distribution). The distribution of all the pattern speeds derived from the MUSE stellar mass maps are also shown in this figure, and it can be seen that they occupy a similar range of speeds ($29^{+14}_{-10}\,{\rm km\,s^{-1}\,kpc^{-1}}$). Thus, poorly constrained pattern speeds are not found at any particular low or high values of $\Omega_{\rm P}$, and so this manual quality flagging is required compared to simple threshold cut.

\subsection{Comparison to the Literature}\label{sec:comparison_to_literature}

Some of these galaxies have previously published pattern speeds\footnote{NGC~0628: \citet{2014MartinezGarcia}; NGC~1300: \citet{1996Lindblad}; NGC~1365: \citet{2016Speights}; NGC~1433: \citet{2008Treuthardt}; NGC~1512: \citet{2009Koribalski}; NGC~1566: \citet{2005Korchagin}; NGC~1672: \citet{1999Diaz}; NGC~3627: \citet{2004RandWallin}; NGC~4254: \citet{2004Egusa}; NGC~4303: \citet{2002Schinnerer}; NGC~4321: \citet{2005Hernandez}}, and we make comparison between our values and these earlier values in Fig.~\ref{fig:pattern_speed_comparison_literature} (scaling to our assumed galaxy distances and inclinations). These pattern speeds come from a variety of methods, but we note that the pattern speeds of NGC~1365, NGC~3627, and NGC~4321 in the literature come from applying the Tremaine-Weinberg method to ISM tracers. For completeness, we show those points with $Q$-values greater than 2, but we stress that these do not constitute true measurements of pattern speeds, and are not used in any of our analysis here or later in this work. In general, our pattern speeds agree well with previously published values, and typically agree within the uncertainties (a median absolute deviation of $\sim$10\%, compared to the larger differences we see comparing to ISM tracers in Sect.~\ref{sec:clumpy_ism}). Given that these come from extremely heterogeneous pattern speed measurements (either in technique or tracer), the fact that we see good agreement across the board is reassuring. We are therefore confident that our stellar mass pattern speeds presented in Table~\ref{table:well_constrained_pattern_speeds} are robust and reliable.

\begin{figure}[t]
\includegraphics[width=\columnwidth]{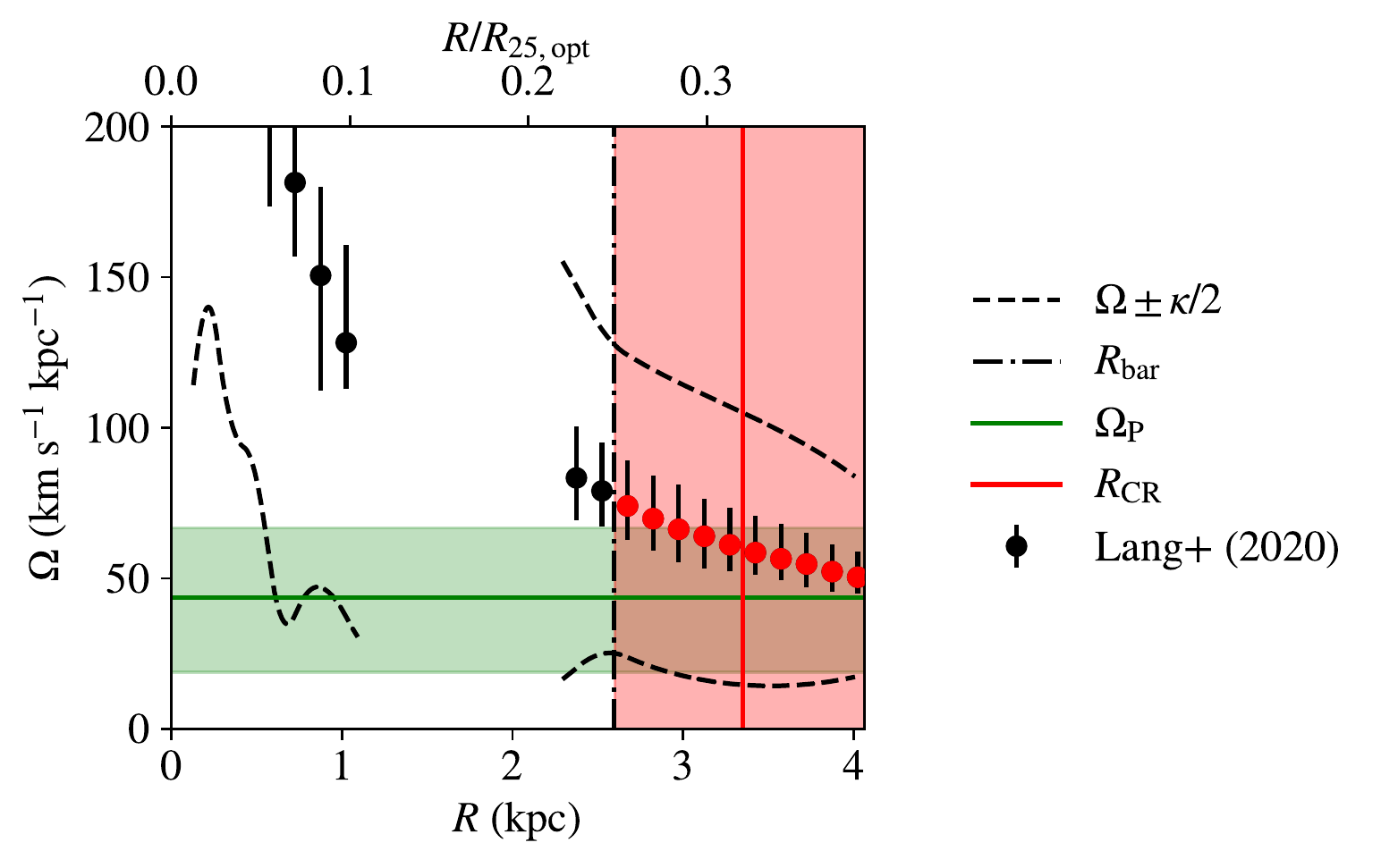}
\caption{Angular speeds ($\Omega$) versus galactic radius for NGC~3351. Angular speeds from \cite{2020Lang} are shown as black points. When one of these $\Omega$ values is consistent with the pattern speed (green line, with associated errors in green), it is highlighted in red. These points are then combined to form our estimate of the co-rotation radius (red line) and associated error (shaded red region). The bar radius is shown as a vertical, black, dot-dash line. We also include dashed, black lines to indicate $\Omega\pm\kappa/2$, which we use to calculate ILR and OLR (these values and associated errors are omitted here to maintain readability, but are included in Table~\ref{table:pattern_speeds}).
\label{fig:ngc3351_corotation_radii}}
\end{figure}

\begin{figure}[t]
\includegraphics[width=\columnwidth]{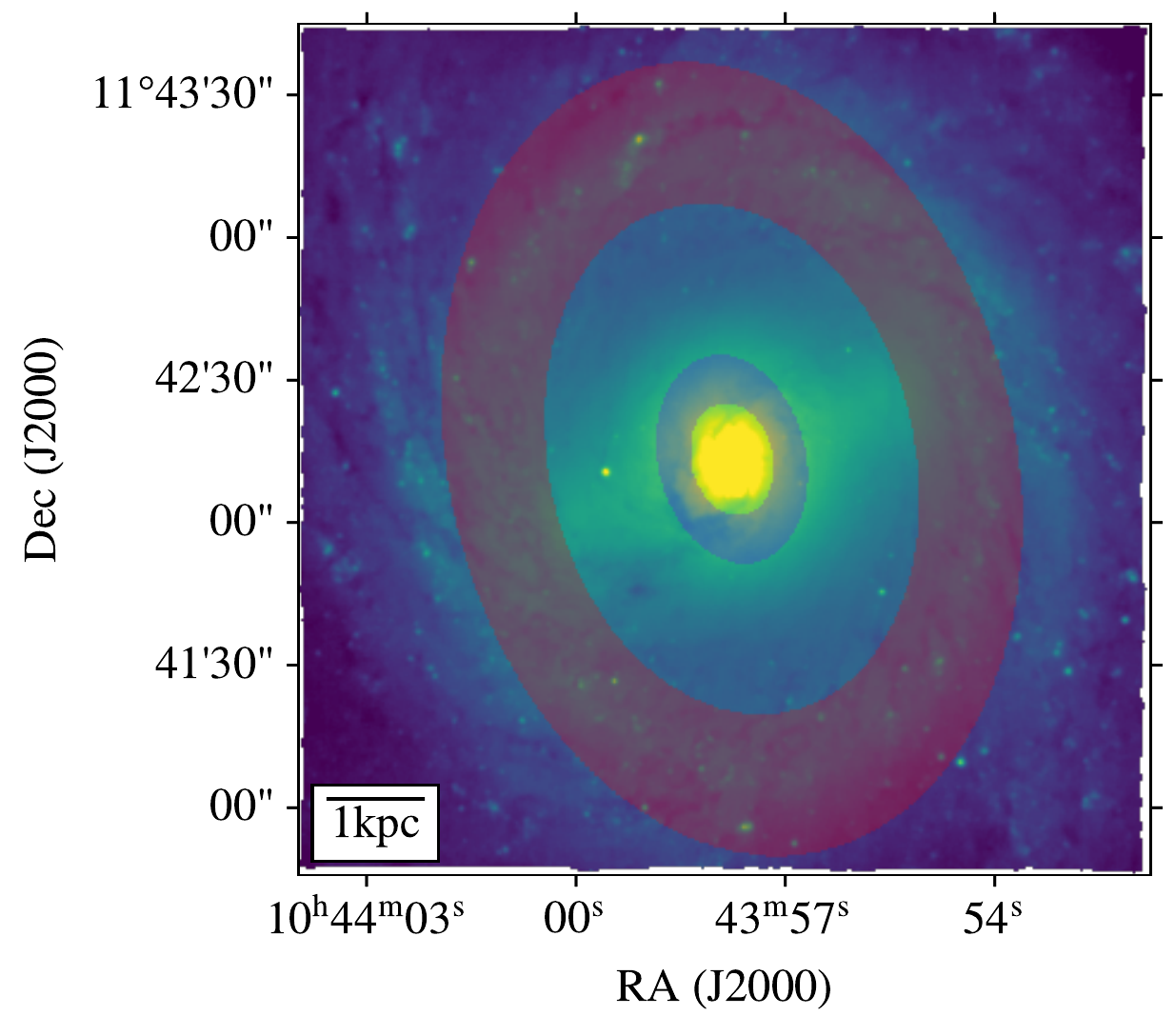}
\caption{Resonances highlighted on MUSE white light map of NGC~3351. Co-rotation (and associated errors) is shown in red, and the inner Lindblad radius in blue. \label{fig:ngc3351_resonance_map}}
\end{figure}

\subsection{Comparisons Between Tracers}\label{sec:tracer_comparison_om_p}

\begin{figure*}[t]
\plottwo{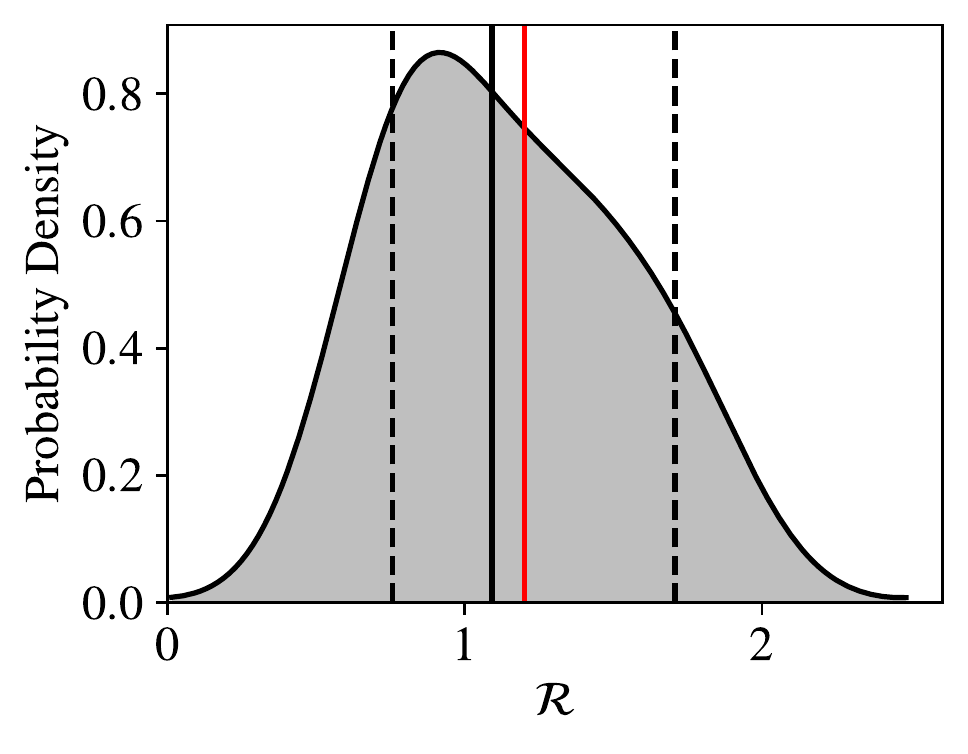}{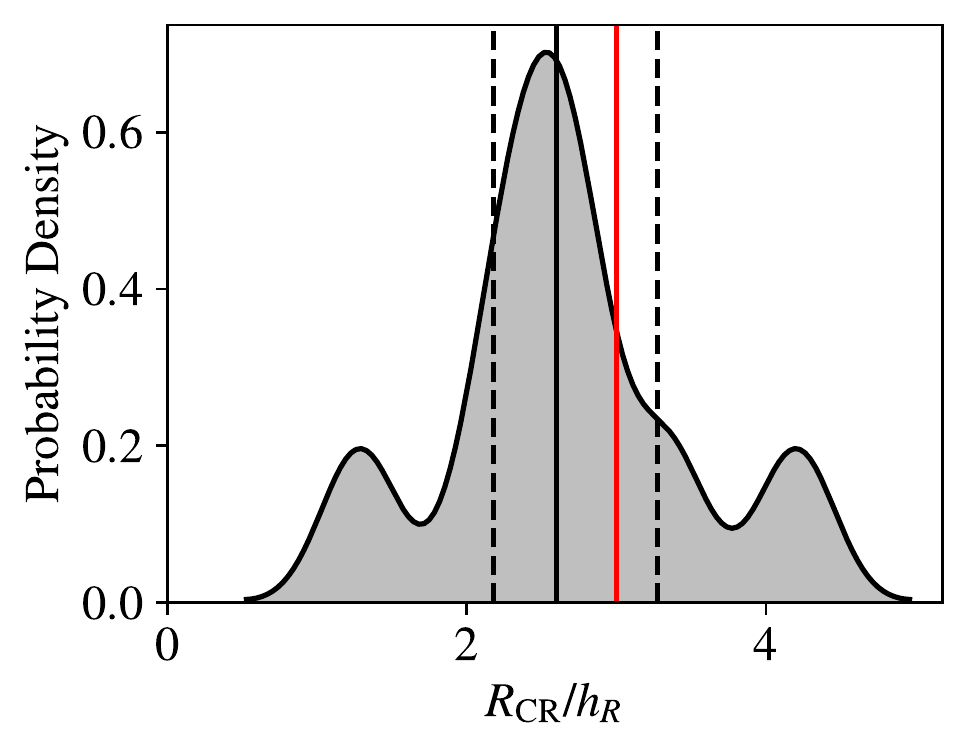}
\caption{{\it Left}: KDE plot showing the distribution of $\mathcal{R}$ (the ratio of co-rotation radius to bar length) for our nine well-constrained stellar mass bar pattern speeds. {\it Right}: KDE plot showing the distribution of the ratio of the co-rotation radius to the galaxy scale length for the same sample of galaxies. In each case, we highlight the median value as a solid black vertical line, and the 16$^{\rm th}$ and 84$^{\rm th}$ percentiles as dashed black lines (these percentiles are calculated directly from the data, rather than from the KDE). The commonly assumed literature values are shown as red, vertical lines. \label{fig:corot_ratios}}
\end{figure*}

As the PHANGS-MUSE sample overlaps entirely with the PHANGS-ALMA sample, we can compare how closely the ISM tracers (H$\alpha$, CO) measured $\Omega_{\rm clump}$ tends to agree with $\Omega_{\rm P}$. We compare values flagged as well-constrained in both the stellar mass, and the ISM tracer in question (for both H$\alpha$ and CO, this leaves us with 9~galaxies). We then define a ``pattern speed difference'' as
\begin{equation}\label{eq:pattern_speed_difference}
    {\rm Pattern~Speed~Difference} = \frac{\Omega_{\rm clump} - \Omega_{\rm P}}{\Omega_{\rm P}},
\end{equation}
and show the distributions for both ISM tracers in Fig.~\ref{fig:ism_mstar_comparison}. Typically, we find that $\Omega_{\rm clump}$ is higher than $\Omega_{\rm P}$, somewhat greater than our $\sim$10\% errors ($\sim$40\% in the case of CO, $\sim$20\% in the case of H$\alpha$). Therefore, whilst applying the Tremaine-Weinberg method to clumpy ISM tracers may yield reasonable results, it is clear from this exercise that $\Omega_{\rm clump}$ and $\Omega_{\rm P}$ are systematically different quantities, and thus $\Omega_{\rm clump}$ should not be used as a proxy for $\Omega_{\rm P}$.

\section{Locations of Major Resonances}\label{sec:resonance_radii}

\begin{figure*}[t]
\plotone{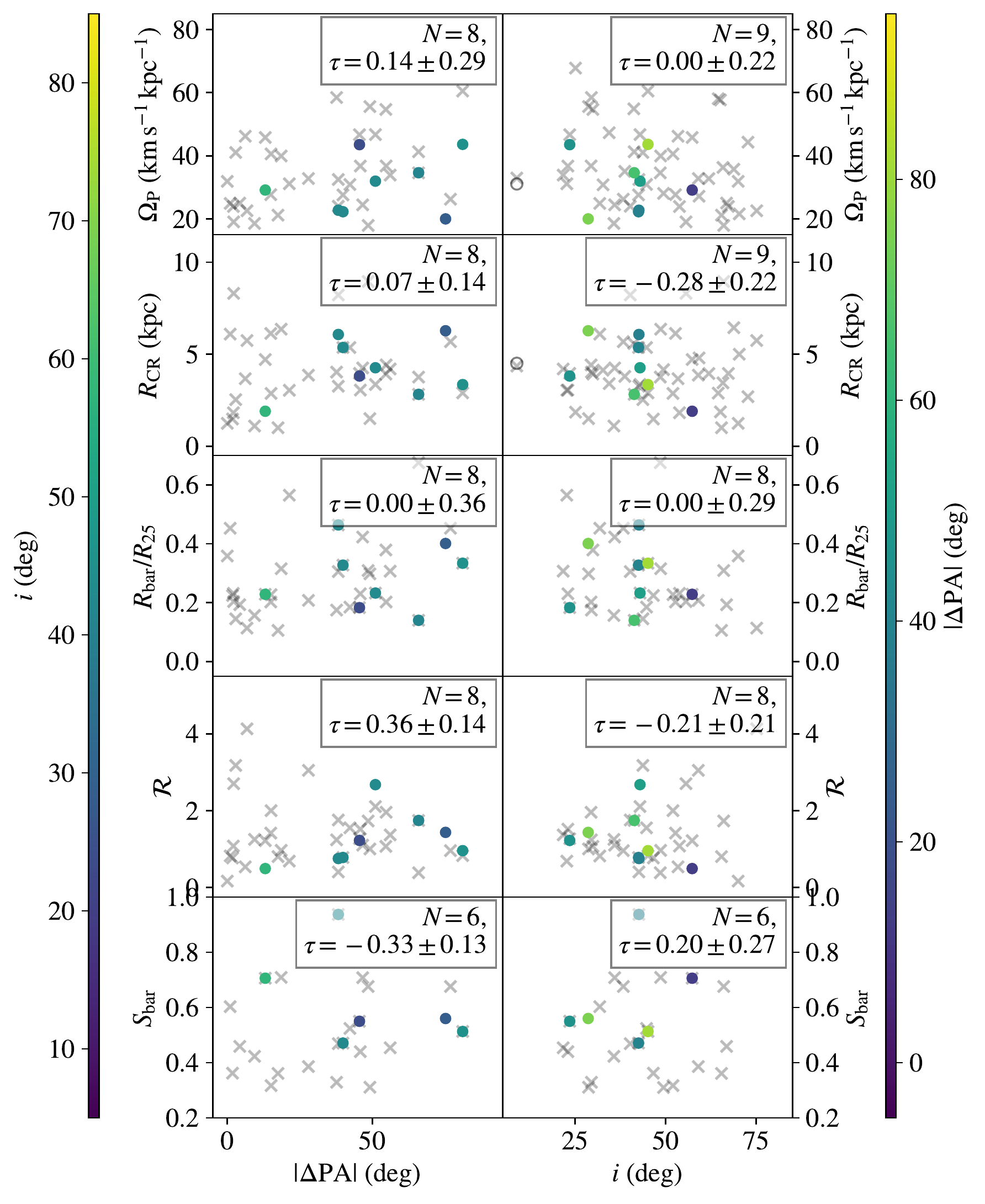}
\caption{Pattern speeds measured from MUSE stellar mass data and global parameters with galaxy orientation parameters. {\it Left}: From top to bottom, the pattern speed, co-rotation radius, bar radius, $\mathcal{R}$, and bar strength versus $\Delta {\rm PA}$ (the difference between bar and galaxy position angles). Points are coloured by the inclination of the galaxy. {\it Right}: Same as the left column, but now versus inclination, with points coloured by $\Delta {\rm PA}$. For galaxies without a bar, we show the pattern speed and co-rotation radius as an unfilled circle. In each case, the number of points, $N$, and the Kendall $\tau$ correlation is given. We also show the well-constrained PHANGS-ALMA sample as grey crosses. \label{fig:correlations_projection}}
\end{figure*}

We next use our pattern speeds to calculate major resonance locations for each galaxy. In this work we focus on the co-rotation radius, \rcr, but we also provide estimates of the outer and inner Lindblad radii (OLR and ILR). These resonance locations are derived with the assumption of circularity, and so we give a single radius for each resonance. To do this, we make use of the rotation curves derived from the PHANGS-ALMA data in \cite{2020Lang}, and rescaled to our assumed distances. These rotation curves approximate the circular velocity, $v_{\rm circ}$, in a number of radial bins, where the galaxy centre, inclination, and position angle are first fit using a Bayesian MCMC analysis. Following this, least-squares fitting uses a harmonic decomposition to model the rotational velocity within a series of radial annuli. However, as these curves are measured from CO, there may be deviations from this circularity, particularly in the centres of galaxies, or for strong spiral or bar streaming motions. We use this observed velocity profile as a proxy for the true circular velocity of the galaxy, and is such is a first-order approximation to calculate the resonance locations. Where non-axisymmetry of the potential dominates, the observed velocity profile will overestimate the true circular velocity, and so this approximation will lead to incorrect resonance locations, but assessing the impact of this requires detailed simulations for each galaxy, and as such is beyond the scope of this work. However, observational works have applied this method previously \citep[see, e.g.][]{2000Schinnerer,2009Fathi}, and find the location of expected structures near these resonance locations, which would indicate this first-order approximation is often close to the true value. As such, we highlight that these resonance locations are approximations, but represent our best estimation of the true value. Throughout this work, we will refer to these resonance locations as the co-rotation radius and ILR/OLR, but bear in mind these caveats.

Typically, the rotation curves derived from the CO maps are very similar to those derived from H$\alpha$ and stars in our sample, and so we are confident our choice of using CO-based rotation curves will not bias our results. By converting the rotation velocity to an angular velocity, $\Omega (R)$, the co-rotation radius is simply where the angular velocity is equal to the pattern speed. 

To calculate \rcr, we use the velocities as fitted to each radial bin. This allows us to propagate through the uncertainties both in the rotation curve and the pattern speed. In some cases, we have multiple regions where the angular velocity crosses the pattern speeds, and we report all of these in Table~\ref{table:pattern_speeds}. In many cases, due to the shape of the rotation curve a number of consecutive points are consistent with being co-rotation, within the errors. For these points, we associate these to the same \rcr\ if they are within $2\sigma$, and we take the mean of them as the nominal values, with the range as the uncertainty. An example of this \rcr\ measurement is shown in Fig.~\ref{fig:ngc3351_corotation_radii}, for NGC~3351. As can be seen, our error propagation leads to reliable estimates of \rcr, but quite large uncertainties, on the order of $\sim$1~kpc (the error is mainly dependent on the shape of the rotation curve). These errors are similar to those obtained using gravitational torque analysis \citep{2016Querejeta}.

We also calculate the OLR and ILR for each galaxy, where they exist. In this case, to avoid issues with numerical derivatives, we use smooth splines fitted to the rotation curves. As these do not account for the errors in the rotation curve fitting, the errors reported in Table~\ref{table:pattern_speeds} only consider the errors in our pattern speed derivation. We suggest taking an uncertainty on these values at least as large as that of \rcr. The locations of these resonances occur when
\begin{equation}
    \Omega_{\rm P} = \Omega(R) \pm \kappa/2
\end{equation} 
where positive is for the OLR, and negative is for the inner ILR. $\kappa$ is the epicyclic frequency, which is given by
\begin{equation}
    \kappa^2 = \frac{2\Omega(R)}{R} \frac{{\rm d}}{{\rm d}R}\left(R^2 \Omega(R)\right).
\end{equation}
These curves are also highlighted in Fig.~\ref{fig:ngc3351_corotation_radii}, and we calculate their average radii and uncertainties in the same way as for \rcr. The locations of the resonances are shown on the MUSE white light map of NGC~3351 in Fig.~\ref{fig:ngc3351_resonance_map}.

With \rcr\ directly measured, we can investigate two commonly assumed ratios for inferring the co-rotation radius when it is not directly accessible. The first is \rcr/$R_{\rm bar}$ (commonly referred to as $\mathcal{R}$), where we calculate the deprojected bar length as
\begin{multline}\label{eq:bar_deproject}
    R_{\rm bar, deproj} = R_{\rm bar, proj} \\
    \times \sqrt{\cos(\Delta {\rm PA})^2 + \left(\sin(\Delta {\rm PA})\sec(i)\right)^2},
\end{multline}
where $R_{\rm bar, proj}$ is the projected bar length from \cite{2015HerreraEndoqui}, $\Delta{\rm PA}$ is the relative alignment of the bar with the galaxy major axis, and $\sec(i)$ is the secant function of the inclination (the inverse cosine). We show the distribution of $\mathcal{R}$ for our sample in the left panel of Fig.~\ref{fig:corot_ratios}. The average value (and spread) for our sample is $1.1^{+0.6}_{-0.4}$ (this spread is calculated from the data, rather than from the KDE plot in Fig. \ref{fig:corot_ratios}). Taking into account the error in co-rotation radius, along with a characteristic uncertainty in the bar length of 20\% \citep{2016DiazGarcia}, we find that around half of our measured values of $\mathcal{R}$ are inconsistent with the commonly assumed value of 1.2 in the literature \citep[e.g.][]{1996Elmegreen,1998Aguerri} to $1\sigma$. This is well within the expectation when including Poisson noise due to small number statistics, and so our values of $\mathcal{R}$ are, on average, consistent with the expected value of 1.2.

Another commonly used ratio for inferring \rcr\ is through the disk scale length, $h_r$. We use the S$^4$G scale lengths calculated in \cite{2015Salo} from multiple component image decompositions of {\it Spitzer} 3.6\micron\, images, and show the ratio of these in the right panel of Fig.~\ref{fig:corot_ratios}. Commonly, $h_r/R_{\rm CR}$ is taken to be 3 \citep[e.g.][]{2003Kranz}, and we find the value for our sample to be $2.6^{+0.7}_{-0.4}$. Again, given the small number statistics we conclude our sample to be consistent with a value for $h_r/R_{\rm CR}$ of 3.

\section{Correlations with Global Galaxy Parameters}\label{sec:correlations}

\begin{longrotatetable}
\begin{figure*}[t]
\includegraphics[width=\columnwidth]{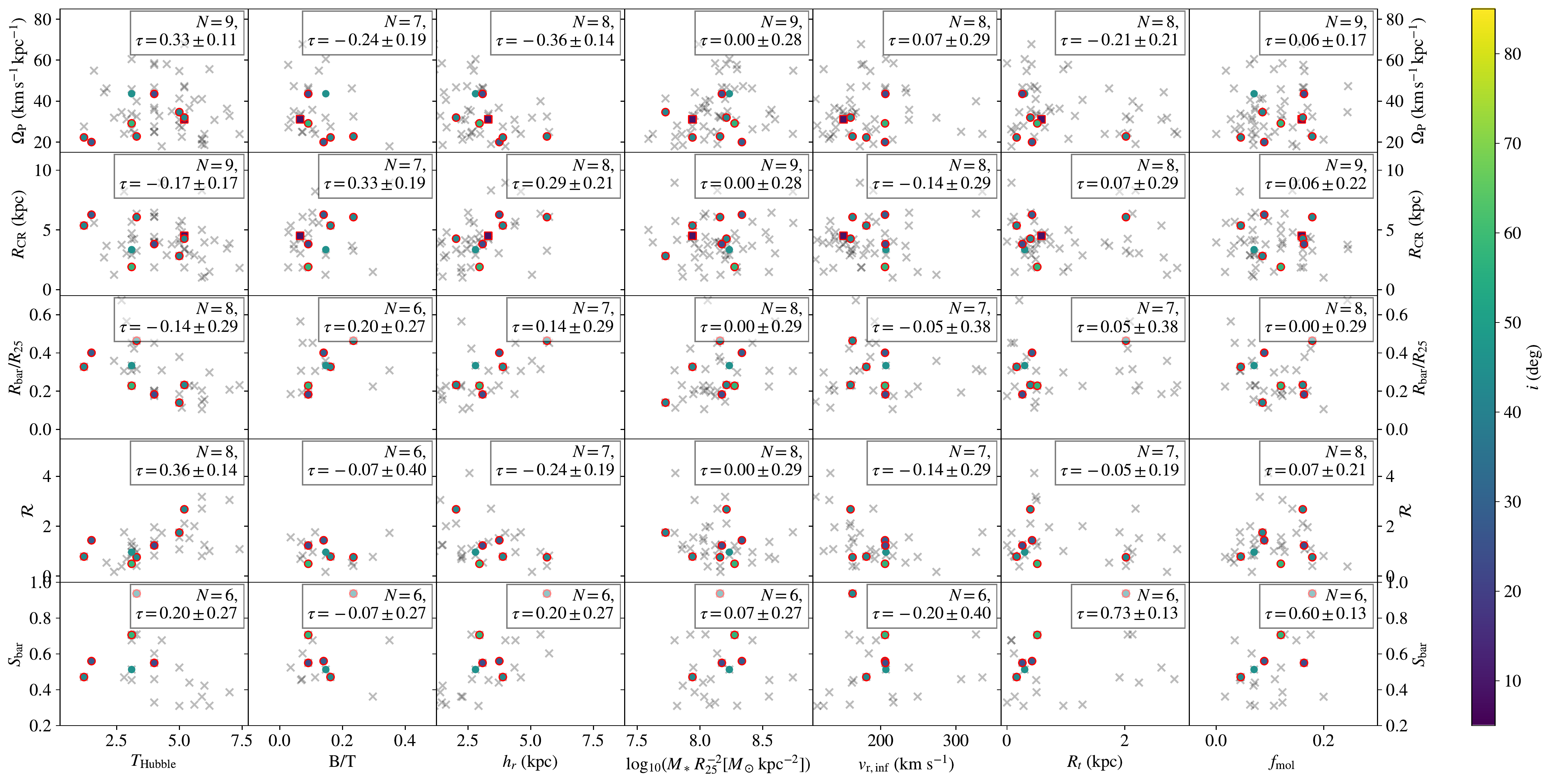}
\caption{From top to bottom: the pattern speed, co-rotation radius, bar radius, $\mathcal{R}$, and bar strength versus, from left to right: the Hubble morphological type, bulge-to-total flux ratio, galaxy disk scale length, stellar mass divided by $R_{25}^2$, asymptotic rotation velocity, galaxy rotation turnover radius, and molecular gas fraction. Points are coloured by the inclination of the galaxy, galaxies with bars shown as circles, and galaxies without as squares, and galaxies with noted spiral arms are highlighted in red. In each subplot, the number of points, $N$, and the Kendall $\tau$ correlation is given, along with its associated uncertainty. We also plot well-constrained PHANGS-ALMA values as grey crosses. \label{fig:correlations_global_parameters}}
\end{figure*}
\end{longrotatetable}

\begin{figure*}[ht]
\plotone{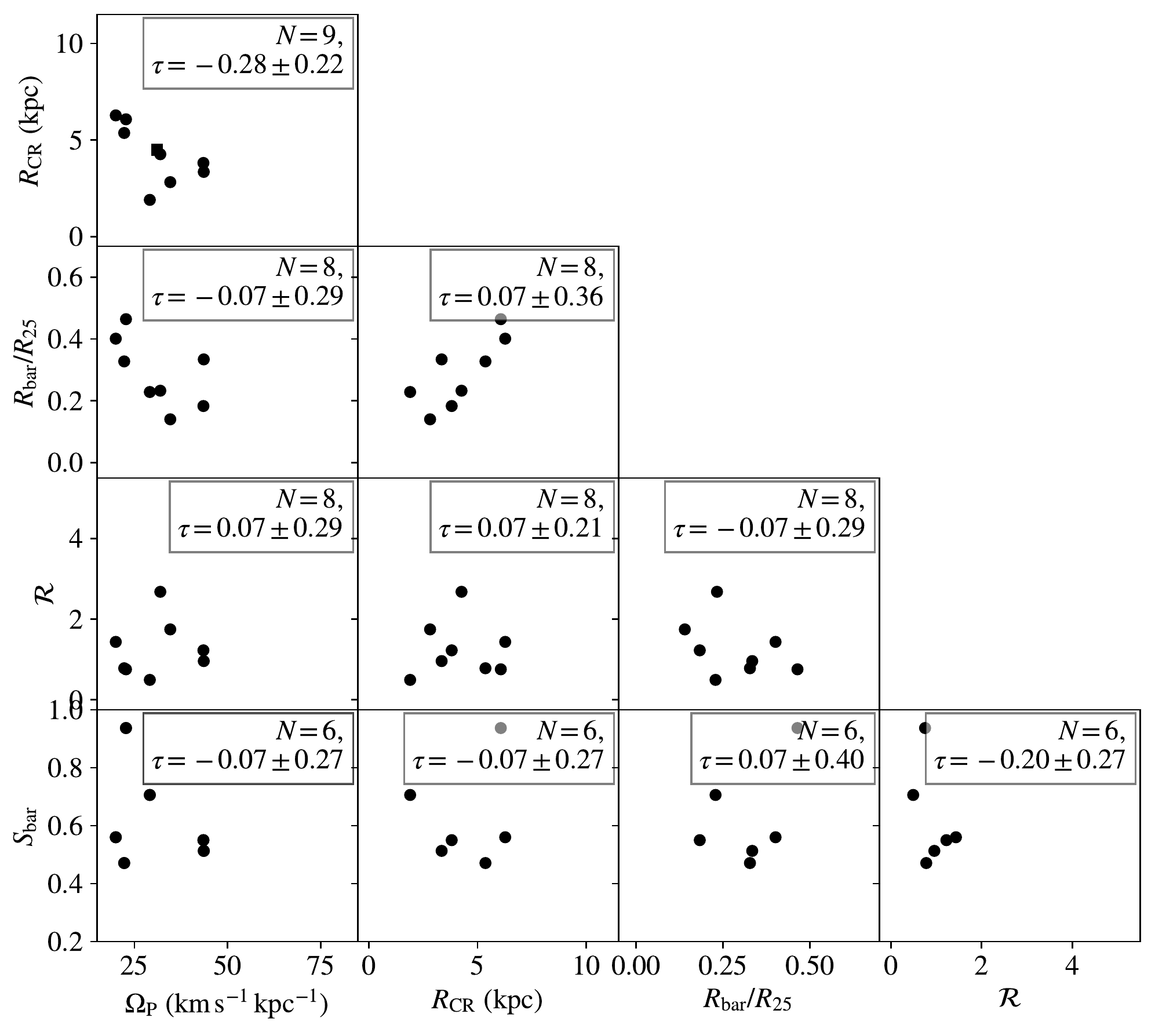}
\caption{Corner plot of the $y$-axis quantities from Figs. \ref{fig:correlations_projection} and \ref{fig:correlations_global_parameters}. Galaxies without bars are indicated as squares (i.e., NGC~0628); galaxies with bars are circles. In each subplot, the number of points, $N$, and the Kendall $\tau$ correlation is given, along with its associated uncertainty. We observe a weak anti-correlation between co-rotation radius and pattern speed, while all other quantities appear uncorrelated. \label{fig:correlations_corner}}
\end{figure*}

With a number of pattern speeds robustly derived, we can statistically study how pattern speeds and resonance locations vary with a number of global parameters of the galaxy. Given that our sample is still small, we see this investigation as exploratory, but provide a basis for future studies to expand on the results we present here. We expect that these further investigations may come from both an observational perspective (i.e. if we see these trends for a larger sample of galaxies), and from simulations (i.e. if we should expect to see these trends in the first place). We first investigate the possibility of any pathological biases arising from the projection of the galaxy to our line of sight, in particular, the inclination and the relative alignment of the bar with the galaxy major axis ($\Delta {\rm PA}$). In this work, we will focus on $\Omega_{\rm P}$, \rcr, $R_{\rm bar}$, $\mathcal{R}$, and the bar strength, $S_{\rm bar}$. We take the bar lengths from \cite{2015HerreraEndoqui}, and the bar strength as the normalised $m = 2$ Fourier density amplitude ($A_2^{\rm max}$) from \cite{2016DiazGarcia}. Following \cite{2020Cuomo}, we only use points that satisfy $\Delta\Omega_{\rm P}/\Omega_{\rm P} \leq 0.5$, where $\Delta\Omega_{\rm P}$ is the error in $\Omega_{\rm P}$, but in this work we find that this removes only one galaxy (NGC~7496). This leaves us with a maximum of 9~galaxies, but depending on the available literature data, the number of galaxies compared may vary slightly with parameter \citep[for example, only 6~galaxies have bar strengths in][]{2016DiazGarcia}. We also show the values from the ALMA points in grey, to highlight any misleading conclusions that may be drawn from using $\Omega_{\rm clump}$ instead of $\Omega_{\rm P}$ (or, equivalently, the bias of using CO rather than stars).

We first consider any pathological effects that may arise from the orientation of the galaxy. Our ability to measure morphological properties of bars may be affected by the inclination, or relative misalignment of the bar with the galaxy itself, and it is important to consider whether this may bias our results. Fig.~\ref{fig:correlations_projection} shows the relationships between $\Omega_{\rm P}$, \rcr, $R_{\rm bar}$, $\mathcal{R}$, and $S_{\rm bar}$ with $\Delta {\rm PA}$ and the galaxy inclination, where we also indicate the \cite{Kendall1938} $\tau$ correlation coefficient. As we are correlating a number of variables against each other, by random chance we may see some ``significant'' correlations in random data, based simply on using $p$-values (i.e. for 100 random distributions, five would be expected to have $p<0.05$; see e.g. \citealt{holm79}, as well as \citealt{2019Kruijssen} for a recent discussion of this). The $p$-value also assumes all errors are equal, which is not the case here. Instead, we take a Monte-Carlo approach to assign a statistical error to the significance of the correlation. For $10{,}000$ realisations, we perturb each value by its associated errors in both $x$ and $y$, and repeat the $\tau$ calculation. We then quote the median value of the correlation, and the $1\sigma$ spread in these values. If the correlation is larger than this 1$\sigma$ error, then the correlation is significant.

Ideally, we should see no correlations with projection parameters. Reassuringly, the only significant correlations we see are between $S_{\rm bar}$ and $\Delta {\rm PA}$, and $\mathcal{R}$ and $\Delta {\rm PA}$. These may give some artificially more significant correlation going forwards, but otherwise we are confident that the projection of the galaxy will not systematically bias our results. If we consider the PHANGS-ALMA points we flag as well-constrained, which cover a wider range of galaxy inclinations, there is clearly an anti-correlation between bar length and inclination and a correlation between bar length and $\Delta {\rm PA}$. As these bars are typically defined by-eye, at high inclination recovering the true bar length becomes increasingly difficult, and this tends to lead to the length being underestimated. For studies that attempt to measure such quantities in highly inclined systems, it is therefore important to take this bias into account.

We next correlate these quantities with a number of global galaxy parameters. These are the Hubble morphological type (from HyperLeda), the bulge-to-total flux ratio at 3.6\micron\ (B/T) and galaxy disk scale length \citep[$h_r$; both from][]{2015Salo}, the total stellar mass ($M_\ast$) from the ${z=0}$ Multiwavelength Galaxy Synthesis \citep[z0MGS;][]{2019Leroy} normalised by $R_{25}$ for a measure of the average stellar surface density of the galaxy. We also take the asymptotic velocity of the galaxy rotation curve ($v_{r, {\rm inf}}$) and turnover radius between the rising and flat part of the rotation curve ($R_t$) from \cite[][see their Eq.~10]{2020Lang}, and the molecular gas fraction \citep[$f_{\rm mol}$;][]{PHANGSSurvey}. These are shown in Fig.~\ref{fig:correlations_global_parameters}, where we also highlight the number of points in each combination of these parameters, along with the Kendall~$\tau$. We colour each point by inclination to highlight any potential dependence of these correlations with galaxy orientation (see Fig.~\ref{fig:correlations_projection}).

Even given our relatively small number of galaxies, we find a number of significant correlations. In terms of the pattern speeds themselves, we find higher $\Omega_{\rm P}$ in later-type galaxies, and galaxies that are more molecular gas-dominated. We also find lower pattern speeds in galaxies that are more bulge-dominated (with a higher bulge-to-total flux ratio), and those with a larger scale length. We find the co-rotation radius to be larger in more bulge-dominated galaxies, and also galaxies with a larger scale length. We see no significant correlations with the $R_{25}$ normalised bar length. We find that $\mathcal{R}$ tends to be higher in later-type galaxies. This appears to be contradictory to numerical simulations, showing that the bar pattern speed slows down over time \citep{2018Wu}. However, as we do not know the age of the bars in these galaxies, we cannot draw this conclusion from these results. Further work to calculate the bar ages in these galaxies may help to answer this question. We also find that $\mathcal{R}$ tends to be lower in galaxies with a larger scale length. We also see that bars tend to be stronger in galaxies with a higher $R_t$, and a higher $f_{\rm mol}$. Whilst we do see correlations between other parameters in this figure, the small number of galaxies means that we cannot robustly draw conclusions about the strength of the correlations between them. We see no strong dependence with inclination in any of these correlations, and we therefore are confident any biases seen in Fig.~\ref{fig:correlations_projection} do not strongly affect the results presented here. Across all of these parameters, our single non-barred galaxy (NGC~0628) occupies the same parameter space as the barred galaxies, as do galaxies with clear spiral arms versus more flocculent morphologies.

The grey points indicate PHANGS-ALMA values that we consider well-constrained (those that the straight line fits to $\langle v \rangle$ versus $\langle x \rangle$ look reasonable, although these are $\Omega_{\rm clump}$ values, rather than $\Omega_{\rm P}$). Notably, these values -- even if they do not measure the true pattern speed of the structure present in the disk -- appear to occupy very similar regions in parameter space as our more robust stellar pattern speed estimates. This serves as a warning that similar measurements made with ISM tracers may give the appearance of meaningful correlations between dynamical structure and global galaxy properties, even when none are present. Indeed, $\Omega_{\rm clump}$ exhibits a trend with stellar mass surface density, whilst we see no significant trends with our stellar mass-based pattern speeds. Based on our studies described in Sect.~\ref{sec:clumpy_ism} (which suggest that $\Omega_{\rm clump}$ is a function of both the shape of the rotation curve of the galaxy as well as the radial extent of the ISM clumps), the trends in Fig.~\ref{fig:correlations_global_parameters} arise as a result of a relation between the galaxy gravitational potential and the way the ISM is distributed within it.

Finally, we also correlate the $y$-axis quantities of Fig.~\ref{fig:correlations_global_parameters} with each other, and this is shown in Fig.~\ref{fig:correlations_corner}. The only significant correlation we see here is unsurprising -- that galaxies with higher pattern speeds have shorter co-rotation radii. We also see that there is variation both in $R_{\rm CR}$ and $R_{\rm bar}/R_{25}$ with $\Omega_{\rm P}$ (although we do not find these correlations individually to be significant). The decrease in $R_{\rm bar}/R_{25}$ appears to be driven mainly by the decrease in $R_{\rm bar}$, and this means we do not see a strong variation in $\mathcal{R}$ across our galaxies.

\section{Discussion}\label{sec:discussion}

This work presents homogeneous measurements of pattern speeds and resonance locations for ten galaxies observed as part of the PHANGS-MUSE survey. This sample size means that we can draw exploratory conclusions about trends in these parameters with the global properties of the galaxy, and, as our methodology for the application of the Tremaine-Weinberg method is generally data-driven, with minimal qualitative checks (except for the quality flagging at the end), we believe that future studies can build on the present work in a homogeneous way.

We have first investigated the scatter in $\mathcal{R}$, the ratio of co-rotation to bar length. Theoretical arguments by \cite{1980Contopoulos} have shown this value should be a little larger than~1, and measurements of $\mathcal{R}$ in simulation \citep{1992Athanassoula,2000DebattistaSellwood} and in observation \citep[e.g.][]{1995MerrifieldKuijken,1999Gerssen,2017Font} have shown $1 < \mathcal{R} < 1.4$. This has led to, when the co-rotation radius cannot be measured from the pattern speed, a common expectation that $R_{\rm CR} \sim 1.2 R_{\rm bar}$. We find for our sample that the median and 16$^{\rm th}$/84$^{\rm th}$ percentile spread in the sample is $\mathcal{R} = 1.3^{+0.4}_{-0.5}$. The scatter on this distribution is consistent
with the typical uncertainty in the measurement of \rcr. Thus, we find that our sample is compatible with a value of $\mathcal{R}=1.2$. We also find this to be the case for the ratio of the co-rotation radius, \rcr, to the disk scale length, $h_r$. This ratio is commonly assumed to be~3 \citep{2003Kranz}, with little scatter. We find a large scatter, with a median and 16$^{\rm th}$/84$^{\rm th}$ percentile spread in the sample of \rcr/$h_r = 2.7^{+0.7}_{-0.1}$. There is some scatter here, but given the small number statistics of this study we find our measurement to be in agreement with the commonly assumed \rcr/$h_r=3$.

\subsection{Relationships between pattern speeds, co-rotation radii, and bar parameters with global galaxy properties}

Next, we discuss the relationships between the pattern speeds, co-rotation radii, and bar parameters in the context of global galaxy parameters (Fig.~\ref{fig:correlations_global_parameters}). Even given our small number of galaxies, we find some significant correlations here. For galaxies with a later Hubble type, we find a larger $\mathcal{R}$ (driven by shorter bar lengths). This is in agreement with, e.g. \cite{2005Erwin}, and so this finding is not surprising. We also find that galaxies with a longer disk scale length tend to have a lower $\mathcal{R}$. Given that we expect galaxies with higher scale lengths to have larger \rcr\ \citep{2003Kranz}, but not necessarily larger bars, this naturally follows. More molecular gas-rich systems tend to have a higher pattern speed. This has been observed in simulations by \cite{2016Ghosh}, and also in some observational studies \citep{2015Aguerri,2019Guo,2020GarmaOehmichen}. Galaxies that have larger bulges tend to have somewhat lower pattern speeds. This is seen in simulations \citep{2019Kataria}, and our results appear to confirm that. Finally, unlike \cite{2020Cuomo}, we do not find a significant correlation between $\Omega_{\rm P}$ and the stellar mass surface density.

\subsection{Relationships between pattern speeds, co-rotation radii, and bar parameters}

Finally, we turn to the inter-relation between our derived parameters with various bar parameters (Fig.~\ref{fig:correlations_corner}). We find only one significant correlation: galaxies with higher pattern speeds tend to have smaller co-rotation radii. Despite this, our sample of galaxies has a roughly constant $\mathcal{R}$. Although, it is also informative to look at parameters where we find no significant correlations -- as in \cite{2020Cuomo}, we find no correlation between $\mathcal{R}$ and $R_{\rm bar}$. Their galaxy sample tends to focus on earlier-type galaxies than the PHANGS sample, and so this may be a trend throughout the entire galaxy population. Unlike previous works \citep[e.g.][]{2005Erwin,2016DiazGarcia,2018Kruk,2019Guo}, we find no trend between bar length and bar strength. We also find no correlation between $\mathcal{R}$ and $S_{\rm bar}$, much like in \cite{2020Cuomo}. Finally, there is a theoretical prediction that there should be an anti-correlation between $S_{\rm bar}$ and $\Omega_{\rm P}$ \citep[e.g.][]{2000DebattistaSellwood, 2003Athanassoula}. As a bar loses angular momentum, it slows down and hence the pattern speed reduces. However, we do not see this in our sample. This may be due to low number statistics, or the variety of initial conditions in these galaxies washing out any correlation we may see between the bar strength and the pattern speed. It may be useful to study this in the context of more similar galaxies (e.g. age, size, morphology), but our sample has insufficient numbers to allow for this sort of binning.

\section{Conclusions}\label{sec:conclusions}

In this work, we have applied the Tremaine-Weinberg method to stellar mass maps obtained as part of the PHANGS-MUSE survey, to measure pattern speeds for a sample of galaxies. This work improves on previous studies in a number of ways. These pattern speeds are calculated homogeneously, with a consistent methodology on a consistent kinematic tracer. We rigorously account for various uncertainties in our measurements of $\Omega_{\rm P}$, and perform a number of tests that allow us to determine whether these pattern speeds are reliable. We find that of our sample of 19 galaxies, ten have well measured pattern speeds. For the nine galaxies that do not have well-measured pattern speeds, in two we detect no evidence of a pattern (IC~5332 and NGC~5068), and for the others we find the velocity field to be extremely messy, meaning the galaxy position angle changes strongly with radius.

Leveraging H$\alpha$ maps from PHANGS-MUSE, as well as CO maps from PHANGS-ALMA, we also critically examine the use of clumpy ISM tracers in determining pattern speeds. We find that the incomplete coverage can lead to a false signal in Tremaine-Weinberg integrals (which we refer to as $\Omega_{\rm clump}$), and that this $\Omega_{\rm clump}$ is systematically different from $\Omega_{\rm P}$ ($\sim$40\% higher for CO, $\sim$20\% higher for H$\alpha$). An important result from this paper, therefore, is that these tracers produce a compromised measure of the pattern speed. We thus caution the reliability of pattern speeds in the literature derived from ISM tracers.

With these pattern speeds, combined with measured CO-based rotation curves for PHANGS galaxies by \cite{2020Lang}, we calculate a number of resonances for the sample: the co-rotation radius, and the outer and inner Lindblad radii. The full list of pattern speeds, quality flags, and resonance locations are given in Table~\ref{table:pattern_speeds}, along with our adopted orientation parameters. We have compared the co-rotation radius both to the bar length and disk scale length, and find that these ratios are consistent with commonly assumed literature values of 1.2, and~3, respectively. Given the spread in our values, we would suggest taking an uncertainty in these ratios of $\sim$30\% and $\sim$15\%, respectively.

We have investigated how our derived parameters depend on a number of global galaxy parameters. We confirm a number of previous findings in our work; later-type galaxies have higher pattern speeds, and larger values of $\mathcal{R}$, that more molecular gas-dominated galaxies have higher pattern speeds, and that more bulge-dominated galaxies have lower pattern speeds. However, we also find an absence of correlations where we may expect them. In particular, we find no correlation between bar strength and pattern speed, nor between total stellar mass surface density and pattern speed. With larger number statistics the small correlations we see here may become more significant, and future work will be able to rule out whether these correlations truly are absent, or simply due to small number statistics.

There are a number of future studies that follow on naturally from this work. Firstly, it is important to note that the Tremaine-Weinberg method can only recover a single pattern speed in a galaxy. We may expect different morphological features to have different pattern speeds, which we are unable to recover in this work. We have flagged those we believe to have strong signals of multiple pattern speeds. Applying a method that allows for radial variation in the pattern speed has been beyond the scope of this work, but will be the focus of future work. Secondly, our work has been applied to data with cloud-scale resolution (1\arcsec, corresponding to $\sim$100\,pc). We have performed tests for a number of slit widths ranging from $\sim$100\,pc to $\sim$1\,kpc, and find that the width of the slit has little effect on the pattern speed measurement. Thus, it should be possible to extend this study to much larger samples of galaxies with surveys such as MaNGA, which have $\sim$kpc resolution. Given that we are only confident in pattern speeds measured using stellar masses and velocities, this work is hindered by low number statistics. The possibility of having reliable pattern speeds for a factor of 100 more galaxies will present a significant increase in statistical power for studying the links between the properties of galaxies and pattern speeds.

\acknowledgments

This work has been carried out as part of the PHANGS collaboration.

The authors would like to thank the anonymous peer reviewer for their constructive comments, that have improved the quality of the paper. TGW would also like to personally thank P. Woolford for discussions and support during the preparation of this manuscript. TGW, ES, H-AP, TS, FS acknowledge funding from the European Research Council (ERC) under the European Union’s Horizon 2020 research and innovation programme (grant agreement No. 694343). IB, FB acknowledge funding from the European Research Council (ERC) under the European Union’s Horizon 2020 research and innovation programme (grant agreement No.726384/Empire). SCOG, RSK, MCS, EJW acknowledge financial support from the German Research Foundation (DFG) via the Collaborative Research Center (SFB 881, Project-ID 138713538) `The Milky Way System' (subprojects A1, B1, B2, B8, and P2). SCOG and RSK also acknowledge financial support from the Heidelberg Cluster of Excellence STRUCTURES in the framework of Germany's Excellence Strategy (grant EXC-2181/1 - 390900948), and from the ERC via the ERC Synergy Grant ECOGAL (grant 855130). JMDK and MC gratefully acknowledges funding from the Deutsche Forschungsgemeinschaft (DFG) through an Emmy Noether Research Group, grant number KR4801/1-1 and the DFG Sachbeihilfe, grant number KR4801/2-1. JMDK gratefully acknowledges funding from the European Research Council (ERC) under the European Union's Horizon 2020 research and innovation programme via the ERC Starting Grant MUSTANG (grant agreement number 714907). The work of AKL and JS is partially supported by the National Science Foundation under Grants No. 1615105, 1615109, and 1653300. ER acknowledges the support of the Natural Sciences and Engineering Research Council of Canada (NSERC), funding reference number RGPIN-2017-03987.

This paper makes use of the following ALMA data, which have been processed as part of the PHANGS-ALMA survey:\\
ADS/JAO.ALMA\#2012.1.00650.S, \linebreak 
ADS/JAO.ALMA\#2013.1.00803.S, \linebreak 
ADS/JAO.ALMA\#2013.1.01161.S, \linebreak 
ADS/JAO.ALMA\#2015.1.00121.S, \linebreak 
ADS/JAO.ALMA\#2015.1.00782.S, \linebreak 
ADS/JAO.ALMA\#2015.1.00925.S, \linebreak 
ADS/JAO.ALMA\#2015.1.00956.S, \linebreak 
ADS/JAO.ALMA\#2016.1.00386.S, \linebreak 
ADS/JAO.ALMA\#2017.1.00392.S, \linebreak 
ADS/JAO.ALMA\#2017.1.00766.S, \linebreak 
ADS/JAO.ALMA\#2017.1.00886.L, \linebreak 
ADS/JAO.ALMA\#2018.1.01321.S, \linebreak 
ADS/JAO.ALMA\#2018.1.01651.S. \linebreak 
ADS/JAO.ALMA\#2018.A.00062.S. \linebreak 
ALMA is a partnership of ESO (representing its member states), NSF (USA), and NINS (Japan), together with NRC (Canada), NSC and ASIAA (Taiwan), and KASI (Republic of Korea), in cooperation with the Republic of Chile. The Joint ALMA Observatory is operated by ESO, AUI/NRAO, and NAOJ. The National Radio Astronomy Observatory is a facility of the National Science Foundation operated under cooperative agreement by Associated Universities, Inc.

Based on observations collected at the European Southern Observatory under ESO programmes 1100.B-0651, 095.C-0473, and 094.C-0623.

This work has made use of {\tt NumPy} \citep{2006Oliphant}, {\tt AstroPy} \citep{2013Astropy,2018Astropy}, {\tt SciPy} \citep{2020SciPy}, {\tt Matplotlib} \citep{2007Hunter}, and {\tt Seaborn} \citep{2017Waskom}. This work has also made use of the HyperLeda\footnote{\url{http://leda.univ-lyon1.fr}} database.

\bibliographystyle{yahapj}
\bibliography{bibliography}

\appendix

\section{Derived Pattern Speeds and Resonance Locations}

\clearpage
\begin{longrotatetable}
\begin{deluxetable*}{ccccccccchhhhhhhhhhhhhccchhhhhhhhhhhhhccchhhhhhhhhhh}
\tablecaption{``Pattern speeds'' ($\Omega_p$ for stellar mass, $\Omega_{\rm clump}$ for ISM tracers) and inferred resonance locations for 83 galaxies. A description of the column names is given in the Table notes.}
\label{table:pattern_speeds}
\tablehead{\colhead{Galaxy} & \colhead{PGC} & \colhead{D} & \colhead{$i$} & \colhead{PA} & \colhead{Bar?} & \colhead{$\Omega_{\rm clump, A}$} & \colhead{Q$_{\rm A}$} & \colhead{$R_{\rm CR1, A}$} & \nocolhead{$R_{\rm CR2, A}$} & \nocolhead{$R_{\rm CR3, A}$} & \nocolhead{$R_{\rm CR4, A}$} & \nocolhead{$R_{\rm ILR1, A}$} & \nocolhead{$R_{\rm ILR2, A}$} & \nocolhead{$R_{\rm ILR3, A}$} & \nocolhead{$R_{\rm ILR4, A}$} & \nocolhead{$R_{\rm ILR5, A}$} & \nocolhead{$R_{\rm ILR6, A}$} & \nocolhead{$R_{\rm ILR7, A}$} & \nocolhead{$R_{\rm OLR1, A}$} & \nocolhead{$R_{\rm OLR2, A}$} & \nocolhead{$R_{\rm OLR3, A}$} & \colhead{$\Omega_{\rm P, MM}$} & \colhead{Q$_{\rm MM}$} & \colhead{$R_{\rm CR1, MM}$} & \nocolhead{$R_{\rm CR2, MM}$} & \nocolhead{$R_{\rm ILR1, MM}$} & \nocolhead{$R_{\rm ILR2, MM}$} & \nocolhead{$R_{\rm ILR3, MM}$} & \nocolhead{$R_{\rm ILR4, MM}$} & \nocolhead{$R_{\rm ILR5, MM}$} & \nocolhead{$R_{\rm ILR6, MM}$} & \nocolhead{$R_{\rm ILR7, MM}$} & \nocolhead{$R_{\rm ILR8, MM}$} & \nocolhead{$R_{\rm ILR9, MM}$} & \nocolhead{$R_{\rm OLR1, MM}$} & \nocolhead{$R_{\rm OLR2, MM}$} & \nocolhead{$R_{\rm OLR3, MM}$} & \colhead{$\Omega_{\rm clump, MH{\alpha}}$} & \colhead{Q$_{\rm MH{\alpha}}$} & \colhead{$R_{\rm CR1, MH{\alpha}}$} & \nocolhead{$R_{\rm CR2, MH{\alpha}}$} & \nocolhead{$R_{\rm ILR1, MH{\alpha}}$} & \nocolhead{$R_{\rm ILR2, MH{\alpha}}$} & \nocolhead{$R_{\rm ILR3, MH{\alpha}}$} & \nocolhead{$R_{\rm ILR4, MH{\alpha}}$} & \nocolhead{$R_{\rm ILR5, MH{\alpha}}$} & \nocolhead{$R_{\rm ILR6, MH{\alpha}}$} & \nocolhead{$R_{\rm ILR7, MH{\alpha}}$} & \nocolhead{$R_{\rm OLR1, MH{\alpha}}$} & \nocolhead{$R_{\rm OLR2, MH{\alpha}}$} & \nocolhead{$R_{\rm OLR3, MH{\alpha}}$}\\ \colhead{ } & \colhead{ } & \colhead{$\mathrm{Mpc}$} & \colhead{$\mathrm{{}^{\circ}}$} & \colhead{$\mathrm{{}^{\circ}}$} & \colhead{ } & \colhead{$\mathrm{\frac{km}{kpc\,s}}$} & \colhead{ } & \colhead{$\mathrm{kpc}$} & \nocolhead{$\mathrm{kpc}$} & \nocolhead{$\mathrm{kpc}$} & \nocolhead{$\mathrm{kpc}$} & \nocolhead{$\mathrm{kpc}$} & \nocolhead{$\mathrm{kpc}$} & \nocolhead{$\mathrm{kpc}$} & \nocolhead{$\mathrm{kpc}$} & \nocolhead{$\mathrm{kpc}$} & \nocolhead{$\mathrm{kpc}$} & \nocolhead{$\mathrm{kpc}$} & \nocolhead{$\mathrm{kpc}$} & \nocolhead{$\mathrm{kpc}$} & \nocolhead{$\mathrm{kpc}$} & \colhead{$\mathrm{\frac{km}{kpc\,s}}$} & \colhead{ } & \colhead{$\mathrm{kpc}$} & \nocolhead{$\mathrm{kpc}$} & \nocolhead{$\mathrm{kpc}$} & \nocolhead{$\mathrm{kpc}$} & \nocolhead{$\mathrm{kpc}$} & \nocolhead{$\mathrm{kpc}$} & \nocolhead{$\mathrm{kpc}$} & \nocolhead{$\mathrm{kpc}$} & \nocolhead{$\mathrm{kpc}$} & \nocolhead{$\mathrm{kpc}$} & \nocolhead{$\mathrm{kpc}$} & \nocolhead{$\mathrm{kpc}$} & \nocolhead{$\mathrm{kpc}$} & \nocolhead{$\mathrm{kpc}$} & \colhead{$\mathrm{\frac{km}{kpc\,s}}$} & \colhead{ } & \colhead{$\mathrm{kpc}$} & \nocolhead{$\mathrm{kpc}$} & \nocolhead{$\mathrm{kpc}$} & \nocolhead{$\mathrm{kpc}$} & \nocolhead{$\mathrm{kpc}$} & \nocolhead{$\mathrm{kpc}$} & \nocolhead{$\mathrm{kpc}$} & \nocolhead{$\mathrm{kpc}$} & \nocolhead{$\mathrm{kpc}$} & \nocolhead{$\mathrm{kpc}$} & \nocolhead{$\mathrm{kpc}$} & \nocolhead{$\mathrm{kpc}$}}
\startdata
ESO097-013 & 50779 & 4.2 & 64.3 & 36.74 & 0 & $-12.8^{+68.7}_{-124.8}$ & 3 & \nodata & \nodata & \nodata & \nodata & \nodata & \nodata & \nodata & \nodata & \nodata & \nodata & \nodata & \nodata & \nodata & \nodata & \nodata & \nodata & \nodata & \nodata & \nodata & \nodata & \nodata & \nodata & \nodata & \nodata & \nodata & \nodata & \nodata & \nodata & \nodata & \nodata & \nodata & \nodata & \nodata & \nodata & \nodata & \nodata & \nodata & \nodata & \nodata & \nodata & \nodata & \nodata & \nodata & \nodata \\
IC1954 & 13090 & 12.0 & 57.1 & 63.4 & 1 & $69.3^{+1.4}_{-1.8}$ & 4 & $0.1\pm0.1$ & $1.5\pm0.2$ & \nodata & \nodata & \nodata & \nodata & \nodata & \nodata & \nodata & \nodata & \nodata & $3.3\pm0.2$ & \nodata & \nodata & \nodata & \nodata & \nodata & \nodata & \nodata & \nodata & \nodata & \nodata & \nodata & \nodata & \nodata & \nodata & \nodata & \nodata & \nodata & \nodata & \nodata & \nodata & \nodata & \nodata & \nodata & \nodata & \nodata & \nodata & \nodata & \nodata & \nodata & \nodata & \nodata & \nodata \\
IC5273 & 70184 & 14.18 & 52.0 & 234.1 & 1 & $40.6^{+2.8}_{-1.9}$ & 1 & $2.9\pm0.8$ & \nodata & \nodata & \nodata & \nodata & \nodata & \nodata & \nodata & \nodata & \nodata & \nodata & \nodata & \nodata & \nodata & \nodata & \nodata & \nodata & \nodata & \nodata & \nodata & \nodata & \nodata & \nodata & \nodata & \nodata & \nodata & \nodata & \nodata & \nodata & \nodata & \nodata & \nodata & \nodata & \nodata & \nodata & \nodata & \nodata & \nodata & \nodata & \nodata & \nodata & \nodata & \nodata & \nodata \\
IC5332 & 71775 & 8.18 & 26.9 & 74.4 & 0 & \nodata & \nodata & \nodata & \nodata & \nodata & \nodata & \nodata & \nodata & \nodata & \nodata & \nodata & \nodata & \nodata & \nodata & \nodata & \nodata & $-9.2^{+16.5}_{-4.4}$ & 3 & \nodata & \nodata & \nodata & \nodata & \nodata & \nodata & \nodata & \nodata & \nodata & \nodata & \nodata & \nodata & \nodata & \nodata & $31.4^{+1.2}_{-1.0}$ & 3 & \nodata & \nodata & \nodata & \nodata & \nodata & \nodata & \nodata & \nodata & \nodata & \nodata & \nodata & \nodata \\
NGC0253 & 2789 & 3.7 & 75.0 & 52.48 & 0 & $51.4^{+32.0}_{-7.1}$ & 3 & \nodata & \nodata & \nodata & \nodata & \nodata & \nodata & \nodata & \nodata & \nodata & \nodata & \nodata & \nodata & \nodata & \nodata & \nodata & \nodata & \nodata & \nodata & \nodata & \nodata & \nodata & \nodata & \nodata & \nodata & \nodata & \nodata & \nodata & \nodata & \nodata & \nodata & \nodata & \nodata & \nodata & \nodata & \nodata & \nodata & \nodata & \nodata & \nodata & \nodata & \nodata & \nodata & \nodata & \nodata \\
NGC0300 & 3238 & 2.09 & 39.8 & 114.3 & 0 & $31.6^{+4.7}_{-4.2}$ & 3 & \nodata & \nodata & \nodata & \nodata & \nodata & \nodata & \nodata & \nodata & \nodata & \nodata & \nodata & \nodata & \nodata & \nodata & \nodata & \nodata & \nodata & \nodata & \nodata & \nodata & \nodata & \nodata & \nodata & \nodata & \nodata & \nodata & \nodata & \nodata & \nodata & \nodata & \nodata & \nodata & \nodata & \nodata & \nodata & \nodata & \nodata & \nodata & \nodata & \nodata & \nodata & \nodata & \nodata & \nodata \\
NGC0628 & 5974 & 9.84 & 8.9 & 20.7 & 0 & $32.9^{+0.2}_{-0.3}$ & 1 & $4.4\pm1.0$ & \nodata & \nodata & \nodata & $0.8\pm0.0$ & $2.2\pm0.0$ & $2.4\pm0.0$ & \nodata & \nodata & \nodata & \nodata & $6.0\pm0.0$ & $6.2\pm0.0$ & \nodata & $31.1^{+4.0}_{-2.9}$ & 1 & $4.5\pm2.0$ & \nodata & $0.8\pm0.0$ & $2.2\pm0.0$ & $2.4\pm0.0$ & \nodata & \nodata & \nodata & \nodata & \nodata & \nodata & $6.0\pm0.0$ & $6.2\pm0.0$ & \nodata & $35.4^{+0.4}_{-0.5}$ & 1 & $3.8\pm1.2$ & \nodata & $0.8\pm0.0$ & $2.3\pm0.0$ & $2.4\pm0.0$ & \nodata & \nodata & \nodata & \nodata & $6.0\pm0.0$ & $6.2\pm0.0$ & \nodata \\
NGC0685 & 6581 & 19.94 & 23.0 & 100.9 & 1 & $36.7^{+2.9}_{-2.3}$ & 1 & $0.1\pm0.1$ & $3.1\pm2.6$ & \nodata & \nodata & \nodata & \nodata & \nodata & \nodata & \nodata & \nodata & \nodata & $5.3\pm0.1$ & \nodata & \nodata & \nodata & \nodata & \nodata & \nodata & \nodata & \nodata & \nodata & \nodata & \nodata & \nodata & \nodata & \nodata & \nodata & \nodata & \nodata & \nodata & \nodata & \nodata & \nodata & \nodata & \nodata & \nodata & \nodata & \nodata & \nodata & \nodata & \nodata & \nodata & \nodata & \nodata \\
NGC1087 & 10496 & 15.85 & 42.9 & 359.1 & 1 & $46.7^{+0.6}_{-0.8}$ & 1 & $3.4\pm0.7$ & \nodata & \nodata & \nodata & \nodata & \nodata & \nodata & \nodata & \nodata & \nodata & \nodata & $5.2\pm0.1$ & \nodata & \nodata & $31.9^{+3.2}_{-1.6}$ & 1 & $4.3\pm1.1$ & \nodata & $0.5\pm0.1$ & \nodata & \nodata & \nodata & \nodata & \nodata & \nodata & \nodata & \nodata & \nodata & \nodata & \nodata & $49.7^{+6.9}_{-7.4}$ & 1 & $3.7\pm1.7$ & \nodata & $0.5\pm0.0$ & \nodata & \nodata & \nodata & \nodata & \nodata & \nodata & $4.8\pm0.5$ & \nodata & \nodata \\
NGC1097 & 10488 & 13.58 & 48.6 & 122.4 & 1 & $39.9^{+2.6}_{-0.8}$ & 2 & $6.4\pm2.0$ & $8.7\pm0.1$ & \nodata & \nodata & $1.6\pm0.0$ & $1.7\pm0.0$ & $2.0\pm0.0$ & $3.3\pm0.0$ & $3.4\pm0.0$ & \nodata & \nodata & $8.1\pm0.0$ & $9.6\pm0.0$ & \nodata & \nodata & \nodata & \nodata & \nodata & \nodata & \nodata & \nodata & \nodata & \nodata & \nodata & \nodata & \nodata & \nodata & \nodata & \nodata & \nodata & \nodata & \nodata & \nodata & \nodata & \nodata & \nodata & \nodata & \nodata & \nodata & \nodata & \nodata & \nodata & \nodata & \nodata \\
NGC1300 & 12412 & 18.99 & 31.8 & 278.0 & 1 & $24.9^{+1.7}_{-0.2}$ & 1 & $6.1\pm2.5$ & \nodata & \nodata & \nodata & $1.4\pm0.0$ & $1.5\pm0.0$ & \nodata & \nodata & \nodata & \nodata & \nodata & \nodata & \nodata & \nodata & $18.9^{+9.4}_{-8.9}$ & 3 & $0.1\pm0.1$ & $6.1\pm2.5$ & $1.4\pm0.1$ & $6.3\pm0.1$ & $7.0\pm0.2$ & $7.5\pm0.2$ & $8.2\pm0.2$ & \nodata & \nodata & \nodata & \nodata & \nodata & \nodata & \nodata & $25.1^{+0.7}_{-0.6}$ & 1 & $6.1\pm2.5$ & \nodata & $1.4\pm0.0$ & $1.4\pm0.0$ & \nodata & \nodata & \nodata & \nodata & \nodata & \nodata & \nodata & \nodata \\
NGC1317 & 12653 & 19.11 & 23.2 & 221.5 & 1 & $94.6^{+43.7}_{-14.9}$ & 3 & $0.2\pm0.2$ & $1.6\pm1.1$ & \nodata & \nodata & $0.7\pm0.2$ & \nodata & \nodata & \nodata & \nodata & \nodata & \nodata & $1.9\pm0.6$ & \nodata & \nodata & \nodata & \nodata & \nodata & \nodata & \nodata & \nodata & \nodata & \nodata & \nodata & \nodata & \nodata & \nodata & \nodata & \nodata & \nodata & \nodata & \nodata & \nodata & \nodata & \nodata & \nodata & \nodata & \nodata & \nodata & \nodata & \nodata & \nodata & \nodata & \nodata & \nodata \\
NGC1365 & 13179 & 19.57 & 55.4 & 201.1 & 1 & $109.1^{+22.0}_{-36.8}$ & 3 & $2.3\pm1.6$ & \nodata & \nodata & \nodata & $0.4\pm0.3$ & $1.4\pm0.2$ & $1.7\pm0.0$ & $2.0\pm0.1$ & $3.8\pm0.0$ & \nodata & \nodata & $1.6\pm0.0$ & $2.8\pm1.0$ & \nodata & $38.1^{+20.1}_{-21.2}$ & 3 & $2.8\pm1.1$ & \nodata & $1.0\pm0.6$ & $2.1\pm0.4$ & $2.9\pm0.2$ & $3.7\pm0.1$ & \nodata & \nodata & \nodata & \nodata & \nodata & $3.8\pm0.0$ & \nodata & \nodata & $26.8^{+5.6}_{-20.7}$ & 2 & \nodata & \nodata & $1.0\pm0.3$ & $1.9\pm0.1$ & $2.7\pm0.6$ & $3.7\pm0.1$ & \nodata & \nodata & \nodata & \nodata & \nodata & \nodata \\
NGC1385 & 13368 & 17.22 & 44.0 & 181.3 & 0 & $27.4^{+4.1}_{-7.1}$ & 3 & $0.1\pm0.1$ & $4.2\pm1.4$ & \nodata & \nodata & $0.6\pm0.1$ & $1.7\pm0.8$ & $3.0\pm0.2$ & $4.1\pm0.1$ & $4.4\pm0.0$ & $5.0\pm0.3$ & \nodata & $4.1\pm0.0$ & $4.4\pm0.0$ & $5.2\pm0.4$ & $25.0^{+5.6}_{-19.3}$ & 3 & $0.1\pm0.1$ & $4.2\pm1.4$ & $2.1\pm2.0$ & $5.0\pm0.6$ & \nodata & \nodata & \nodata & \nodata & \nodata & \nodata & \nodata & $4.1\pm0.0$ & $4.4\pm0.0$ & $5.2\pm0.4$ & $21.9^{+10.4}_{-5.9}$ & 3 & $0.1\pm0.1$ & $4.0\pm1.6$ & $0.6\pm0.2$ & $1.7\pm0.8$ & $3.1\pm0.2$ & $4.0\pm0.1$ & $4.4\pm0.0$ & $5.1\pm0.4$ & \nodata & $4.1\pm0.0$ & $4.4\pm0.0$ & $5.1\pm0.4$ \\
NGC1433 & 13586 & 12.11 & 28.6 & 199.7 & 1 & $105.6^{+4.6}_{-8.1}$ & 3 & $0.1\pm0.1$ & $0.3\pm0.1$ & $1.8\pm0.8$ & \nodata & \nodata & \nodata & \nodata & \nodata & \nodata & \nodata & \nodata & $4.7\pm0.0$ & \nodata & \nodata & $20.0^{+2.4}_{-1.9}$ & 1 & $6.3\pm0.4$ & \nodata & $0.3\pm0.0$ & $0.6\pm0.0$ & $0.8\pm0.0$ & $0.8\pm0.0$ & $1.2\pm0.0$ & $1.4\pm0.0$ & $3.9\pm0.1$ & $4.4\pm0.0$ & $4.6\pm0.0$ & \nodata & \nodata & \nodata & $41.2^{+0.7}_{-0.7}$ & 1 & $0.1\pm0.1$ & $3.7\pm1.1$ & $0.2\pm0.0$ & $0.6\pm0.0$ & $0.7\pm0.0$ & $0.9\pm0.0$ & $1.2\pm0.0$ & $1.4\pm0.0$ & $4.4\pm0.0$ & \nodata & \nodata & \nodata \\
NGC1511 & 14236 & 15.28 & 72.7 & 297.0 & 0 & $44.3^{+1.6}_{-1.0}$ & 1 & $0.3\pm0.3$ & $0.9\pm0.1$ & $2.7\pm0.9$ & \nodata & \nodata & \nodata & \nodata & \nodata & \nodata & \nodata & \nodata & \nodata & \nodata & \nodata & \nodata & \nodata & \nodata & \nodata & \nodata & \nodata & \nodata & \nodata & \nodata & \nodata & \nodata & \nodata & \nodata & \nodata & \nodata & \nodata & \nodata & \nodata & \nodata & \nodata & \nodata & \nodata & \nodata & \nodata & \nodata & \nodata & \nodata & \nodata & \nodata & \nodata \\
NGC1512 & 14391 & 17.13 & 42.5 & 261.9 & 1 & $27.6^{+2.3}_{-2.0}$ & 1 & $5.4\pm2.2$ & \nodata & \nodata & \nodata & \nodata & \nodata & \nodata & \nodata & \nodata & \nodata & \nodata & \nodata & \nodata & \nodata & $22.3^{+5.1}_{-5.5}$ & 1 & $5.4\pm2.2$ & \nodata & $7.2\pm0.2$ & \nodata & \nodata & \nodata & \nodata & \nodata & \nodata & \nodata & \nodata & \nodata & \nodata & \nodata & $25.0^{+2.3}_{-2.2}$ & 1 & $5.4\pm2.2$ & \nodata & \nodata & \nodata & \nodata & \nodata & \nodata & \nodata & \nodata & \nodata & \nodata & \nodata \\
NGC1546 & 14723 & 17.69 & 70.3 & 147.8 & 0 & $74.5^{+1.8}_{-2.1}$ & 3 & $2.3\pm0.4$ & \nodata & \nodata & \nodata & \nodata & \nodata & \nodata & \nodata & \nodata & \nodata & \nodata & $3.9\pm0.2$ & \nodata & \nodata & \nodata & \nodata & \nodata & \nodata & \nodata & \nodata & \nodata & \nodata & \nodata & \nodata & \nodata & \nodata & \nodata & \nodata & \nodata & \nodata & \nodata & \nodata & \nodata & \nodata & \nodata & \nodata & \nodata & \nodata & \nodata & \nodata & \nodata & \nodata & \nodata & \nodata \\
NGC1559 & 14814 & 19.44 & 65.4 & 244.5 & 1 & $21.2^{+3.1}_{-1.2}$ & 1 & $0.2\pm0.2$ & $1.0\pm0.4$ & $3.7\pm0.9$ & $6.7\pm1.6$ & $0.6\pm0.0$ & $7.0\pm0.0$ & $7.2\pm0.0$ & \nodata & \nodata & \nodata & \nodata & $7.0\pm0.0$ & $7.3\pm0.0$ & $8.2\pm0.0$ & \nodata & \nodata & \nodata & \nodata & \nodata & \nodata & \nodata & \nodata & \nodata & \nodata & \nodata & \nodata & \nodata & \nodata & \nodata & \nodata & \nodata & \nodata & \nodata & \nodata & \nodata & \nodata & \nodata & \nodata & \nodata & \nodata & \nodata & \nodata & \nodata & \nodata \\
NGC1566 & 14897 & 17.69 & 29.5 & 214.7 & 1 & $58.5^{+7.2}_{-5.7}$ & 1 & $4.0\pm2.5$ & \nodata & \nodata & \nodata & $0.8\pm0.1$ & $1.4\pm0.1$ & \nodata & \nodata & \nodata & \nodata & \nodata & $5.8\pm0.8$ & $8.1\pm0.2$ & \nodata & $29.4^{+9.1}_{-7.2}$ & 3 & $6.8\pm2.7$ & \nodata & $1.6\pm0.1$ & $1.9\pm0.1$ & $4.1\pm0.6$ & $5.2\pm0.2$ & $6.6\pm0.4$ & $7.3\pm0.0$ & $7.7\pm0.0$ & \nodata & \nodata & $7.0\pm0.3$ & $7.8\pm0.1$ & $8.8\pm0.6$ & $64.7^{+6.4}_{-7.1}$ & 1 & $3.8\pm2.3$ & \nodata & $0.8\pm0.1$ & $1.4\pm0.1$ & \nodata & \nodata & \nodata & \nodata & \nodata & $4.4\pm0.1$ & $5.1\pm0.1$ & $5.8\pm0.6$ \\
NGC1637 & 15821 & 11.7 & 31.1 & 20.61 & 1 & $90.5^{+10.2}_{-29.5}$ & 3 & \nodata & \nodata & \nodata & \nodata & \nodata & \nodata & \nodata & \nodata & \nodata & \nodata & \nodata & \nodata & \nodata & \nodata & \nodata & \nodata & \nodata & \nodata & \nodata & \nodata & \nodata & \nodata & \nodata & \nodata & \nodata & \nodata & \nodata & \nodata & \nodata & \nodata & \nodata & \nodata & \nodata & \nodata & \nodata & \nodata & \nodata & \nodata & \nodata & \nodata & \nodata & \nodata & \nodata & \nodata \\
NGC1672 & 15941 & 19.4 & 42.6 & 134.3 & 1 & $32.5^{+2.0}_{-1.3}$ & 2 & $3.3\pm1.0$ & \nodata & \nodata & \nodata & $1.2\pm0.0$ & \nodata & \nodata & \nodata & \nodata & \nodata & \nodata & $4.1\pm0.0$ & $4.5\pm0.0$ & \nodata & $22.7^{+0.6}_{-0.6}$ & 1 & $6.1\pm1.6$ & \nodata & $1.2\pm0.0$ & $4.1\pm0.0$ & $4.4\pm0.0$ & \nodata & \nodata & \nodata & \nodata & \nodata & \nodata & \nodata & \nodata & \nodata & $28.1^{+0.1}_{-0.1}$ & 1 & $4.0\pm0.5$ & \nodata & \nodata & \nodata & \nodata & \nodata & \nodata & \nodata & \nodata & $4.4\pm0.0$ & \nodata & \nodata \\
NGC1792 & 16709 & 16.2 & 65.1 & 318.9 & 0 & $57.7^{+2.8}_{-4.2}$ & 1 & $1.9\pm0.5$ & \nodata & \nodata & \nodata & $0.4\pm0.1$ & \nodata & \nodata & \nodata & \nodata & \nodata & \nodata & $4.8\pm0.6$ & \nodata & \nodata & \nodata & \nodata & \nodata & \nodata & \nodata & \nodata & \nodata & \nodata & \nodata & \nodata & \nodata & \nodata & \nodata & \nodata & \nodata & \nodata & \nodata & \nodata & \nodata & \nodata & \nodata & \nodata & \nodata & \nodata & \nodata & \nodata & \nodata & \nodata & \nodata & \nodata \\
NGC1809 & 16599 & 19.95 & 57.6 & 138.2 & 0 & $47.4^{+3.5}_{-6.1}$ & 4 & $1.4\pm1.1$ & \nodata & \nodata & \nodata & \nodata & \nodata & \nodata & \nodata & \nodata & \nodata & \nodata & \nodata & \nodata & \nodata & \nodata & \nodata & \nodata & \nodata & \nodata & \nodata & \nodata & \nodata & \nodata & \nodata & \nodata & \nodata & \nodata & \nodata & \nodata & \nodata & \nodata & \nodata & \nodata & \nodata & \nodata & \nodata & \nodata & \nodata & \nodata & \nodata & \nodata & \nodata & \nodata & \nodata \\
NGC2090 & 17819 & 11.75 & 64.5 & 192.46 & 0 & $58.1^{+4.0}_{-4.6}$ & 1 & $0.1\pm0.1$ & $2.9\pm0.4$ & \nodata & \nodata & \nodata & \nodata & \nodata & \nodata & \nodata & \nodata & \nodata & \nodata & \nodata & \nodata & \nodata & \nodata & \nodata & \nodata & \nodata & \nodata & \nodata & \nodata & \nodata & \nodata & \nodata & \nodata & \nodata & \nodata & \nodata & \nodata & \nodata & \nodata & \nodata & \nodata & \nodata & \nodata & \nodata & \nodata & \nodata & \nodata & \nodata & \nodata & \nodata & \nodata \\
NGC2283 & 19562 & 13.68 & 43.7 & -4.1 & 1 & $41.0^{+1.9}_{-3.1}$ & 1 & $0.1\pm0.1$ & $2.5\pm0.3$ & \nodata & \nodata & $2.1\pm0.0$ & $2.4\pm0.0$ & \nodata & \nodata & \nodata & \nodata & \nodata & $2.4\pm0.0$ & \nodata & \nodata & \nodata & \nodata & \nodata & \nodata & \nodata & \nodata & \nodata & \nodata & \nodata & \nodata & \nodata & \nodata & \nodata & \nodata & \nodata & \nodata & \nodata & \nodata & \nodata & \nodata & \nodata & \nodata & \nodata & \nodata & \nodata & \nodata & \nodata & \nodata & \nodata & \nodata \\
NGC2566 & 23303 & 23.44 & 48.5 & 312.0 & 1 & $34.6^{+2.0}_{-4.8}$ & 1 & $3.8\pm2.2$ & \nodata & \nodata & \nodata & $0.5\pm0.0$ & $0.6\pm0.0$ & $1.0\pm0.1$ & $1.4\pm0.0$ & $1.8\pm0.1$ & $4.4\pm0.1$ & $4.9\pm0.0$ & $4.8\pm0.0$ & \nodata & \nodata & \nodata & \nodata & \nodata & \nodata & \nodata & \nodata & \nodata & \nodata & \nodata & \nodata & \nodata & \nodata & \nodata & \nodata & \nodata & \nodata & \nodata & \nodata & \nodata & \nodata & \nodata & \nodata & \nodata & \nodata & \nodata & \nodata & \nodata & \nodata & \nodata & \nodata \\
NGC2775 & 25861 & 23.15 & 41.2 & 156.5 & 0 & $54.9^{+0.1}_{-0.1}$ & 1 & $5.6\pm0.2$ & \nodata & \nodata & \nodata & \nodata & \nodata & \nodata & \nodata & \nodata & \nodata & \nodata & \nodata & \nodata & \nodata & \nodata & \nodata & \nodata & \nodata & \nodata & \nodata & \nodata & \nodata & \nodata & \nodata & \nodata & \nodata & \nodata & \nodata & \nodata & \nodata & \nodata & \nodata & \nodata & \nodata & \nodata & \nodata & \nodata & \nodata & \nodata & \nodata & \nodata & \nodata & \nodata & \nodata \\
NGC2835 & 26259 & 12.38 & 41.3 & 1.0 & 1 & $41.3^{+0.3}_{-0.8}$ & 1 & $2.8\pm0.6$ & \nodata & \nodata & \nodata & $0.2\pm0.0$ & $0.5\pm0.0$ & \nodata & \nodata & \nodata & \nodata & \nodata & \nodata & \nodata & \nodata & $34.6^{+4.3}_{-3.1}$ & 1 & $2.8\pm0.6$ & \nodata & $0.2\pm0.1$ & $0.5\pm0.1$ & \nodata & \nodata & \nodata & \nodata & \nodata & \nodata & \nodata & \nodata & \nodata & \nodata & $37.6^{+0.4}_{-0.4}$ & 1 & $3.2\pm0.2$ & \nodata & $0.2\pm0.0$ & $0.5\pm0.0$ & \nodata & \nodata & \nodata & \nodata & \nodata & \nodata & \nodata & \nodata \\
NGC2903 & 27077 & 10.0 & 66.8 & 203.7 & 1 & $25.0^{+1.8}_{-0.5}$ & 1 & \nodata & \nodata & \nodata & \nodata & $0.6\pm0.0$ & $1.9\pm0.1$ & \nodata & \nodata & \nodata & \nodata & \nodata & \nodata & \nodata & \nodata & \nodata & \nodata & \nodata & \nodata & \nodata & \nodata & \nodata & \nodata & \nodata & \nodata & \nodata & \nodata & \nodata & \nodata & \nodata & \nodata & \nodata & \nodata & \nodata & \nodata & \nodata & \nodata & \nodata & \nodata & \nodata & \nodata & \nodata & \nodata & \nodata & \nodata \\
NGC2997 & 27978 & 14.06 & 33.0 & 108.1 & 1 & $36.0^{+0.7}_{-0.3}$ & 4 & $7.2\pm3.8$ & \nodata & \nodata & \nodata & $0.6\pm0.0$ & $1.1\pm0.0$ & $1.3\pm0.0$ & $1.6\pm0.0$ & $2.0\pm0.0$ & \nodata & \nodata & $10.2\pm0.0$ & $10.7\pm0.0$ & $11.0\pm0.0$ & \nodata & \nodata & \nodata & \nodata & \nodata & \nodata & \nodata & \nodata & \nodata & \nodata & \nodata & \nodata & \nodata & \nodata & \nodata & \nodata & \nodata & \nodata & \nodata & \nodata & \nodata & \nodata & \nodata & \nodata & \nodata & \nodata & \nodata & \nodata & \nodata & \nodata \\
NGC3059 & 28298 & 20.23 & 29.4 & -14.8 & 1 & $36.8^{+5.2}_{-7.8}$ & 1 & $0.1\pm0.1$ & $4.4\pm2.5$ & \nodata & \nodata & $0.3\pm0.1$ & $0.7\pm0.1$ & $2.3\pm0.1$ & \nodata & \nodata & \nodata & \nodata & $3.9\pm0.1$ & $4.4\pm0.1$ & $5.9\pm1.0$ & \nodata & \nodata & \nodata & \nodata & \nodata & \nodata & \nodata & \nodata & \nodata & \nodata & \nodata & \nodata & \nodata & \nodata & \nodata & \nodata & \nodata & \nodata & \nodata & \nodata & \nodata & \nodata & \nodata & \nodata & \nodata & \nodata & \nodata & \nodata & \nodata & \nodata \\
NGC3137 & 29530 & 16.37 & 70.3 & -0.3 & 0 & $21.6^{+0.6}_{-0.7}$ & 1 & $5.0\pm0.5$ & \nodata & \nodata & \nodata & $0.3\pm0.0$ & \nodata & \nodata & \nodata & \nodata & \nodata & \nodata & \nodata & \nodata & \nodata & \nodata & \nodata & \nodata & \nodata & \nodata & \nodata & \nodata & \nodata & \nodata & \nodata & \nodata & \nodata & \nodata & \nodata & \nodata & \nodata & \nodata & \nodata & \nodata & \nodata & \nodata & \nodata & \nodata & \nodata & \nodata & \nodata & \nodata & \nodata & \nodata & \nodata \\
NGC3351 & 32007 & 9.96 & 45.1 & 193.2 & 1 & $60.5^{+7.2}_{-12.6}$ & 1 & $2.9\pm1.2$ & \nodata & \nodata & \nodata & $0.6\pm0.1$ & $0.9\pm0.2$ & \nodata & \nodata & \nodata & \nodata & \nodata & \nodata & \nodata & \nodata & $43.6^{+11.6}_{-12.4}$ & 1 & $3.4\pm0.8$ & \nodata & $0.8\pm0.3$ & $2.6\pm0.3$ & \nodata & \nodata & \nodata & \nodata & \nodata & \nodata & \nodata & \nodata & \nodata & \nodata & $89.7^{+12.6}_{-16.3}$ & 3 & $2.5\pm1.7$ & \nodata & $0.1\pm0.0$ & $0.4\pm0.1$ & \nodata & \nodata & \nodata & \nodata & \nodata & $3.5\pm0.5$ & \nodata & \nodata \\
NGC3489 & 33160 & 11.86 & 63.68 & 70.0 & 0 & $193.9^{+78.9}_{-56.8}$ & 4 & \nodata & \nodata & \nodata & \nodata & \nodata & \nodata & \nodata & \nodata & \nodata & \nodata & \nodata & \nodata & \nodata & \nodata & \nodata & \nodata & \nodata & \nodata & \nodata & \nodata & \nodata & \nodata & \nodata & \nodata & \nodata & \nodata & \nodata & \nodata & \nodata & \nodata & \nodata & \nodata & \nodata & \nodata & \nodata & \nodata & \nodata & \nodata & \nodata & \nodata & \nodata & \nodata & \nodata & \nodata \\
NGC3507 & 33390 & 23.55 & 21.7 & 55.8 & 1 & $33.9^{+2.3}_{-1.6}$ & 1 & $0.1\pm0.1$ & $1.0\pm0.3$ & $4.2\pm2.8$ & \nodata & $0.6\pm0.0$ & $0.9\pm0.0$ & $2.1\pm0.1$ & \nodata & \nodata & \nodata & \nodata & \nodata & \nodata & \nodata & \nodata & \nodata & \nodata & \nodata & \nodata & \nodata & \nodata & \nodata & \nodata & \nodata & \nodata & \nodata & \nodata & \nodata & \nodata & \nodata & \nodata & \nodata & \nodata & \nodata & \nodata & \nodata & \nodata & \nodata & \nodata & \nodata & \nodata & \nodata & \nodata & \nodata \\
NGC3511 & 33385 & 13.94 & 75.1 & 256.8 & 1 & $22.6^{+1.8}_{-0.9}$ & 1 & $5.7\pm1.5$ & \nodata & \nodata & \nodata & $0.2\pm0.0$ & $0.6\pm0.0$ & $1.4\pm0.4$ & \nodata & \nodata & \nodata & \nodata & \nodata & \nodata & \nodata & \nodata & \nodata & \nodata & \nodata & \nodata & \nodata & \nodata & \nodata & \nodata & \nodata & \nodata & \nodata & \nodata & \nodata & \nodata & \nodata & \nodata & \nodata & \nodata & \nodata & \nodata & \nodata & \nodata & \nodata & \nodata & \nodata & \nodata & \nodata & \nodata & \nodata \\
NGC3521 & 33550 & 13.24 & 68.8 & 343.0 & 0 & $35.7^{+0.4}_{-0.7}$ & 1 & $6.4\pm0.4$ & \nodata & \nodata & \nodata & $0.5\pm0.0$ & $1.3\pm0.0$ & \nodata & \nodata & \nodata & \nodata & \nodata & \nodata & \nodata & \nodata & \nodata & \nodata & \nodata & \nodata & \nodata & \nodata & \nodata & \nodata & \nodata & \nodata & \nodata & \nodata & \nodata & \nodata & \nodata & \nodata & \nodata & \nodata & \nodata & \nodata & \nodata & \nodata & \nodata & \nodata & \nodata & \nodata & \nodata & \nodata & \nodata & \nodata \\
NGC3596 & 34298 & 11.0 & 25.1 & 78.4 & 0 & $67.8^{+0.6}_{-1.1}$ & 1 & $1.9\pm1.1$ & \nodata & \nodata & \nodata & $0.2\pm0.0$ & $0.3\pm0.0$ & \nodata & \nodata & \nodata & \nodata & \nodata & \nodata & \nodata & \nodata & \nodata & \nodata & \nodata & \nodata & \nodata & \nodata & \nodata & \nodata & \nodata & \nodata & \nodata & \nodata & \nodata & \nodata & \nodata & \nodata & \nodata & \nodata & \nodata & \nodata & \nodata & \nodata & \nodata & \nodata & \nodata & \nodata & \nodata & \nodata & \nodata & \nodata \\
NGC3599 & 34326 & 19.86 & 23.0 & 41.9 & 0 & $55.6^{+16.0}_{-13.0}$ & 4 & \nodata & \nodata & \nodata & \nodata & \nodata & \nodata & \nodata & \nodata & \nodata & \nodata & \nodata & \nodata & \nodata & \nodata & \nodata & \nodata & \nodata & \nodata & \nodata & \nodata & \nodata & \nodata & \nodata & \nodata & \nodata & \nodata & \nodata & \nodata & \nodata & \nodata & \nodata & \nodata & \nodata & \nodata & \nodata & \nodata & \nodata & \nodata & \nodata & \nodata & \nodata & \nodata & \nodata & \nodata \\
NGC3621 & 34554 & 7.06 & 65.8 & 343.8 & 0 & $36.3^{+0.4}_{-0.4}$ & 2 & $3.4\pm0.2$ & \nodata & \nodata & \nodata & \nodata & \nodata & \nodata & \nodata & \nodata & \nodata & \nodata & \nodata & \nodata & \nodata & \nodata & \nodata & \nodata & \nodata & \nodata & \nodata & \nodata & \nodata & \nodata & \nodata & \nodata & \nodata & \nodata & \nodata & \nodata & \nodata & \nodata & \nodata & \nodata & \nodata & \nodata & \nodata & \nodata & \nodata & \nodata & \nodata & \nodata & \nodata & \nodata & \nodata \\
NGC3626 & 34684 & 20.05 & 46.6 & 165.2 & 1 & $170.2^{+11.6}_{-4.1}$ & 1 & $1.5\pm0.4$ & \nodata & \nodata & \nodata & $0.2\pm0.1$ & \nodata & \nodata & \nodata & \nodata & \nodata & \nodata & \nodata & \nodata & \nodata & \nodata & \nodata & \nodata & \nodata & \nodata & \nodata & \nodata & \nodata & \nodata & \nodata & \nodata & \nodata & \nodata & \nodata & \nodata & \nodata & \nodata & \nodata & \nodata & \nodata & \nodata & \nodata & \nodata & \nodata & \nodata & \nodata & \nodata & \nodata & \nodata & \nodata \\
NGC3627 & 34695 & 11.32 & 57.3 & 173.1 & 1 & $45.8^{+1.0}_{-2.5}$ & 1 & $4.7\pm0.5$ & \nodata & \nodata & \nodata & $0.6\pm0.0$ & $1.7\pm0.0$ & \nodata & \nodata & \nodata & \nodata & \nodata & $6.1\pm0.2$ & $6.7\pm0.1$ & $7.5\pm0.3$ & $29.1^{+20.6}_{-8.1}$ & 2 & $1.9\pm0.1$ & $5.2\pm3.1$ & $0.9\pm0.4$ & $1.7\pm0.2$ & $2.3\pm0.1$ & $2.8\pm0.1$ & $4.7\pm0.5$ & $6.0\pm0.3$ & $6.9\pm0.2$ & $8.0\pm0.0$ & \nodata & $6.4\pm1.6$ & \nodata & \nodata & $50.7^{+5.6}_{-6.4}$ & 1 & $4.2\pm1.4$ & \nodata & $0.6\pm0.0$ & $1.7\pm0.1$ & $2.8\pm0.0$ & \nodata & \nodata & \nodata & \nodata & $5.1\pm0.2$ & $6.3\pm0.6$ & $7.5\pm0.4$ \\
NGC4207 & 39206 & 15.78 & 64.5 & 121.9 & 0 & $64.6^{+4.6}_{-14.8}$ & 4 & $1.2\pm1.2$ & \nodata & \nodata & \nodata & \nodata & \nodata & \nodata & \nodata & \nodata & \nodata & \nodata & \nodata & \nodata & \nodata & \nodata & \nodata & \nodata & \nodata & \nodata & \nodata & \nodata & \nodata & \nodata & \nodata & \nodata & \nodata & \nodata & \nodata & \nodata & \nodata & \nodata & \nodata & \nodata & \nodata & \nodata & \nodata & \nodata & \nodata & \nodata & \nodata & \nodata & \nodata & \nodata & \nodata \\
NGC4254 & 39578 & 13.0 & 34.4 & 68.1 & 0 & $47.3^{+1.8}_{-1.7}$ & 1 & $3.4\pm0.3$ & \nodata & \nodata & \nodata & $0.3\pm0.0$ & \nodata & \nodata & \nodata & \nodata & \nodata & \nodata & $6.5\pm0.9$ & \nodata & \nodata & $52.4^{+6.2}_{-4.5}$ & 3 & $3.1\pm0.6$ & \nodata & $0.3\pm0.0$ & \nodata & \nodata & \nodata & \nodata & \nodata & \nodata & \nodata & \nodata & $3.8\pm0.2$ & $6.0\pm1.4$ & \nodata & $36.1^{+0.5}_{-0.3}$ & 3 & $4.4\pm0.6$ & \nodata & $0.3\pm0.0$ & \nodata & \nodata & \nodata & \nodata & \nodata & \nodata & $8.9\pm0.0$ & \nodata & \nodata \\
NGC4293 & 39907 & 15.76 & 65.0 & 48.3 & 1 & $185.2^{+6.3}_{-4.8}$ & 4 & $0.2\pm0.2$ & \nodata & \nodata & \nodata & \nodata & \nodata & \nodata & \nodata & \nodata & \nodata & \nodata & $0.8\pm0.1$ & \nodata & \nodata & \nodata & \nodata & \nodata & \nodata & \nodata & \nodata & \nodata & \nodata & \nodata & \nodata & \nodata & \nodata & \nodata & \nodata & \nodata & \nodata & \nodata & \nodata & \nodata & \nodata & \nodata & \nodata & \nodata & \nodata & \nodata & \nodata & \nodata & \nodata & \nodata & \nodata \\
NGC4298 & 39950 & 13.0 & 59.2 & 313.9 & 0 & $27.2^{+0.7}_{-0.2}$ & 1 & $4.8\pm0.6$ & \nodata & \nodata & \nodata & $0.2\pm0.0$ & $0.7\pm0.0$ & \nodata & \nodata & \nodata & \nodata & \nodata & \nodata & \nodata & \nodata & \nodata & \nodata & \nodata & \nodata & \nodata & \nodata & \nodata & \nodata & \nodata & \nodata & \nodata & \nodata & \nodata & \nodata & \nodata & \nodata & \nodata & \nodata & \nodata & \nodata & \nodata & \nodata & \nodata & \nodata & \nodata & \nodata & \nodata & \nodata & \nodata & \nodata \\
NGC4303 & 40001 & 16.99 & 23.5 & 312.4 & 1 & $46.7^{+2.2}_{-8.2}$ & 1 & $4.0\pm2.3$ & \nodata & \nodata & \nodata & $0.1\pm0.0$ & $0.7\pm0.0$ & $2.0\pm0.2$ & $6.1\pm0.0$ & \nodata & \nodata & \nodata & $6.1\pm0.0$ & \nodata & \nodata & $43.5^{+5.3}_{-10.0}$ & 1 & $3.8\pm2.5$ & \nodata & $0.1\pm0.0$ & $0.7\pm0.1$ & $1.8\pm0.5$ & $3.5\pm0.1$ & $6.1\pm0.0$ & \nodata & \nodata & \nodata & \nodata & $6.1\pm0.1$ & \nodata & \nodata & $48.4^{+2.5}_{-6.7}$ & 1 & $3.9\pm2.4$ & \nodata & $0.2\pm0.0$ & $0.7\pm0.0$ & \nodata & \nodata & \nodata & \nodata & \nodata & $6.1\pm0.1$ & \nodata & \nodata \\
NGC4321 & 40153 & 15.21 & 38.5 & 156.2 & 1 & $26.6^{+5.7}_{-9.6}$ & 3 & $6.3\pm2.3$ & \nodata & \nodata & \nodata & $1.2\pm0.3$ & $3.3\pm1.7$ & \nodata & \nodata & \nodata & \nodata & \nodata & \nodata & \nodata & \nodata & $43.4^{+3.1}_{-9.1}$ & 4 & $5.6\pm3.0$ & \nodata & $0.9\pm0.0$ & \nodata & \nodata & \nodata & \nodata & \nodata & \nodata & \nodata & \nodata & $7.3\pm1.0$ & \nodata & \nodata & $34.7^{+8.1}_{-3.4}$ & 3 & $5.1\pm2.6$ & $8.3\pm0.3$ & $0.1\pm0.0$ & $0.9\pm0.0$ & \nodata & \nodata & \nodata & \nodata & \nodata & $7.3\pm1.1$ & \nodata & \nodata \\
NGC4424 & 40809 & 16.2 & 58.2 & 88.3 & 0 & $17.2^{+1.4}_{-2.2}$ & 4 & \nodata & \nodata & \nodata & \nodata & \nodata & \nodata & \nodata & \nodata & \nodata & \nodata & \nodata & \nodata & \nodata & \nodata & \nodata & \nodata & \nodata & \nodata & \nodata & \nodata & \nodata & \nodata & \nodata & \nodata & \nodata & \nodata & \nodata & \nodata & \nodata & \nodata & \nodata & \nodata & \nodata & \nodata & \nodata & \nodata & \nodata & \nodata & \nodata & \nodata & \nodata & \nodata & \nodata & \nodata \\
NGC4457 & 41101 & 15.0 & 17.4 & 78.7 & 0 & $122.4^{+1.0}_{-0.4}$ & 3 & $0.1\pm0.1$ & $1.9\pm0.8$ & \nodata & \nodata & $0.4\pm0.0$ & $1.0\pm0.0$ & \nodata & \nodata & \nodata & \nodata & \nodata & $2.4\pm0.0$ & \nodata & \nodata & \nodata & \nodata & \nodata & \nodata & \nodata & \nodata & \nodata & \nodata & \nodata & \nodata & \nodata & \nodata & \nodata & \nodata & \nodata & \nodata & \nodata & \nodata & \nodata & \nodata & \nodata & \nodata & \nodata & \nodata & \nodata & \nodata & \nodata & \nodata & \nodata & \nodata \\
NGC4459 & 41104 & 15.85 & 46.95 & 108.75 & 0 & $89.1^{+366.8}_{-40.6}$ & 3 & \nodata & \nodata & \nodata & \nodata & \nodata & \nodata & \nodata & \nodata & \nodata & \nodata & \nodata & \nodata & \nodata & \nodata & \nodata & \nodata & \nodata & \nodata & \nodata & \nodata & \nodata & \nodata & \nodata & \nodata & \nodata & \nodata & \nodata & \nodata & \nodata & \nodata & \nodata & \nodata & \nodata & \nodata & \nodata & \nodata & \nodata & \nodata & \nodata & \nodata & \nodata & \nodata & \nodata & \nodata \\
NGC4476 & 41255 & 17.54 & 60.14 & 27.38 & 0 & $174.6^{+33.9}_{-25.1}$ & 4 & \nodata & \nodata & \nodata & \nodata & \nodata & \nodata & \nodata & \nodata & \nodata & \nodata & \nodata & \nodata & \nodata & \nodata & \nodata & \nodata & \nodata & \nodata & \nodata & \nodata & \nodata & \nodata & \nodata & \nodata & \nodata & \nodata & \nodata & \nodata & \nodata & \nodata & \nodata & \nodata & \nodata & \nodata & \nodata & \nodata & \nodata & \nodata & \nodata & \nodata & \nodata & \nodata & \nodata & \nodata \\
NGC4477 & 41260 & 15.76 & 33.51 & 25.68 & 0 & $1440.3^{+439.2}_{-623.3}$ & 4 & \nodata & \nodata & \nodata & \nodata & \nodata & \nodata & \nodata & \nodata & \nodata & \nodata & \nodata & \nodata & \nodata & \nodata & \nodata & \nodata & \nodata & \nodata & \nodata & \nodata & \nodata & \nodata & \nodata & \nodata & \nodata & \nodata & \nodata & \nodata & \nodata & \nodata & \nodata & \nodata & \nodata & \nodata & \nodata & \nodata & \nodata & \nodata & \nodata & \nodata & \nodata & \nodata & \nodata & \nodata \\
NGC4496A & 41471 & 14.86 & 53.8 & 51.1 & 1 & $23.9^{+2.3}_{-1.2}$ & 1 & $1.8\pm1.8$ & \nodata & \nodata & \nodata & \nodata & \nodata & \nodata & \nodata & \nodata & \nodata & \nodata & \nodata & \nodata & \nodata & \nodata & \nodata & \nodata & \nodata & \nodata & \nodata & \nodata & \nodata & \nodata & \nodata & \nodata & \nodata & \nodata & \nodata & \nodata & \nodata & \nodata & \nodata & \nodata & \nodata & \nodata & \nodata & \nodata & \nodata & \nodata & \nodata & \nodata & \nodata & \nodata & \nodata \\
NGC4535 & 41812 & 15.77 & 44.7 & 179.7 & 1 & $30.9^{+0.5}_{-0.3}$ & 1 & $5.4\pm2.8$ & \nodata & \nodata & \nodata & $1.9\pm0.0$ & $2.2\pm0.0$ & \nodata & \nodata & \nodata & \nodata & \nodata & \nodata & \nodata & \nodata & $21.0^{+7.7}_{-1.3}$ & 3 & $5.4\pm3.0$ & \nodata & $0.9\pm0.0$ & $1.9\pm0.1$ & $2.2\pm0.0$ & $2.8\pm0.0$ & \nodata & \nodata & \nodata & \nodata & \nodata & \nodata & \nodata & \nodata & $30.7^{+0.9}_{-0.3}$ & 1 & $5.4\pm2.8$ & \nodata & $1.9\pm0.0$ & $2.2\pm0.0$ & \nodata & \nodata & \nodata & \nodata & \nodata & \nodata & \nodata & \nodata \\
NGC4536 & 41823 & 16.25 & 66.0 & 305.6 & 1 & $17.9^{+2.6}_{-4.1}$ & 1 & $9.0\pm2.2$ & \nodata & \nodata & \nodata & $0.9\pm0.0$ & $1.2\pm0.1$ & $4.0\pm1.3$ & $10.0\pm0.2$ & \nodata & \nodata & \nodata & $10.1\pm0.3$ & \nodata & \nodata & \nodata & \nodata & \nodata & \nodata & \nodata & \nodata & \nodata & \nodata & \nodata & \nodata & \nodata & \nodata & \nodata & \nodata & \nodata & \nodata & \nodata & \nodata & \nodata & \nodata & \nodata & \nodata & \nodata & \nodata & \nodata & \nodata & \nodata & \nodata & \nodata & \nodata \\
NGC4540 & 41876 & 15.76 & 28.7 & 12.8 & 1 & $55.6^{+11.9}_{-9.0}$ & 1 & $1.5\pm1.5$ & \nodata & \nodata & \nodata & \nodata & \nodata & \nodata & \nodata & \nodata & \nodata & \nodata & $1.9\pm0.1$ & $2.5\pm0.1$ & $2.9\pm0.1$ & \nodata & \nodata & \nodata & \nodata & \nodata & \nodata & \nodata & \nodata & \nodata & \nodata & \nodata & \nodata & \nodata & \nodata & \nodata & \nodata & \nodata & \nodata & \nodata & \nodata & \nodata & \nodata & \nodata & \nodata & \nodata & \nodata & \nodata & \nodata & \nodata & \nodata \\
NGC4548 & 41934 & 16.22 & 38.3 & 138.0 & 1 & $26.2^{+2.7}_{-2.5}$ & 2 & $5.7\pm0.7$ & \nodata & \nodata & \nodata & $0.3\pm0.0$ & $0.7\pm0.0$ & $3.3\pm0.1$ & $4.8\pm0.0$ & $5.2\pm0.1$ & $6.3\pm0.0$ & \nodata & $6.3\pm0.0$ & \nodata & \nodata & \nodata & \nodata & \nodata & \nodata & \nodata & \nodata & \nodata & \nodata & \nodata & \nodata & \nodata & \nodata & \nodata & \nodata & \nodata & \nodata & \nodata & \nodata & \nodata & \nodata & \nodata & \nodata & \nodata & \nodata & \nodata & \nodata & \nodata & \nodata & \nodata & \nodata \\
NGC4569 & 42089 & 15.76 & 70.0 & 18.0 & 1 & $31.9^{+10.5}_{-11.9}$ & 1 & $0.1\pm0.1$ & $1.2\pm0.1$ & $4.7\pm2.4$ & \nodata & $0.6\pm0.2$ & $1.1\pm0.2$ & $1.8\pm0.1$ & $2.7\pm0.3$ & $5.3\pm0.2$ & \nodata & \nodata & \nodata & \nodata & \nodata & \nodata & \nodata & \nodata & \nodata & \nodata & \nodata & \nodata & \nodata & \nodata & \nodata & \nodata & \nodata & \nodata & \nodata & \nodata & \nodata & \nodata & \nodata & \nodata & \nodata & \nodata & \nodata & \nodata & \nodata & \nodata & \nodata & \nodata & \nodata & \nodata & \nodata \\
NGC4571 & 42100 & 14.0 & 32.7 & 217.5 & 0 & $30.8^{+1.2}_{-1.0}$ & 1 & $4.1\pm0.7$ & \nodata & \nodata & \nodata & \nodata & \nodata & \nodata & \nodata & \nodata & \nodata & \nodata & \nodata & \nodata & \nodata & \nodata & \nodata & \nodata & \nodata & \nodata & \nodata & \nodata & \nodata & \nodata & \nodata & \nodata & \nodata & \nodata & \nodata & \nodata & \nodata & \nodata & \nodata & \nodata & \nodata & \nodata & \nodata & \nodata & \nodata & \nodata & \nodata & \nodata & \nodata & \nodata & \nodata \\
NGC4579 & 42168 & 21.0 & 40.22 & 91.3 & 1 & $24.0^{+5.5}_{-6.5}$ & 2 & $8.2\pm1.8$ & \nodata & \nodata & \nodata & $0.3\pm0.0$ & $1.2\pm0.0$ & $2.7\pm0.2$ & $3.8\pm0.2$ & $5.3\pm0.7$ & $9.1\pm0.3$ & \nodata & \nodata & \nodata & \nodata & \nodata & \nodata & \nodata & \nodata & \nodata & \nodata & \nodata & \nodata & \nodata & \nodata & \nodata & \nodata & \nodata & \nodata & \nodata & \nodata & \nodata & \nodata & \nodata & \nodata & \nodata & \nodata & \nodata & \nodata & \nodata & \nodata & \nodata & \nodata & \nodata & \nodata \\
NGC4596 & 42401 & 15.76 & 36.56 & 120.0 & 0 & $1290.0^{+604.7}_{-754.1}$ & 4 & \nodata & \nodata & \nodata & \nodata & \nodata & \nodata & \nodata & \nodata & \nodata & \nodata & \nodata & \nodata & \nodata & \nodata & \nodata & \nodata & \nodata & \nodata & \nodata & \nodata & \nodata & \nodata & \nodata & \nodata & \nodata & \nodata & \nodata & \nodata & \nodata & \nodata & \nodata & \nodata & \nodata & \nodata & \nodata & \nodata & \nodata & \nodata & \nodata & \nodata & \nodata & \nodata & \nodata & \nodata \\
NGC4654 & 42857 & 21.98 & 55.6 & 123.2 & 1 & $19.0^{+0.8}_{-0.2}$ & 1 & $8.3\pm1.8$ & \nodata & \nodata & \nodata & \nodata & \nodata & \nodata & \nodata & \nodata & \nodata & \nodata & $10.0\pm0.0$ & $10.4\pm0.0$ & \nodata & \nodata & \nodata & \nodata & \nodata & \nodata & \nodata & \nodata & \nodata & \nodata & \nodata & \nodata & \nodata & \nodata & \nodata & \nodata & \nodata & \nodata & \nodata & \nodata & \nodata & \nodata & \nodata & \nodata & \nodata & \nodata & \nodata & \nodata & \nodata & \nodata & \nodata \\
NGC4689 & 43186 & 15.0 & 38.7 & 164.1 & 0 & $35.1^{+0.7}_{-0.8}$ & 1 & $3.8\pm0.5$ & \nodata & \nodata & \nodata & \nodata & \nodata & \nodata & \nodata & \nodata & \nodata & \nodata & \nodata & \nodata & \nodata & \nodata & \nodata & \nodata & \nodata & \nodata & \nodata & \nodata & \nodata & \nodata & \nodata & \nodata & \nodata & \nodata & \nodata & \nodata & \nodata & \nodata & \nodata & \nodata & \nodata & \nodata & \nodata & \nodata & \nodata & \nodata & \nodata & \nodata & \nodata & \nodata & \nodata \\
NGC4694 & 43241 & 15.76 & 60.7 & 143.3 & 0 & $95.9^{+12.0}_{-7.3}$ & 4 & \nodata & \nodata & \nodata & \nodata & \nodata & \nodata & \nodata & \nodata & \nodata & \nodata & \nodata & \nodata & \nodata & \nodata & \nodata & \nodata & \nodata & \nodata & \nodata & \nodata & \nodata & \nodata & \nodata & \nodata & \nodata & \nodata & \nodata & \nodata & \nodata & \nodata & \nodata & \nodata & \nodata & \nodata & \nodata & \nodata & \nodata & \nodata & \nodata & \nodata & \nodata & \nodata & \nodata & \nodata \\
NGC4731 & 43507 & 13.28 & 64.0 & 255.4 & 1 & $22.7^{+5.0}_{-10.6}$ & 4 & \nodata & \nodata & \nodata & \nodata & \nodata & \nodata & \nodata & \nodata & \nodata & \nodata & \nodata & \nodata & \nodata & \nodata & \nodata & \nodata & \nodata & \nodata & \nodata & \nodata & \nodata & \nodata & \nodata & \nodata & \nodata & \nodata & \nodata & \nodata & \nodata & \nodata & \nodata & \nodata & \nodata & \nodata & \nodata & \nodata & \nodata & \nodata & \nodata & \nodata & \nodata & \nodata & \nodata & \nodata \\
NGC4781 & 43902 & 11.31 & 59.0 & 290.0 & 1 & $32.8^{+0.9}_{-0.8}$ & 1 & $3.9\pm0.7$ & \nodata & \nodata & \nodata & \nodata & \nodata & \nodata & \nodata & \nodata & \nodata & \nodata & \nodata & \nodata & \nodata & \nodata & \nodata & \nodata & \nodata & \nodata & \nodata & \nodata & \nodata & \nodata & \nodata & \nodata & \nodata & \nodata & \nodata & \nodata & \nodata & \nodata & \nodata & \nodata & \nodata & \nodata & \nodata & \nodata & \nodata & \nodata & \nodata & \nodata & \nodata & \nodata & \nodata \\
NGC4826 & 44182 & 4.41 & 59.1 & 293.6 & 0 & $236.9^{+48.6}_{-26.0}$ & 3 & $0.8\pm0.3$ & \nodata & \nodata & \nodata & $0.2\pm0.1$ & \nodata & \nodata & \nodata & \nodata & \nodata & \nodata & $1.2\pm0.2$ & \nodata & \nodata & \nodata & \nodata & \nodata & \nodata & \nodata & \nodata & \nodata & \nodata & \nodata & \nodata & \nodata & \nodata & \nodata & \nodata & \nodata & \nodata & \nodata & \nodata & \nodata & \nodata & \nodata & \nodata & \nodata & \nodata & \nodata & \nodata & \nodata & \nodata & \nodata & \nodata \\
NGC4941 & 45165 & 15.0 & 53.4 & 202.2 & 1 & $46.2^{+0.2}_{-0.2}$ & 1 & $3.7\pm0.2$ & \nodata & \nodata & \nodata & $0.3\pm0.0$ & $1.3\pm0.0$ & \nodata & \nodata & \nodata & \nodata & \nodata & \nodata & \nodata & \nodata & \nodata & \nodata & \nodata & \nodata & \nodata & \nodata & \nodata & \nodata & \nodata & \nodata & \nodata & \nodata & \nodata & \nodata & \nodata & \nodata & \nodata & \nodata & \nodata & \nodata & \nodata & \nodata & \nodata & \nodata & \nodata & \nodata & \nodata & \nodata & \nodata & \nodata \\
NGC4951 & 45246 & 15.0 & 70.2 & 91.2 & 0 & $64.0^{+1.6}_{-1.3}$ & 3 & $0.1\pm0.1$ & $1.8\pm0.2$ & \nodata & \nodata & \nodata & \nodata & \nodata & \nodata & \nodata & \nodata & \nodata & $3.2\pm0.0$ & $4.1\pm0.0$ & \nodata & \nodata & \nodata & \nodata & \nodata & \nodata & \nodata & \nodata & \nodata & \nodata & \nodata & \nodata & \nodata & \nodata & \nodata & \nodata & \nodata & \nodata & \nodata & \nodata & \nodata & \nodata & \nodata & \nodata & \nodata & \nodata & \nodata & \nodata & \nodata & \nodata & \nodata \\
NGC5042 & 46126 & 16.78 & 49.4 & 190.6 & 0 & $28.1^{+0.5}_{-0.5}$ & 1 & $4.2\pm0.9$ & \nodata & \nodata & \nodata & $0.2\pm0.0$ & $0.4\pm0.0$ & $1.2\pm0.0$ & $1.4\pm0.0$ & \nodata & \nodata & \nodata & \nodata & \nodata & \nodata & \nodata & \nodata & \nodata & \nodata & \nodata & \nodata & \nodata & \nodata & \nodata & \nodata & \nodata & \nodata & \nodata & \nodata & \nodata & \nodata & \nodata & \nodata & \nodata & \nodata & \nodata & \nodata & \nodata & \nodata & \nodata & \nodata & \nodata & \nodata & \nodata & \nodata \\
NGC5068 & 46400 & 5.16 & 35.7 & 342.4 & 1 & $18.5^{+0.3}_{-3.3}$ & 1 & $0.2\pm0.2$ & $1.1\pm0.4$ & $1.9\pm0.2$ & \nodata & $0.2\pm0.0$ & $2.0\pm0.0$ & \nodata & \nodata & \nodata & \nodata & \nodata & $0.6\pm0.0$ & \nodata & \nodata & $16.3^{+2.0}_{-9.4}$ & 3 & $1.1\pm1.1$ & \nodata & $0.2\pm0.0$ & $0.5\pm0.0$ & $1.1\pm0.1$ & $1.8\pm0.2$ & \nodata & \nodata & \nodata & \nodata & \nodata & $0.6\pm0.1$ & \nodata & \nodata & $15.8^{+4.5}_{-2.0}$ & 1 & $0.2\pm0.2$ & $1.4\pm0.8$ & $0.2\pm0.0$ & $2.0\pm0.0$ & \nodata & \nodata & \nodata & \nodata & \nodata & $0.6\pm0.0$ & \nodata & \nodata \\
NGC5128 & 46957 & 3.69 & 45.33 & 32.17 & 0 & $644.1^{+376.2}_{-1416.9}$ & 3 & \nodata & \nodata & \nodata & \nodata & \nodata & \nodata & \nodata & \nodata & \nodata & \nodata & \nodata & \nodata & \nodata & \nodata & \nodata & \nodata & \nodata & \nodata & \nodata & \nodata & \nodata & \nodata & \nodata & \nodata & \nodata & \nodata & \nodata & \nodata & \nodata & \nodata & \nodata & \nodata & \nodata & \nodata & \nodata & \nodata & \nodata & \nodata & \nodata & \nodata & \nodata & \nodata & \nodata & \nodata \\
NGC5134 & 46938 & 19.92 & 22.7 & 311.6 & 1 & $31.1^{+4.0}_{-1.3}$ & 1 & $0.3\pm0.3$ & $3.0\pm1.9$ & \nodata & \nodata & $0.6\pm0.0$ & $1.1\pm0.1$ & \nodata & \nodata & \nodata & \nodata & \nodata & \nodata & \nodata & \nodata & \nodata & \nodata & \nodata & \nodata & \nodata & \nodata & \nodata & \nodata & \nodata & \nodata & \nodata & \nodata & \nodata & \nodata & \nodata & \nodata & \nodata & \nodata & \nodata & \nodata & \nodata & \nodata & \nodata & \nodata & \nodata & \nodata & \nodata & \nodata & \nodata & \nodata \\
NGC5248 & 48130 & 14.87 & 47.4 & 109.2 & 1 & $48.0^{+6.6}_{-10.8}$ & 3 & $4.7\pm2.4$ & \nodata & \nodata & \nodata & $0.4\pm0.0$ & $0.8\pm0.1$ & $2.2\pm0.4$ & \nodata & \nodata & \nodata & \nodata & $5.9\pm1.1$ & \nodata & \nodata & \nodata & \nodata & \nodata & \nodata & \nodata & \nodata & \nodata & \nodata & \nodata & \nodata & \nodata & \nodata & \nodata & \nodata & \nodata & \nodata & \nodata & \nodata & \nodata & \nodata & \nodata & \nodata & \nodata & \nodata & \nodata & \nodata & \nodata & \nodata & \nodata & \nodata \\
NGC5530 & 51106 & 12.27 & 61.9 & 305.4 & 0 & $32.8^{+0.8}_{-0.3}$ & 1 & $0.1\pm0.1$ & $4.0\pm0.5$ & \nodata & \nodata & \nodata & \nodata & \nodata & \nodata & \nodata & \nodata & \nodata & \nodata & \nodata & \nodata & \nodata & \nodata & \nodata & \nodata & \nodata & \nodata & \nodata & \nodata & \nodata & \nodata & \nodata & \nodata & \nodata & \nodata & \nodata & \nodata & \nodata & \nodata & \nodata & \nodata & \nodata & \nodata & \nodata & \nodata & \nodata & \nodata & \nodata & \nodata & \nodata & \nodata \\
NGC5643 & 51969 & 12.68 & 29.9 & 318.7 & 1 & $54.7^{+5.6}_{-5.3}$ & 1 & $0.1\pm0.1$ & $3.9\pm1.9$ & \nodata & \nodata & $0.2\pm0.0$ & $0.3\pm0.0$ & $0.6\pm0.0$ & $0.8\pm0.0$ & $1.0\pm0.1$ & \nodata & \nodata & $5.7\pm0.7$ & \nodata & \nodata & \nodata & \nodata & \nodata & \nodata & \nodata & \nodata & \nodata & \nodata & \nodata & \nodata & \nodata & \nodata & \nodata & \nodata & \nodata & \nodata & \nodata & \nodata & \nodata & \nodata & \nodata & \nodata & \nodata & \nodata & \nodata & \nodata & \nodata & \nodata & \nodata & \nodata \\
NGC6300 & 60001 & 11.58 & 49.6 & 105.4 & 1 & $41.7^{+4.1}_{-4.8}$ & 3 & $4.4\pm1.3$ & \nodata & \nodata & \nodata & $1.7\pm0.0$ & \nodata & \nodata & \nodata & \nodata & \nodata & \nodata & \nodata & \nodata & \nodata & \nodata & \nodata & \nodata & \nodata & \nodata & \nodata & \nodata & \nodata & \nodata & \nodata & \nodata & \nodata & \nodata & \nodata & \nodata & \nodata & \nodata & \nodata & \nodata & \nodata & \nodata & \nodata & \nodata & \nodata & \nodata & \nodata & \nodata & \nodata & \nodata & \nodata \\
NGC6744 & 62836 & 9.39 & 52.7 & 14.0 & 1 & $27.7^{+0.0}_{-0.1}$ & 1 & $6.1\pm0.4$ & \nodata & \nodata & \nodata & \nodata & \nodata & \nodata & \nodata & \nodata & \nodata & \nodata & \nodata & \nodata & \nodata & \nodata & \nodata & \nodata & \nodata & \nodata & \nodata & \nodata & \nodata & \nodata & \nodata & \nodata & \nodata & \nodata & \nodata & \nodata & \nodata & \nodata & \nodata & \nodata & \nodata & \nodata & \nodata & \nodata & \nodata & \nodata & \nodata & \nodata & \nodata & \nodata & \nodata \\
NGC7456 & 70304 & 15.7 & 67.3 & 16.0 & 0 & $23.9^{+0.6}_{-0.8}$ & 1 & $4.0\pm0.9$ & \nodata & \nodata & \nodata & \nodata & \nodata & \nodata & \nodata & \nodata & \nodata & \nodata & \nodata & \nodata & \nodata & \nodata & \nodata & \nodata & \nodata & \nodata & \nodata & \nodata & \nodata & \nodata & \nodata & \nodata & \nodata & \nodata & \nodata & \nodata & \nodata & \nodata & \nodata & \nodata & \nodata & \nodata & \nodata & \nodata & \nodata & \nodata & \nodata & \nodata & \nodata & \nodata & \nodata \\
NGC7496 & 70588 & 18.72 & 35.9 & 193.7 & 1 & $24.4^{+3.1}_{-3.0}$ & 1 & $4.3\pm2.0$ & \nodata & \nodata & \nodata & $0.4\pm0.0$ & $0.7\pm0.0$ & $1.2\pm0.2$ & $1.8\pm0.2$ & \nodata & \nodata & \nodata & \nodata & \nodata & \nodata & $16.8^{+6.5}_{-12.1}$ & 1 & $4.3\pm2.0$ & \nodata & $0.2\pm0.1$ & $0.7\pm0.1$ & $3.5\pm2.7$ & \nodata & \nodata & \nodata & \nodata & \nodata & \nodata & \nodata & \nodata & \nodata & $23.0^{+7.5}_{-4.3}$ & 3 & $4.0\pm2.2$ & \nodata & $0.4\pm0.0$ & $0.7\pm0.0$ & $1.5\pm0.5$ & \nodata & \nodata & \nodata & \nodata & $6.1\pm0.0$ & \nodata & \nodata \\
NGC7743 & 72263 & 20.32 & 37.1 & 86.24 & 0 & $-13.9^{+31.0}_{-17.8}$ & 4 & \nodata & \nodata & \nodata & \nodata & \nodata & \nodata & \nodata & \nodata & \nodata & \nodata & \nodata & \nodata & \nodata & \nodata & \nodata & \nodata & \nodata & \nodata & \nodata & \nodata & \nodata & \nodata & \nodata & \nodata & \nodata & \nodata & \nodata & \nodata & \nodata & \nodata & \nodata & \nodata & \nodata & \nodata & \nodata & \nodata & \nodata & \nodata & \nodata & \nodata & \nodata & \nodata & \nodata & \nodata
\enddata
\tablecomments{For each galaxy, we list its NGC number, PGC number, distance, inclination, position angle, and whether it has a bar (1 for yes, 0 for no). We use a subscript to refer to the tracer in question -- MM for MUSE stellar mass, MH$\alpha$ for MUSE-H$\alpha$ measurements, and A for ALMA CO measurements. We also show the quality flags (indicated as Q), and co-rotation radii ($R_{\rm CR}$), and outer/inner Lindblad radii ($R_{\rm O/ILR}$). For brevity, only the first co-rotation radius is shown, and Lindblad radii are hidden.}\end{deluxetable*}

\end{longrotatetable}



\section{Tremaine-Weinberg Plots for all MUSE Galaxies}\label{app:all_tw_plots}

\begin{figure*}[t]
\plotone{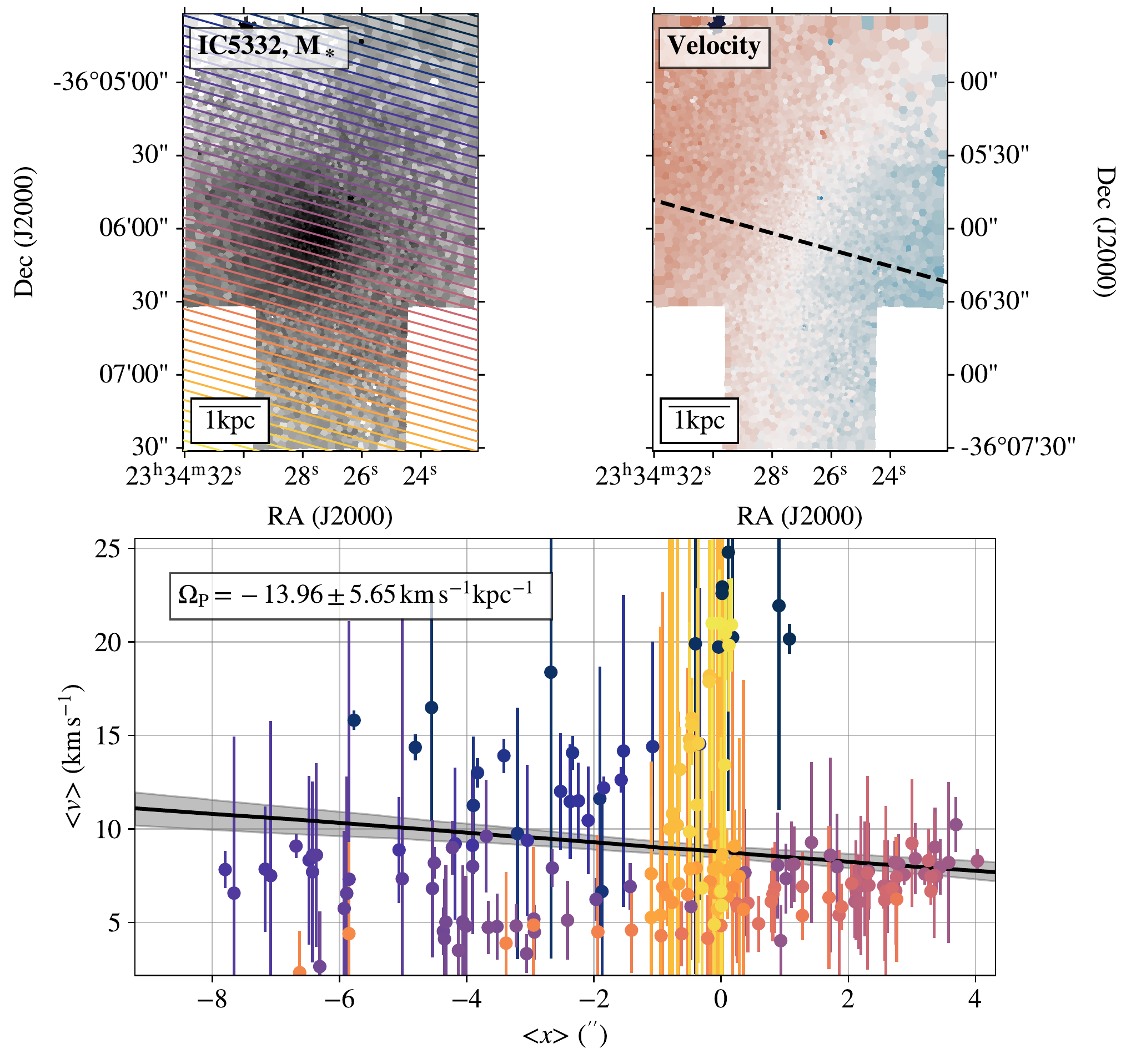}
\caption{As Fig. \ref{fig:ngc3351_tw_integral}, but for IC5332. For this galaxy , $Q=3$. \label{fig:app_ic5332}}
\end{figure*}

\begin{figure*}[t]
\plotone{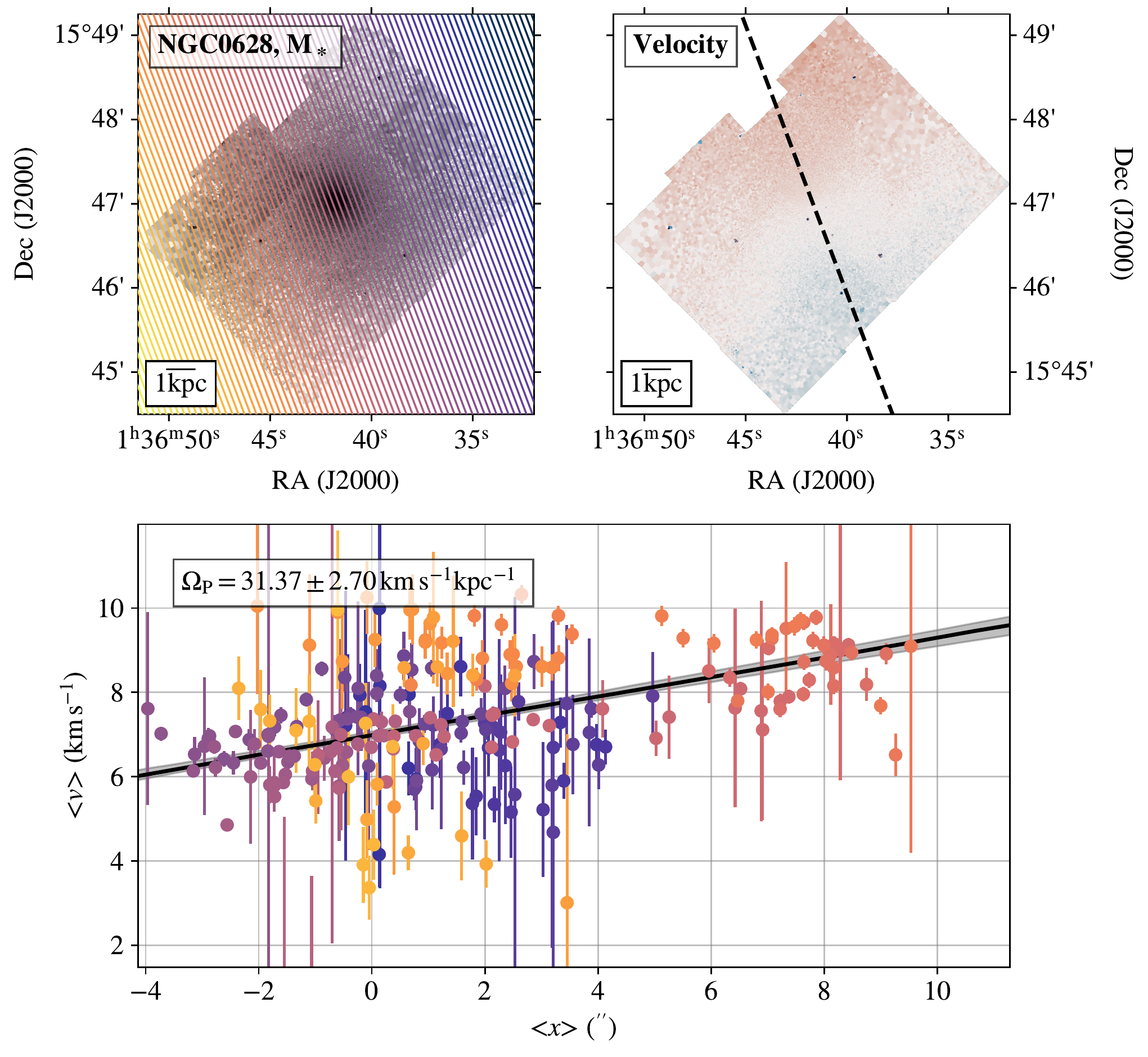}
\caption{As Fig. \ref{fig:ngc3351_tw_integral}, but for NGC0628. For this galaxy , $Q=1$. \label{fig:app_ngc0628}}
\end{figure*}

\begin{figure*}[t]
\plotone{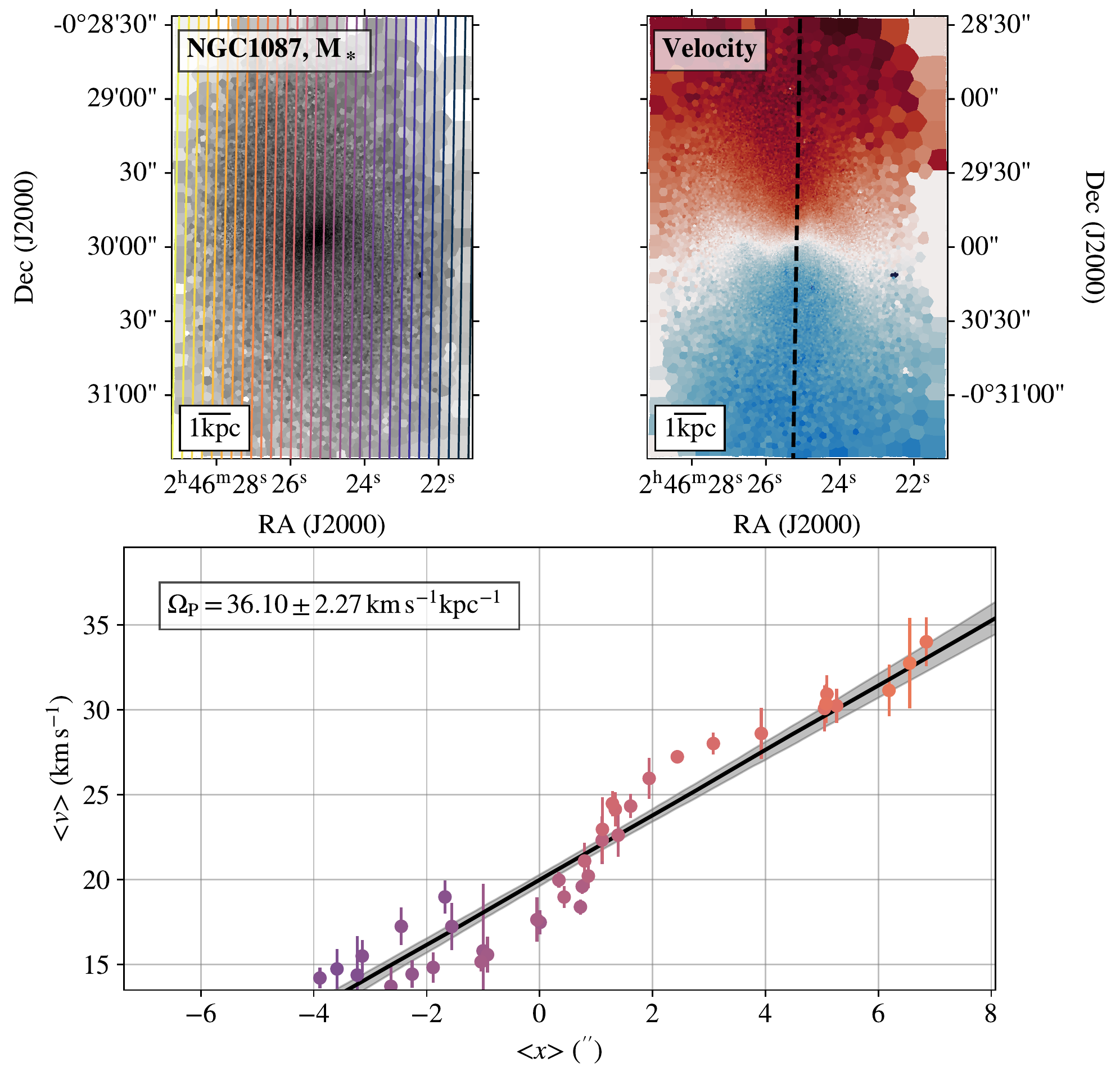}
\caption{As Fig. \ref{fig:ngc3351_tw_integral}, but for NGC1087. For this galaxy , $Q=1$. \label{fig:app_ngc1087}}
\end{figure*}

\begin{figure*}[t]
\plotone{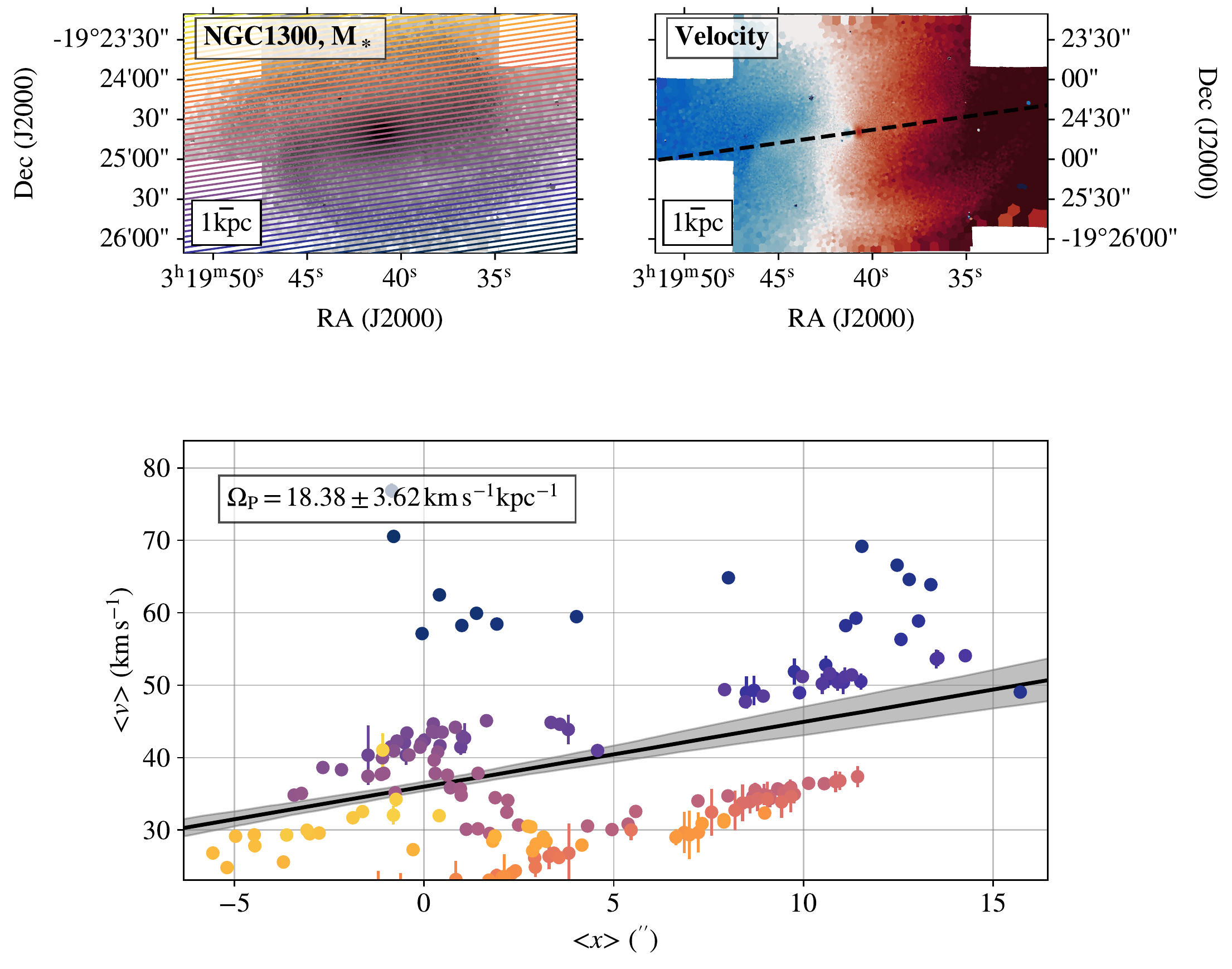}
\caption{As Fig. \ref{fig:ngc3351_tw_integral}, but for NGC1300. For this galaxy , $Q=3$. \label{fig:app_ngc1300}}
\end{figure*}

\begin{figure*}[t]
\plotone{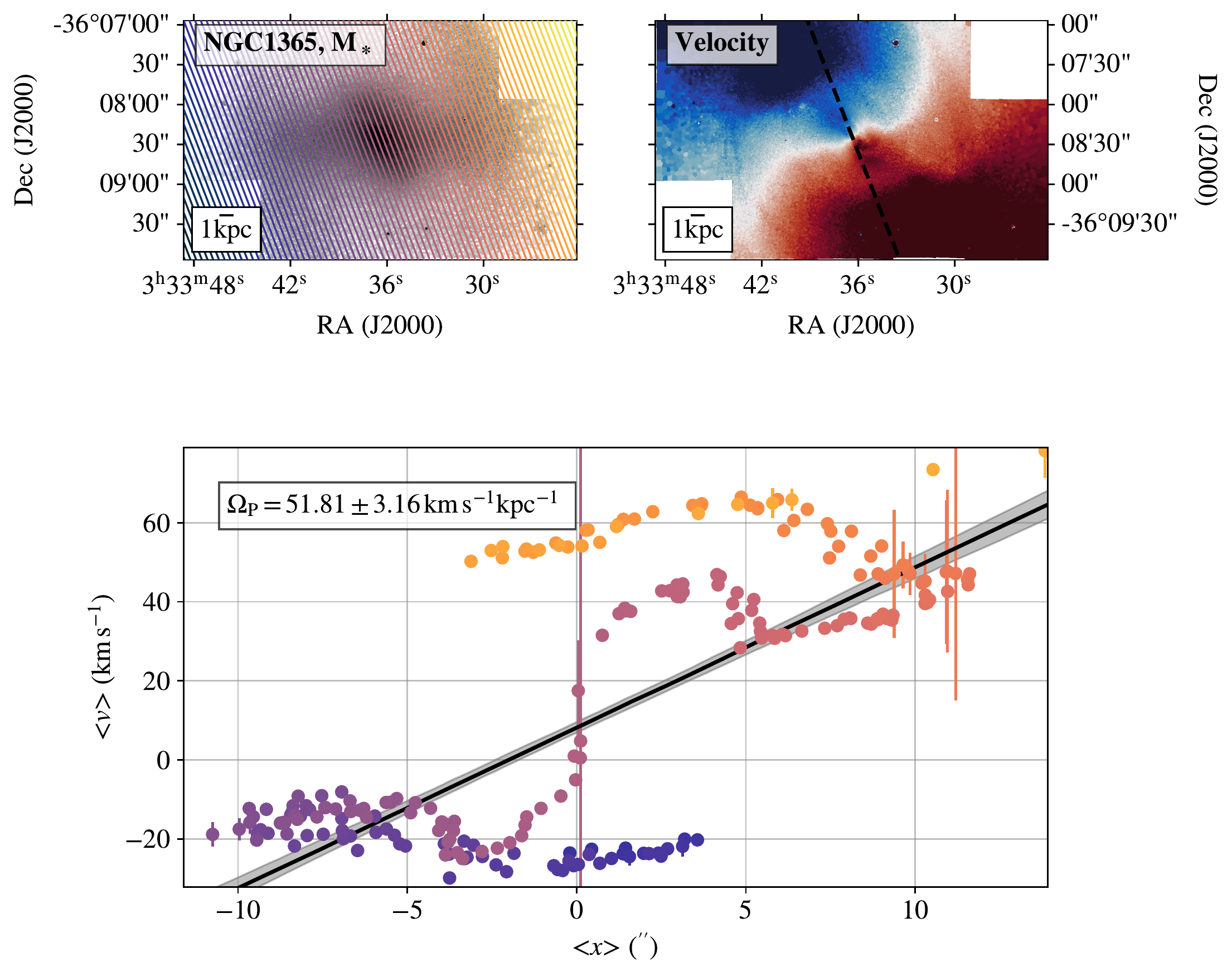}
\caption{As Fig. \ref{fig:ngc3351_tw_integral}, but for NGC1365. For this galaxy , $Q=3$. \label{fig:app_ngc1365}}
\end{figure*}

\begin{figure*}[t]
\plotone{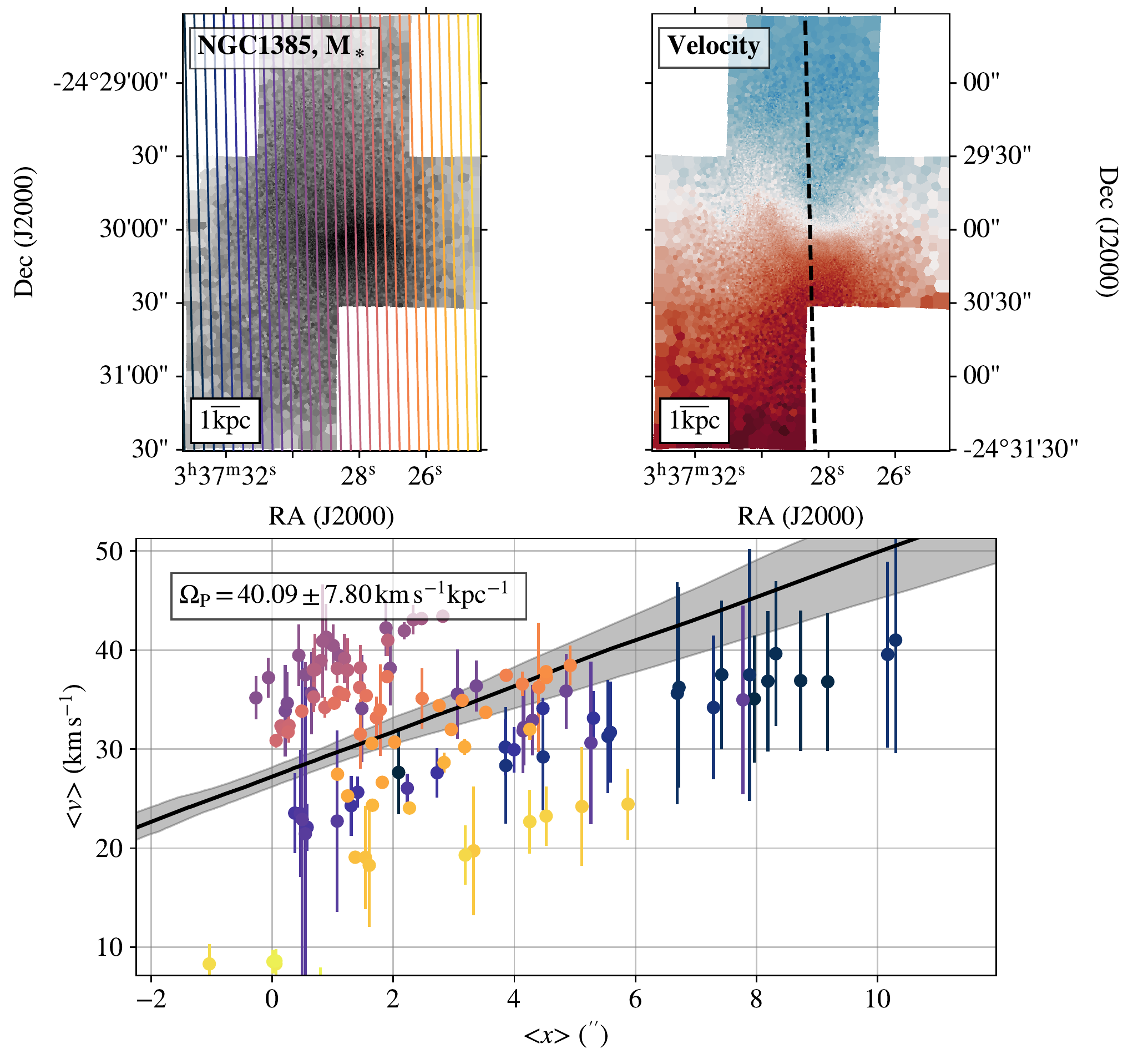}
\caption{As Fig. \ref{fig:ngc3351_tw_integral}, but for NGC1385. For this galaxy , $Q=3$. \label{fig:app_ngc1385}}
\end{figure*}

\begin{figure*}[t]
\plotone{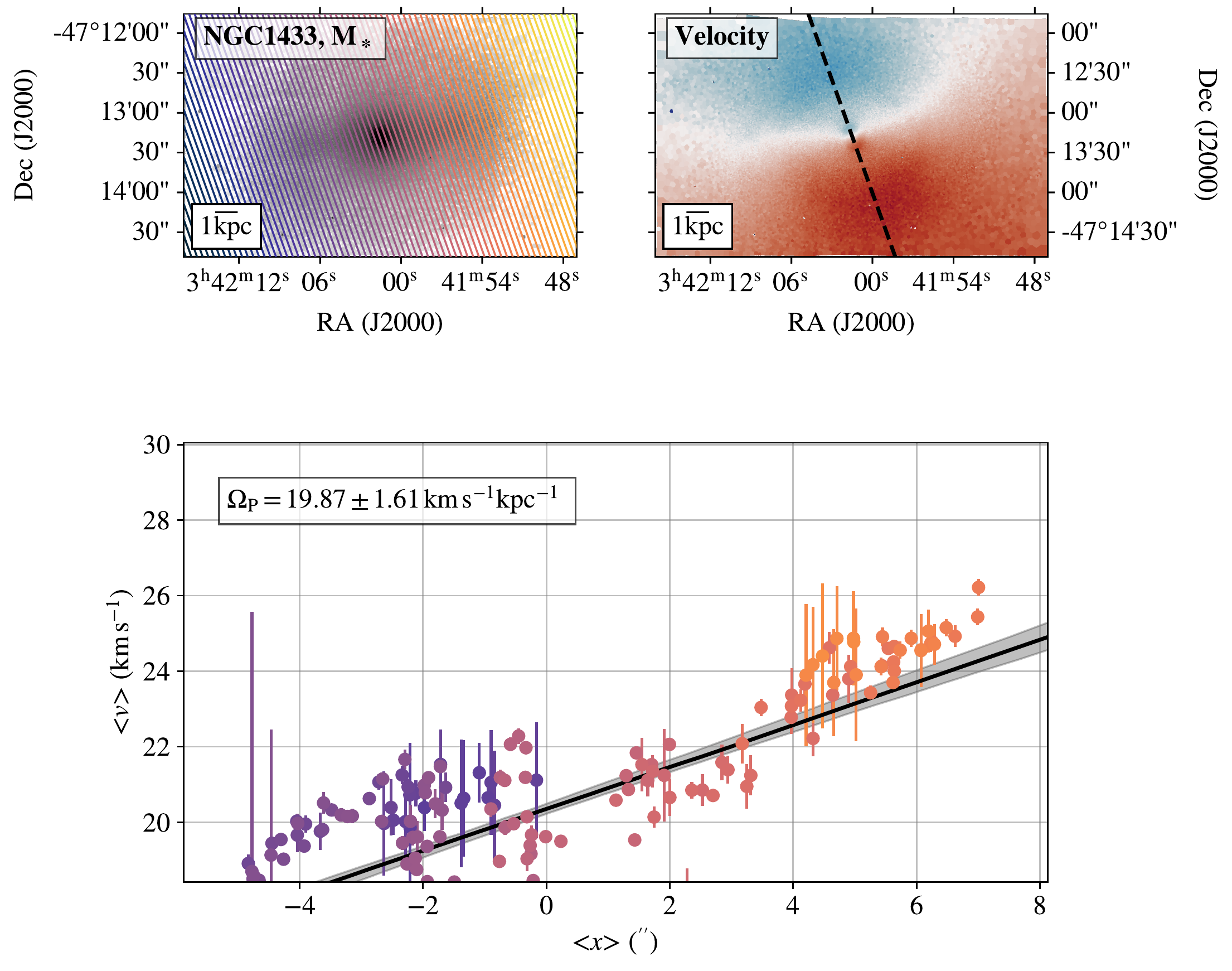}
\caption{As Fig. \ref{fig:ngc3351_tw_integral}, but for NGC1433. For this galaxy , $Q=1$. \label{fig:app_ngc1433}}
\end{figure*}

\begin{figure*}[t]
\plotone{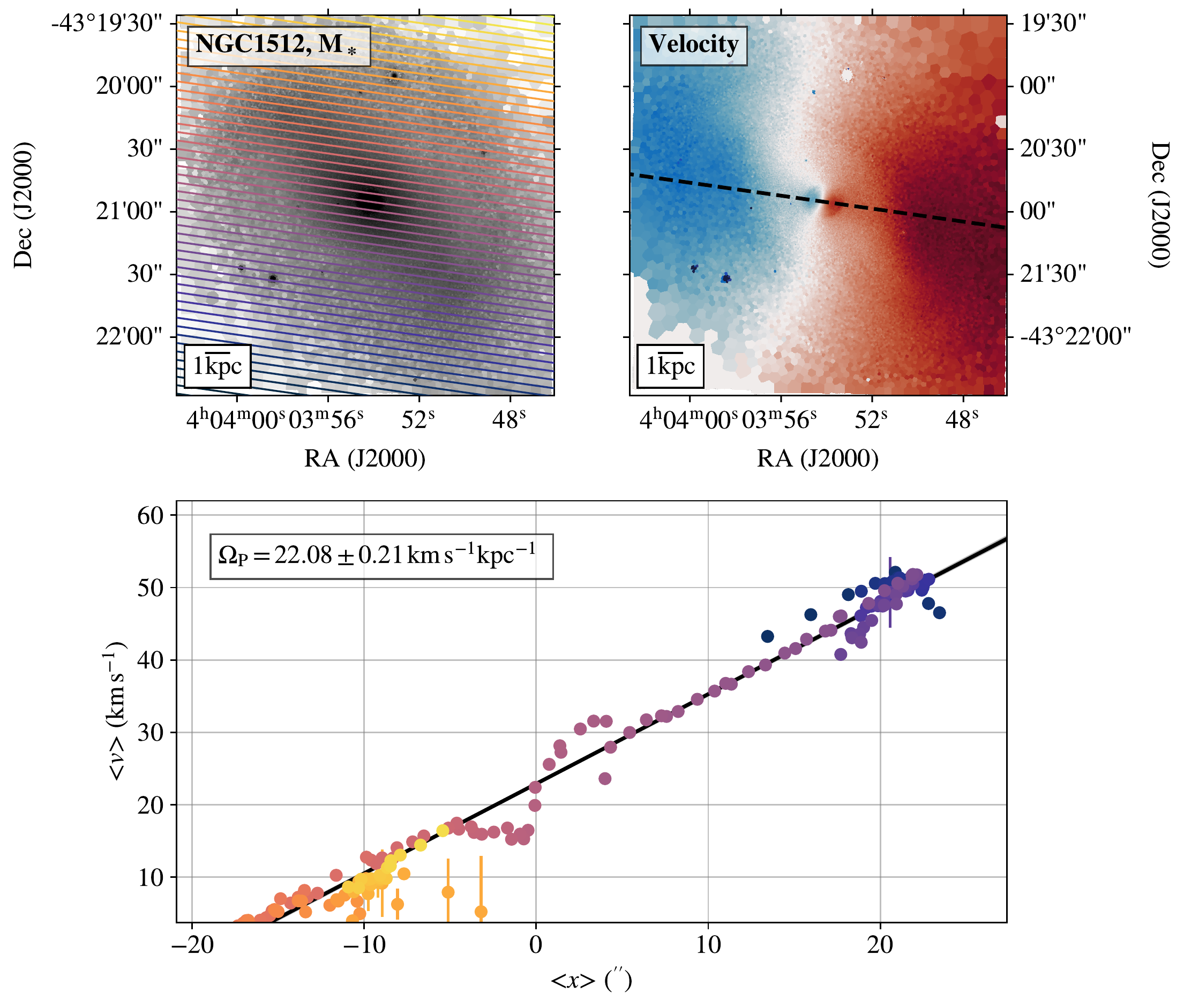}
\caption{As Fig. \ref{fig:ngc3351_tw_integral}, but for NGC1512. For this galaxy , $Q=1$. \label{fig:app_ngc1512}}
\end{figure*}

\begin{figure*}[t]
\plotone{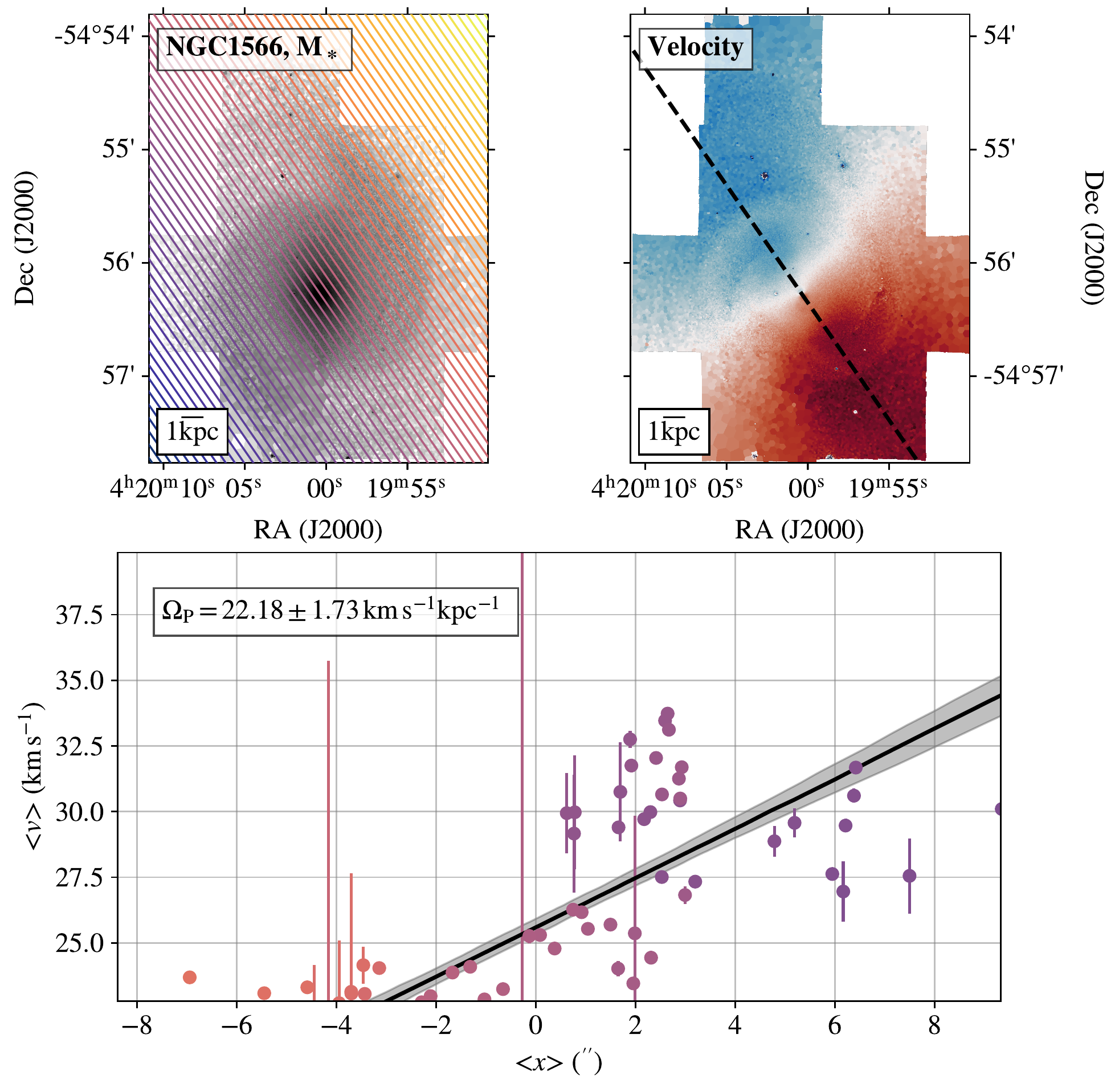}
\caption{As Fig. \ref{fig:ngc3351_tw_integral}, but for NGC1566. For this galaxy , $Q=3$. \label{fig:app_ngc1566}}
\end{figure*}

\begin{figure*}[t]
\plotone{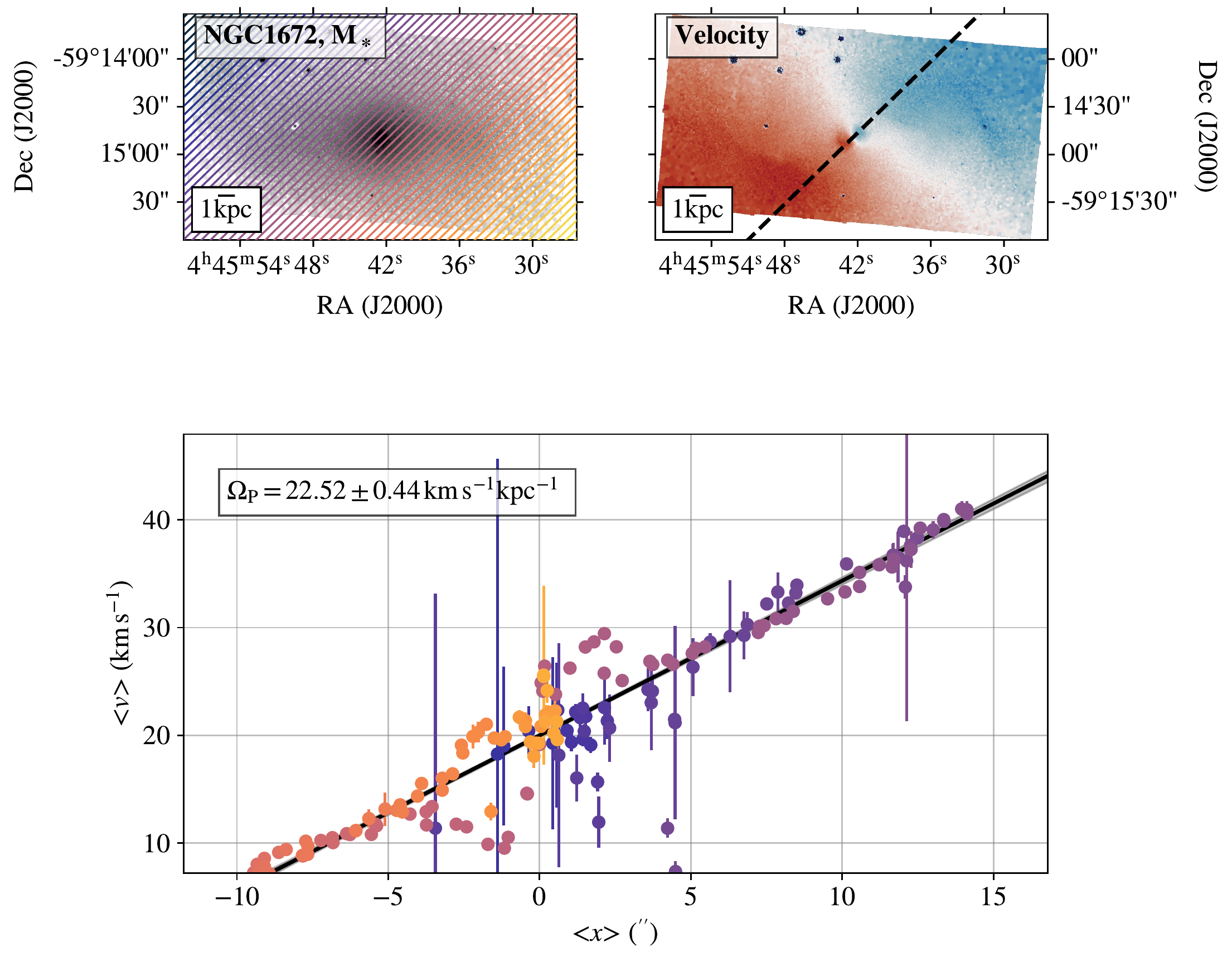}
\caption{As Fig. \ref{fig:ngc3351_tw_integral}, but for NGC1672. For this galaxy , $Q=1$. \label{fig:app_ngc1672}}
\end{figure*}

\begin{figure*}[t]
\plotone{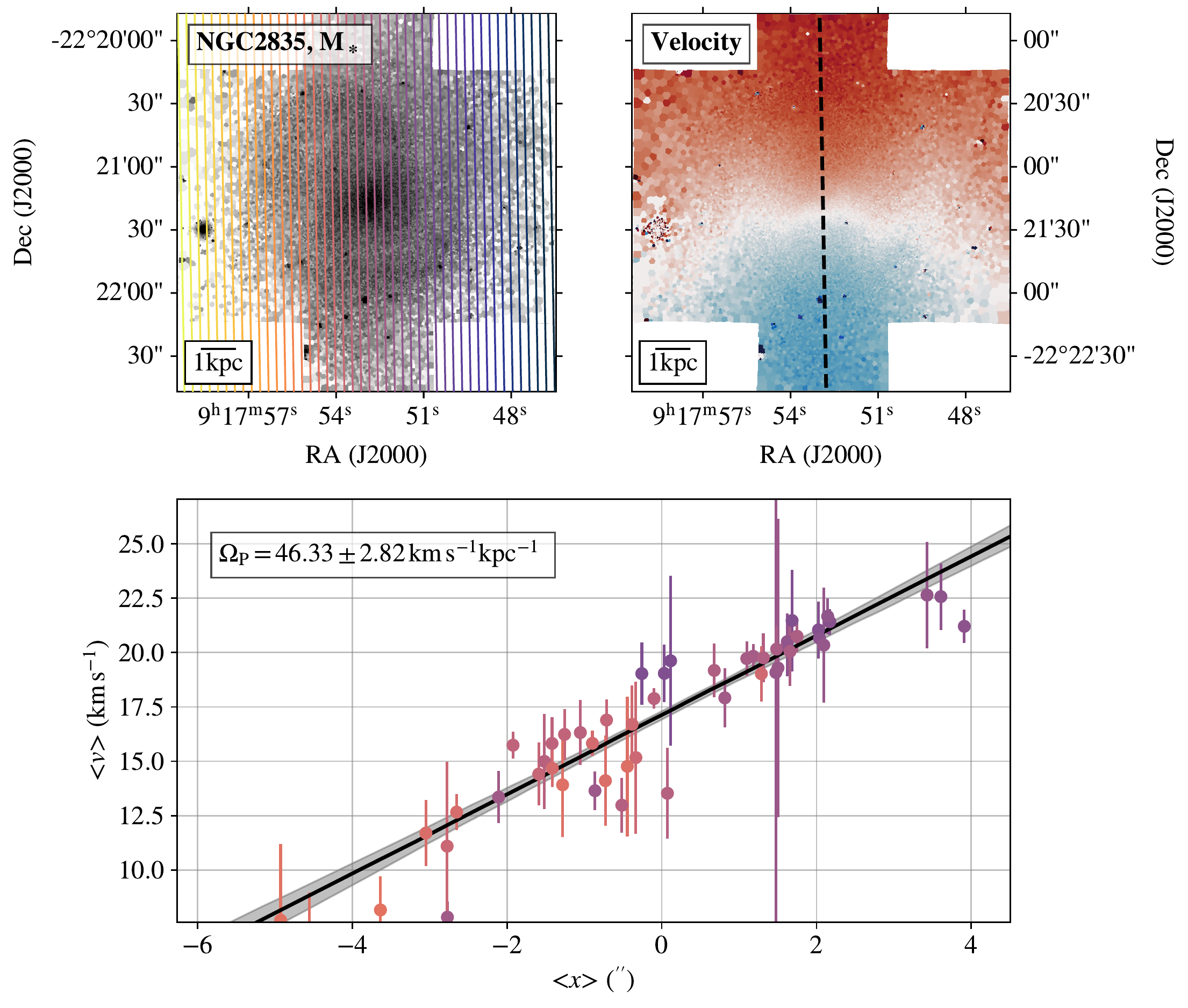}
\caption{As Fig. \ref{fig:ngc3351_tw_integral}, but for NGC2835. For this galaxy , $Q=1$. \label{fig:app_ngc2835}}
\end{figure*}

\begin{figure*}[t]
\plotone{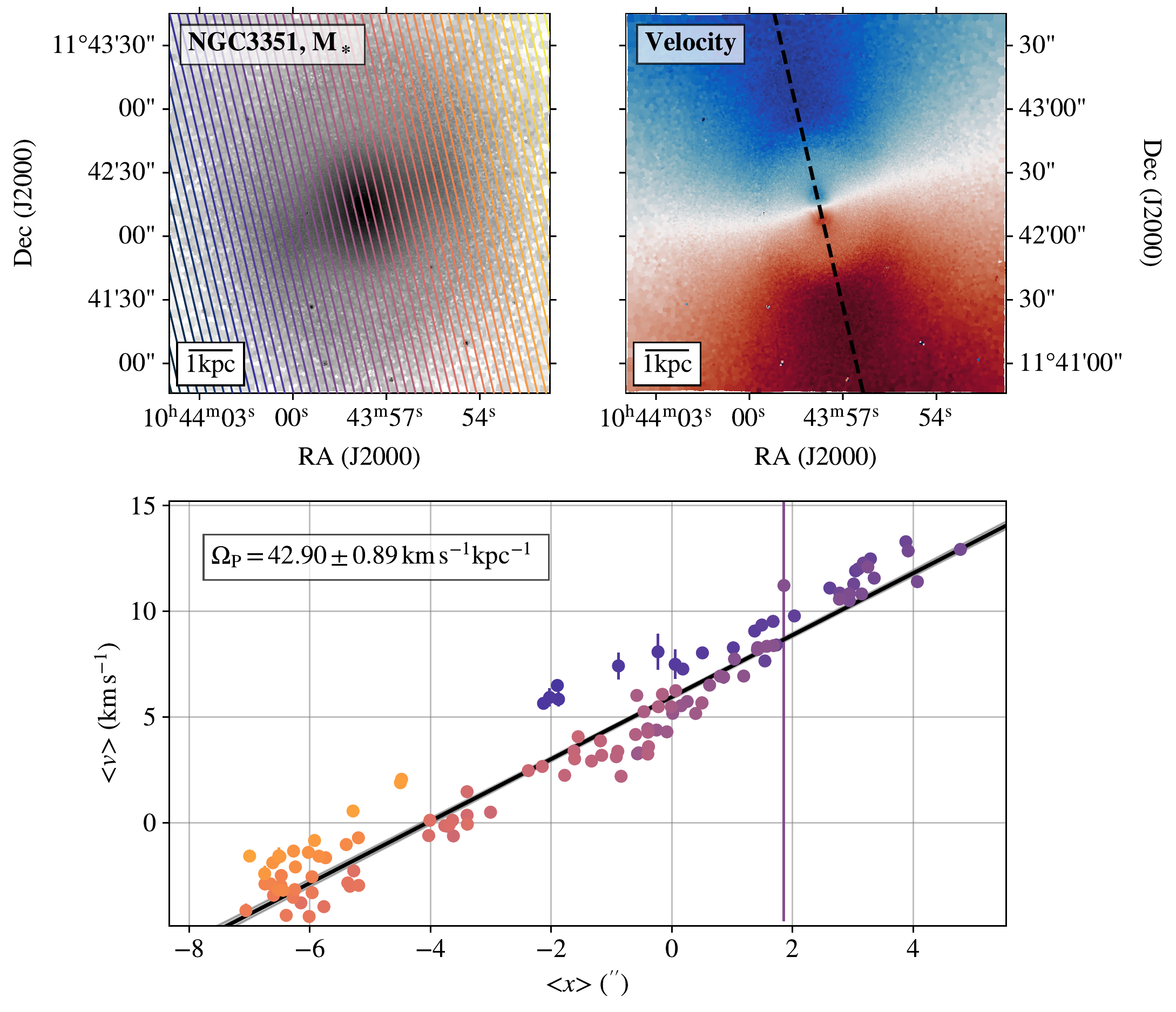}
\caption{As Fig. \ref{fig:ngc3351_tw_integral}, but for NGC3351. For this galaxy , $Q=1$. \label{fig:app_ngc3351}}
\end{figure*}

\begin{figure*}[t]
\plotone{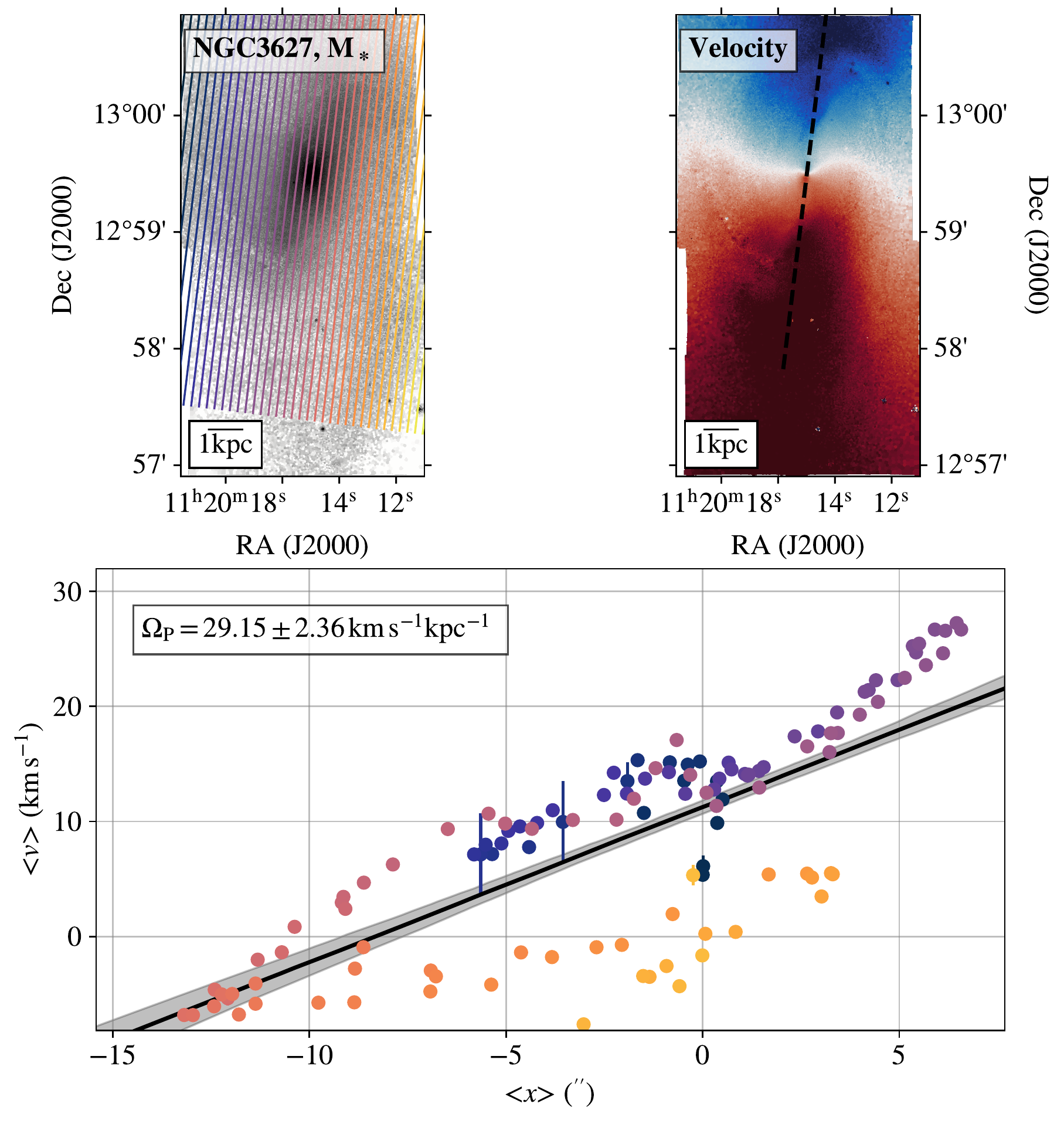}
\caption{As Fig. \ref{fig:ngc3351_tw_integral}, but for NGC3627. For this galaxy , $Q=2$. \label{fig:app_ngc3627}}
\end{figure*}

\begin{figure*}[t]
\plotone{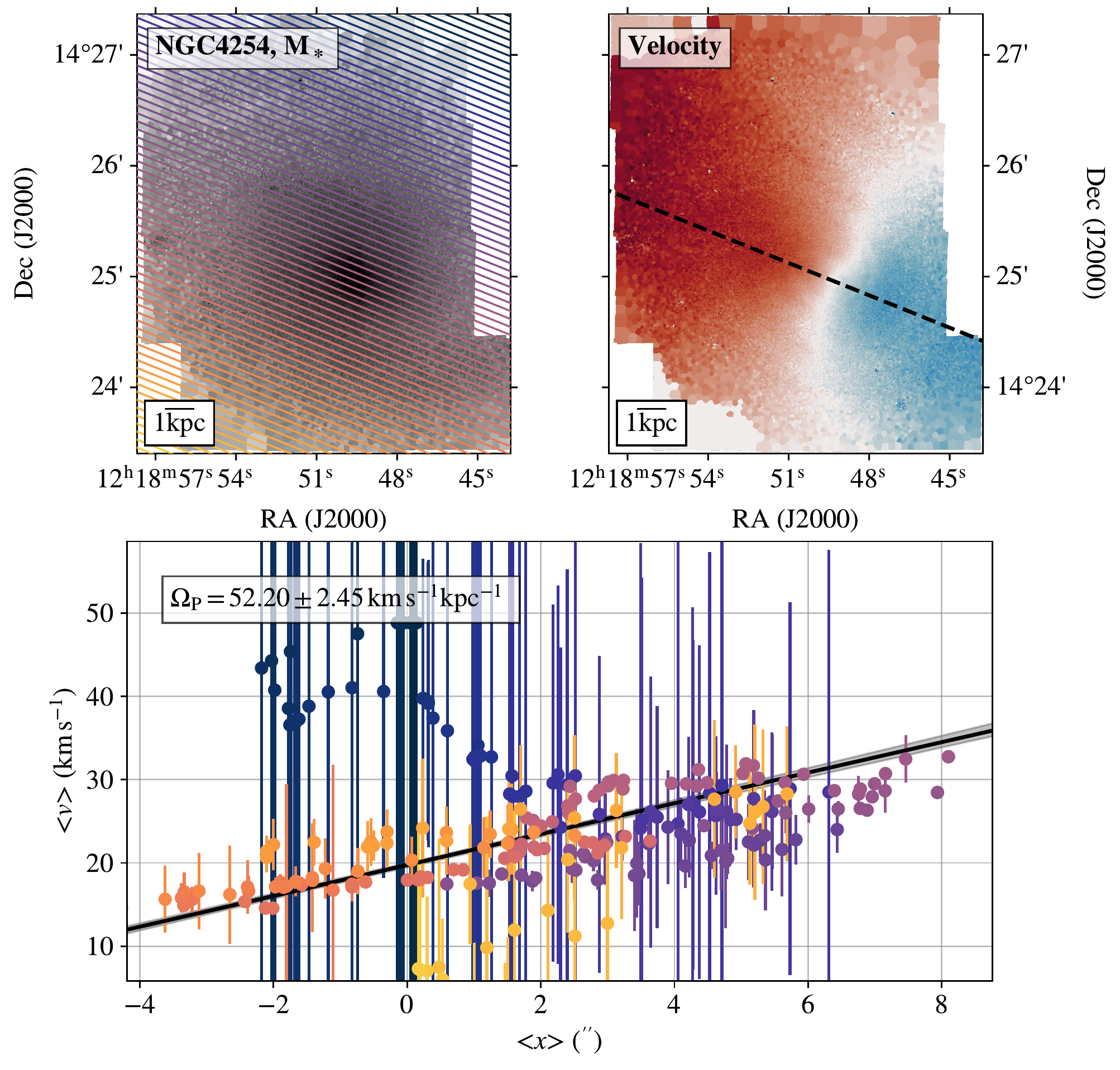}
\caption{As Fig. \ref{fig:ngc3351_tw_integral}, but for NGC4254. For this galaxy , $Q=3$. \label{fig:app_ngc4254}}
\end{figure*}

\begin{figure*}[t]
\plotone{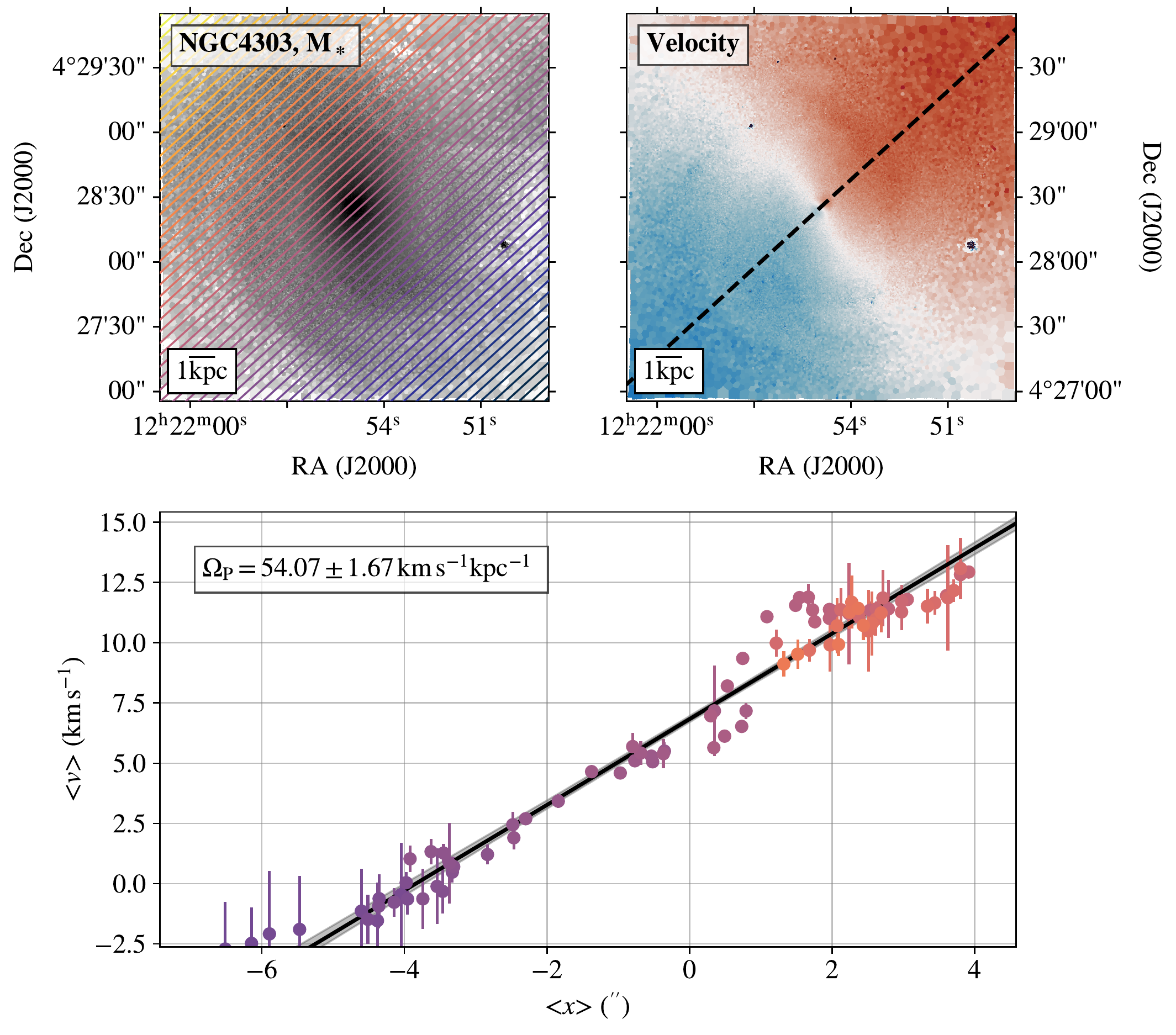}
\caption{As Fig. \ref{fig:ngc3351_tw_integral}, but for NGC4303. For this galaxy , $Q=1$. \label{fig:app_ngc4303}}
\end{figure*}

\begin{figure*}[t]
\plotone{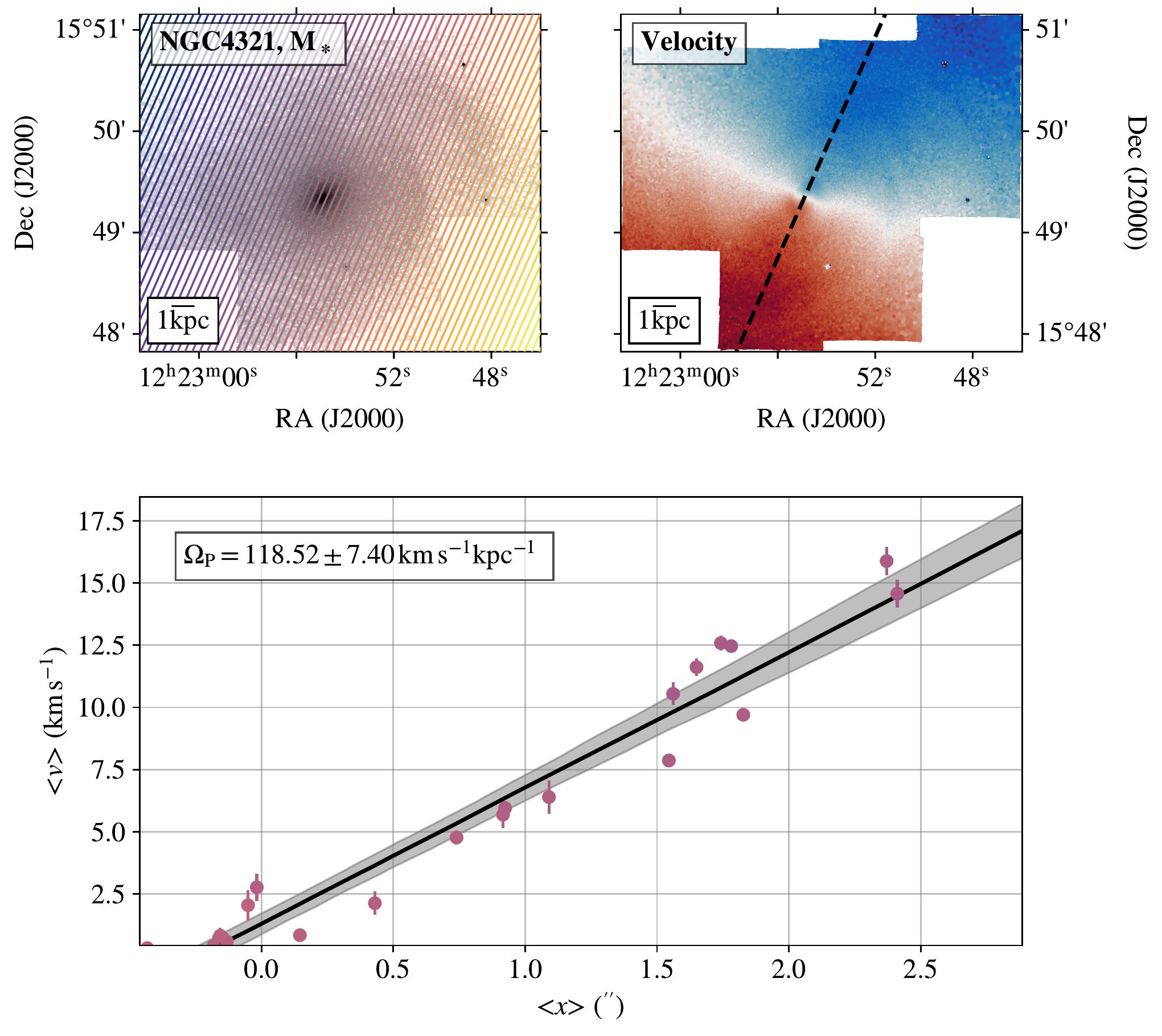}
\caption{As Fig. \ref{fig:ngc3351_tw_integral}, but for NGC4321. For this galaxy , $Q=4$. \label{fig:app_ngc4321}}
\end{figure*}

\begin{figure*}[t]
\plotone{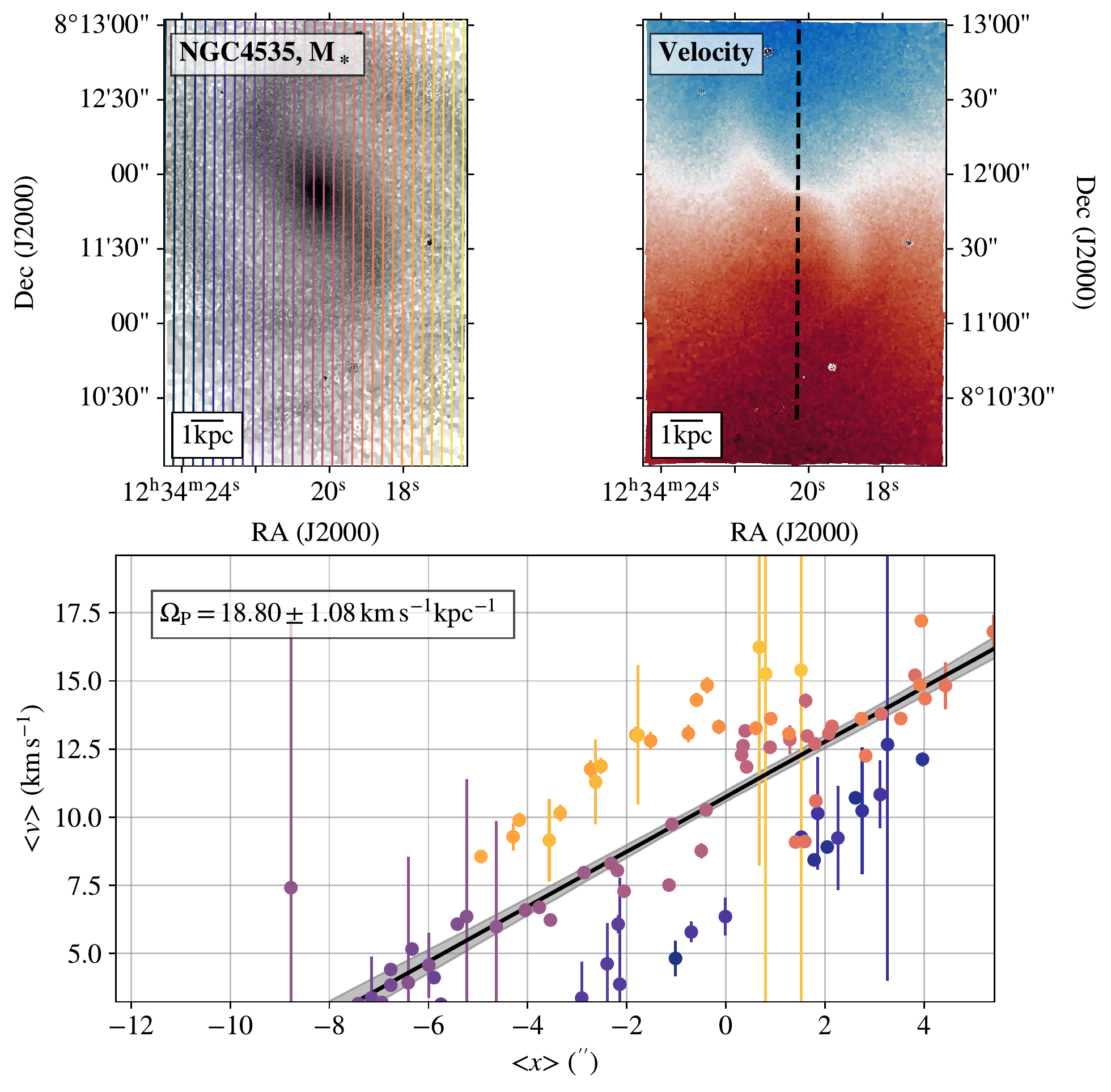}
\caption{As Fig. \ref{fig:ngc3351_tw_integral}, but for NGC4535. For this galaxy , $Q=3$. \label{fig:app_ngc4535}}
\end{figure*}

\begin{figure*}[t]
\plotone{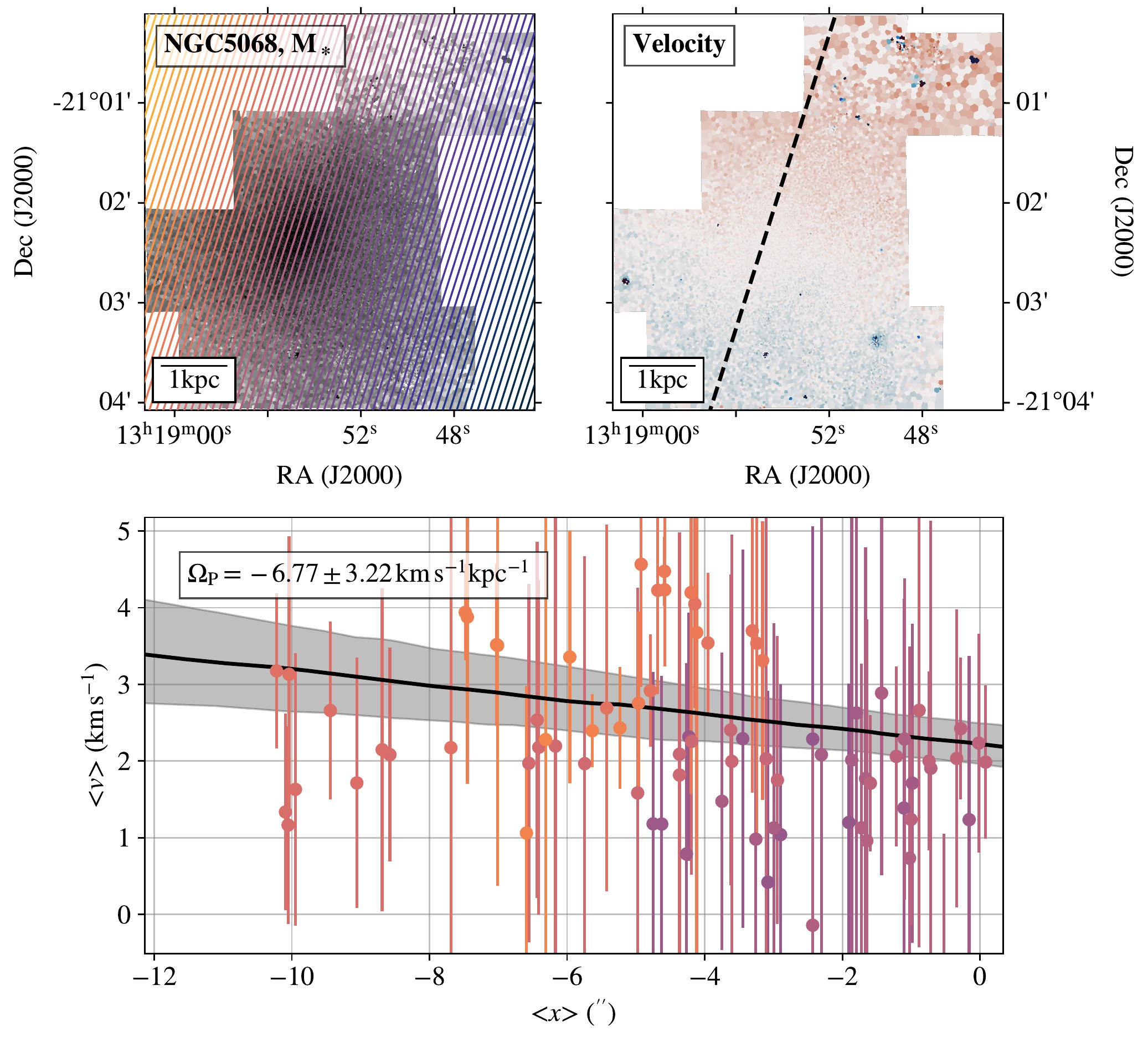}
\caption{As Fig. \ref{fig:ngc3351_tw_integral}, but for NGC5068. For this galaxy , $Q=3$. \label{fig:app_ngc5068}}
\end{figure*}

\begin{figure*}[t]
\plotone{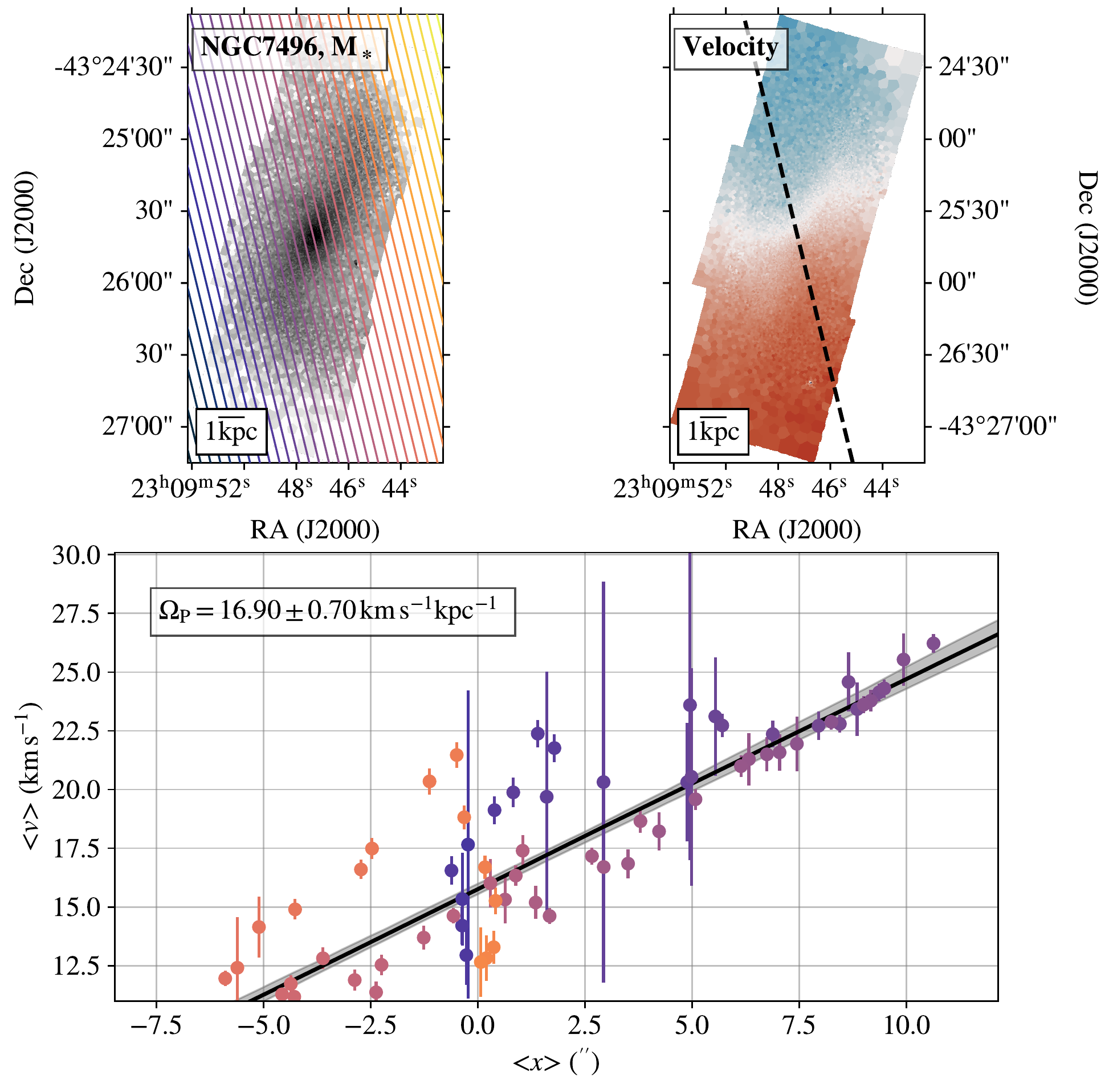}
\caption{As Fig. \ref{fig:ngc3351_tw_integral}, but for NGC7496. For this galaxy , $Q=1$. \label{fig:app_ngc7496}}
\end{figure*}

\end{document}